\documentclass[a4paper,11pt]{article}
\usepackage{jheppub} 
\usepackage{lineno}
\usepackage{feynmp-auto}
\usepackage{subfigure}
\usepackage{subcaption}
\usepackage{mathabx}
\usepackage{nicematrix}
\usepackage{fancyvrb}
\usepackage{snapshot}
\VerbatimFootnotes
\def\l{\langle}
\def\r{\rangle}

\title{\boldmath Multi-Higgs Boson Production with Anomalous Interactions at Current and Future Proton Colliders}

\author[1]{Andreas Papaefstathiou,}
\author[2,3]{Gilberto Tetlalmatzi-Xolocotzi,}
\affiliation[1]{Department of Physics, Kennesaw State University, 830 Polytechnic Lane, Marietta, GA 30060, USA}

\affiliation[2]{Universit\`e Paris-Saclay, CNRS/IN2P3, IJCLab, 91405 Orsay, France}
\affiliation[3]{Theoretische Physik 1, Center for Particle Physics Siegen (CPPS), Universit \"at Siegen, Walter-Flex-Str. 3, 57068 Siegen, Germany}

\emailAdd{apapaefs@cern.ch}
\emailAdd{gtx@physik.uni-siegen.de}

\abstract{We investigate multi-Higgs boson production at proton colliders, in a framework involving anomalous interactions, focusing on triple Higgs boson production. We consider modifications to the Higgs boson self-couplings, to the Yukawa interactions, as well as new contact interactions of Higgs bosons with either quarks or gluons. To this end, we have developed a \texttt{MadGraph5\_aMC@NLO} loop model, publicly available at~\cite{gitlabrepo}, designed to incorporate the relevant operators in the production of multiple Higgs bosons (and beyond). We have performed cross section fits at various energies over the anomalous interactions, and have derived constraints on the most relevant anomalous coefficients, through detailed phenomenological analyses at proton-proton collision energies of 13.6~TeV and 100~TeV, employing the 6 $b$-jet final state.} 

\begin{document}
\begin{flushright}

SI-HEP-2023-32\\

P3H-23-100

\end{flushright}

\maketitle
\flushbottom

\section{Introduction}\label{sec:intro}

Measuring the interactions of the Higgs boson~\cite{Higgs:1964pj,Englert:1964et,Guralnik:1964eu,ATLAS:2012yve,CMS:2012qbp} with the rest of the Standard Model (SM), with itself (i.e.\ its \textit{self-}interactions), and with potential new phenomena, is a crucial task for future collider experiments. There are multiple theoretical justifications for the importance of the Higgs field and how it interacts; for example, the Higgs doublet bilinear, $H^\dagger H$, is the only dimension-2 ($D=2$) singlet operator within the SM. This kind of operator readily allows for $D=3$ or $D=4$ interactions with one, or two, new scalar singlet fields, that may constitute part of a ``hidden sector'' of hitherto undetected particles. Therefore, the Higgs boson could be our primary window to this, potentially ``dark'', sector, that could comprise part, or all, of dark matter. Another reason of the importance of the Higgs boson's interactions is that they may illuminate the mechanism of baryogenesis -- in particular, if this process occurred during the electroweak phase transition, the period during which the Higgs field acquires a vacuum expectation value (VEV), then the Higgs field may be one of the protagonists of electroweak baryogenesis (see, e.g.~\cite{Dorsch:2013wja,No:2013wsa,Curtin:2014jma,Profumo:2014opa,White:2016nbo,Basler:2016obg,Bernon:2017jgv,Kurup:2017dzf,Dorsch:2017nza,Ramsey-Musolf:2019lsf,Li:2019tfd,Papaefstathiou:2020iag,Papaefstathiou:2021glr,Papaefstathiou:2022oyi,Zuk:2022qwx}). A central r\^ole in this mechanism is played by the Higgs potential, which, within the SM, is introduced in a rather ad-hoc manner. The form of this potential can, at least in principle, be directly probed by measuring the Higgs boson's self-interactions through multi-Higgs boson production processes: the triple self-coupling can be probed through Higgs boson pair production, and the quartic through triple production. 

Novel Higgs boson interactions, to new or to SM particles, may arise at the electroweak scale, but they may also appear at higher scales, $\mathcal{O}(\mathrm{few~TeV})$. If this is the case, we can parametrize our ignorance using a higher-dimensional effective field theory (EFT), see, e.g.~\cite{Buchmuller:1985jz,Grzadkowski:2010es,Elias-Miro:2013mua}. Neglecting lepton-number violating operators, the lowest-dimensionality EFT that can be written down consists of $D=6$ operators. Upon electroweak symmetry breaking, the Higgs boson would acquire a VEV and these operators would result in several new interactions, as well modifications of the SM interactions of the Higgs boson. Several of the operators in the physical basis of the Higgs boson scalar ($h$) would then have coefficients that are correlated, according to $D=6$ EFT. These correlations, however, may be broken by even higher-dimensional operators (e.g.\ $D=8$), particularly if the new phenomena are closer to the electroweak scale. Therefore, it may be beneficial to lean towards a more agnostic, and hence more phenomenological, approach and, while still remaining inspired by $D=6$ EFT, consider fully uncorrelated, ``anomalous'' interactions of the Higgs boson with the SM. These kind of interactions can also be achieved within the framework of the electroweak chiral Lagrangian including a light Higgs boson, where free coefficients for all Higgs couplings are obtained~\cite{Feruglio:1992wf,Bagger:1993zf,Koulovassilopoulos:1993pw,Burgess:1999ha,Wang:2006im,Grinstein:2007iv,Alonso:2012px,Buchalla:2013eza,Buchalla:2012qq,Buchalla:2013rka,Buchalla:2013eza,Buchalla:2017jlu}. This framework provides a consistent EFT justification of the usual ``$\kappa$'' formalism. The description of the Higgs boson's interactions in terms of uncorrelated coefficients is the approach that we pursue here. 

Within the realm of self-coupling measurements, and beyond, Higgs boson pair production has received considerable attention from both the experimental (e.g.~\cite{CMS:2017yfv,CMS:2017hea,CMS:2017rpp,CMS:2018tla,CMS:2018vjd,CMS:2018sxu,CMS:2018ipl,CMS:2020tkr,CMS:2022cpr,CMS:2022hgz,CMS:2022kdx,CMS:2022omp,ATLAS:2014pjm,ATLAS:2015zug,ATLAS:2015sxd,ATLAS:2016paq,ATLAS:2018rnh,ATLAS:2018dpp,ATLAS:2018hqk,ATLAS:2018uni,ATLAS:2018fpd,ATLAS:2018ili,ATLAS:2019qdc,ATLAS:2019vwv,ATLAS:2020jgy,ATLAS:2021ifb,ATLAS:2022xzm,ATLAS:2022jtk,ATLAS:2023qzf,ATLAS:2023gzn}) and theoretical (e.g.~\cite{Baur:2002rb,Baur:2002qd,Baur:2003gp,Dolan:2012ac,Papaefstathiou:2012qe,Cao:2013si,Goertz:2013kp,Arbey:2013jla,deFlorian:2013uza,Gupta:2013zza,Ellwanger:2013ova,Barr:2013tda,Maierhofer:2013sha,deFlorian:2013jea,Dolan:2013rja,Goertz:2013eka,Frederix:2014hta,Baglio:2014nea,FerreiradeLima:2014qkf,deFlorian:2014rta,Hespel:2014sla,Barger:2014taa,Godunov:2014waa,Liu:2014rba,Maltoni:2014eza,Chen:2014ask,Barr:2014sga,MartinLozano:2015vtq,Papaefstathiou:2015iba,Dawson:2015oha,Kotwal:2015rba,Lu:2015jza,Carvalho:2015ttv,Cao:2015oxx,Batell:2015koa,Dawson:2015haa,Cao:2015oaa,Kanemura:2016tan,Contino:2016spe,Cao:2016zob,Banerjee:2016nzb,Huang:2017jws,Nakamura:2017irk,Lewis:2017dme,DiLuzio:2017tfn,Grober:2017gut,Zurita:2017sfg,Arganda:2017wjh,Adhikary:2017jtu,Bauer:2017cov,Maltoni:2018ttu,Borowka:2018pxx,Goncalves:2018qas,Chang:2018uwu,Basler:2018dac,Adhikary:2018ise,DiMicco:2019ngk,Li:2019uyy,Cheung:2020xij}) point of view, particularly following the discovery of the Higgs boson. Owing to its much smaller SM cross section at hadron colliders~\cite{Maltoni:2014eza}, triple Higgs boson has understandably received much less attention. To the extent of our knowledge, it was first investigated in~\cite{Plehn:2005nk}, where it was demonstrated that the prospects for the measurement of the quartic self-coupling will be very challenging, both at the LHC and a future `VLHC' at a center-of-mass energy of 200 TeV. Subsequent studies of triple Higgs boson production at future colliders~\cite{Papaefstathiou:2015paa,Chen:2015gva,Fuks:2015hna,Papaefstathiou:2017hsb,Fuks:2017zkg,Liu:2018peg,Papaefstathiou:2019ofh,deFlorian:2019app,Chiesa:2020awd,Abdughani:2020xfo}, with the knowledge of the value of the Higgs boson mass, and the prospects for Higgs boson pair production measurements at hand, further quantified the difficulty of this measurement, nevertheless demonstrating that \textit{some} useful information may be obtained via the process. The situation at LHC energies is much more dire, as expected, requiring either new phenomena for sufficient enhancement~\cite{Papaefstathiou:2020lyp}, or very large new contributions to the Higgs boson's self-interactions~\cite{Stylianou:2023xit}. Up to now, all of the phenomenological studies addressing triple Higgs boson production, apart from discussions of the cross section modifications in an EFT in~\cite{Degrande:2020evl}, have considered only the effects of modifications of the self-couplings. However, due to their loop-induced nature, proceeding via heavy quark loops, multi-Higgs boson production processes are highly sensitive to modifications of the heavy quark Yukawa couplings, as well as to additional quark- or gluon-Higgs boson contact interactions that generically arise in EFTs. Such effects have been studied within the context of Higgs boson pair production, see, e.g.~\cite{Goertz:2014qta,Azatov:2015oxa}. In the present article we study, for the first time to the best of our knowledge, the combined effect of modifications to the self-couplings, Yukawa interactions, and new heavy-quark- or gluon-Higgs boson contact interactions in triple Higgs boson production, within the aforementioned EFT-inspired anomalous coupling picture. To achieve this, we have constructed a specialized anomalous-coupling model within the \texttt{MadGraph5\_aMC@NLO} event generator that incorporates the full interference effects between contributing diagrams.

The paper is organized as follows: in section~\ref{sec:phenolagrangian} we introduce the phenomenological Lagrangian that we employ in our study, inspired by an EFT extension of the  SM. In section~\ref{sec:mc} we discuss the details of the Monte Carlo implementation of the \texttt{MadGraph5\_aMC@NLO} model, briefly addressing its usage for Higgs boson pair and triple production. In section~\ref{sec:pheno} we discuss our phenomenological analysis of triple Higgs boson production, focusing on cross section fits, existing constraints on the anomalous couplings, and the extraction of limits on the most relevant coefficients through the 6 $b$-jet final state. We present our conclusions and outlook in section~\ref{sec:conclusions}. In addition, in appendix~\ref{app:d6eftL} we describe the $D=6$ EFT Lagrangian relevant to Higgs boson production processes, i.e.\ the the linear EFT approach, and in appendix~\ref{app:chiral} we discuss the nonlinear approach through the electroweak chiral Lagrangian. In appendix~\ref{app:validation} we discuss aspects of the validation of the Monte Carlo model. In appendix~\ref{app:feyndiags} we discuss relevant Feynman diagrams that appear within our framework, up to two operator insertions. Finally, in appendix~\ref{app:triplelimits} we discuss limits obtained with up to three powers of anomalous operators (at the matrix element-squared level) instead of the two presented in the main text, and in appendix~\ref{app:fits} we present supplementary cross section fits at various proton collision energies.

\section{Phenomenological Lagrangian}\label{sec:phenolagrangian}

To allow for additional freedom, we have modified the $D=6$ EFT Lagrangian relevant to Higgs boson processes, derived in detail in appendix~\ref{app:d6eftL} (eq.~\ref{eq:Lgghmulti}), breaking the correlations between several of the interactions that originate in the $D=6$ EFT. This yields more flexibility when exploring multi-Higgs boson production experimentally. Nevertheless, $D=6$ EFT can be trivially restored by reinstating the appropriate relations between the coefficients (see below). We also provide additional EFT justification for this approach, inspired by the electroweak chiral Lagrangian including a light Higgs boson in appendix~\ref{app:chiral}.

To achieve this, we have defined new anomalous couplings $d_3$ and $d_4$, as modifications of the triple and quartic self-couplings, $c_{g1}$ and $c_{g2}$ as interactions of two gluons and one or two Higgs bosons, respectively, and $c_{f1}$, $c_{f2}$ and $c_{f3}$ as interactions between fermion-anti-fermion pairs, and one, two or three Higgs bosons, respectively, leading to the following phenomenological Lagrangian:
\begin{equation}\label{eq:LgghhPheno}
\begin{split}
\mathcal{L}_\mathrm{Pheno} = &- \frac{ m_h^2 } { 2 v } \left( 1 + d_3 \right) h^3 - \frac{ m_h^2 } { 8 v^2 }  \left( 1  + d_4 \right) h^4
\\
&+ \frac{\alpha_s } {4 \pi } \left( c_{g1} \frac{h}{v} + c_{g2} \frac{h^2} {2v^2} \right) G_{\mu\nu}^a G^{\mu\nu}_a
\\
&- \left[ \frac{m_t}{v} \left( 1 + c_{t1}   \right) \bar{t}_L t_R h +  \frac{m_b}{v} \left( 1 + c_{b1}   \right) \bar{b}_L b_R h + \text{h.c.} \right]
\\
&- \left[ \frac{m_t}{v^2}\left( \frac{3 c_{t2}}{2} \right) \bar{t}_L t_R h^2 +  \frac{m_b}{v^2}\left( \frac{3 c_{b2}}{2}  \right) \bar{b}_L b_R h^2 +  \text{h.c.} \right] 
\\
&- \left[ \frac{m_t}{v^3}\left( \frac{c_{t3}}{2} \right) \bar{t}_L t_R h^3 +  \frac{m_b}{v^3}\left( \frac{c_{b3}}{2}  \right) \bar{b}_L b_R h^3 +  \text{h.c.} \right].
\end{split}
\end{equation}
To recover the $D=6$ EFT Lagrangian relevant to the Higgs sector (i.e.\ eq.~\ref{eq:Lgghmulti}), one would need to simply set: $d_3 = c_6$, $d_4 = 6 c_6$, $c_{g1} = c_{g2}$, $c_{f1} = c_{f2} = c_{f3}$. 

The implementation of this study further modifies the above phenomenological Lagrangian to match more closely the LHC experimental collaboration definitions:
\begin{equation}\label{eq:LgghhPhenoExp}
\begin{split}
\mathcal{L}_\mathrm{PhenoExp} = &- \lambda_\mathrm{SM} v \left( 1 + d_3 \right) h^3 - \frac{ \lambda_\mathrm{SM} } { 4 }  \left( 1  + d_4 \right) h^4
\\
&+ \frac{\alpha_s } {12 \pi } \left( c_{g1} \frac{h}{v} - c_{g2} \frac{h^2} {2v^2} \right) G_{\mu\nu}^a G^{\mu\nu}_a
\\
&- \left[ \frac{m_t}{v} \left( 1 + c_{t1}   \right) \bar{t}_L t_R h +  \frac{m_b}{v} \left( 1 + c_{b1}   \right) \bar{b}_L b_R h + \text{h.c.} \right]
\\
&- \left[ \frac{m_t}{v^2} c_{t2} \bar{t}_L t_R h^2 +  \frac{m_b}{v^2} c_{b2} \bar{b}_L b_R h^2 +  \text{h.c.} \right] 
\\
&- \left[ \frac{m_t}{v^3}\left( \frac{c_{t3}}{2} \right) \bar{t}_L t_R h^3 +  \frac{m_b}{v^3}\left( \frac{c_{b3}}{2}  \right) \bar{b}_L b_R h^3 +  \text{h.c.} \right],
\end{split}
\end{equation}
where we have taken $\lambda_\mathrm{SM} \equiv m_h^2 / 2v^2$. 

The CMS parametrization is then obtained by setting: $\kappa_\lambda = (1+d_3)$, $k_t = c_{t1}$, $c_2 = c_{t2}$, $c_{g} = c_{g1}$, $c_{gg} = c_{2g}$ and the ATLAS parametrization by $c_{hhh} = (1+d_3)$, $c_{ggh} = 2 c_{g1}/3$, $c_{gghh} = - c_{g2}/3$. The Lagrangian of eq.~\ref{eq:LgghhPhenoExp} encapsulates the form of the interactions that we employ for the rest of our phenomenological analysis. 

\section{Monte Carlo Event Generation}\label{sec:mc}

\subsection{Loop $\times$ Tree-Level Interference}
The implementation of the Lagrangian of eq.~\ref{eq:LgghhPhenoExp} in \texttt{MadGraph5\_aMC@NLO} (\texttt{MG5\_aMC})~\cite{Alwall:2011uj,Hirschi:2015iia} follows closely the instructions for proposed code modifications found in~\cite{loopxtree}.\footnote{As suggested by Valentin Hirschi.} These modifications essentially introduce tree-level diagrams in the form of ``UV counter-terms", that are generated along with any loop-level diagrams, allowing the calculation of interference terms between them. The model of the present article, created through this procedure, has been fully validated by direct comparison to an implementation of Higgs boson pair production in $D=6$ EFT in the \texttt{HERWIG 7} Monte Carlo, and by taking the limit of a heavy scalar boson for those vertices that do not appear in that process. See appendix~\ref{app:validation} for further details of the latter effort. The necessary modifications to the \texttt{MG5\_aMC} codebase,\footnote{At present available for versions 2.9.15 and 3.5.0.} as well as the model can be found in the public gitlab repository at~\cite{gitlabrepo}.\footnote{It is interesting to note here that there exists a more comprehensive \texttt{MG5\_aMC} treatment of one-loop computations in the standard-model effective field theory at $D=6$ (dubbed ``smeft@nlo'')~\cite{Degrande:2020evl}, which should directly map to the $D=6$ limit of the present article.}

\subsection{Higgs Boson Pair Production}

Higgs boson pair production contains a subset of all diagrams relevant to triple Higgs boson production, minus one Higgs boson insertion, and therefore we begin by investigating this process, so as to provide a simple example. 

\texttt{MG5\_aMC} allows for calculation of the interference terms via one insertion of the operator at the squared matrix-element level ($|\mathcal{M}|^2$):
\begin{verbatim}
generate g g > h h [noborn=MHEFT QCD] MHEFT^2<=1
\end{verbatim}
i.e. schematically: $|\mathcal{M}|^2 \sim \bullet 1 + \sqbullet c_i$. The classes of diagrams that are included in addition to the SM ones (fig.~\ref{fig:hh_sm}), are those that appear in fig.~\ref{fig:eft1insert}, found in appendix~\ref{app:feyndiags}.

Alternatively, one can obtain: the interference terms of one or two insertions with the SM, and the squares of one-insertion terms (equivalent to two powers of the anomalous couplings) via:
\begin{verbatim}
generate g g > h h [noborn=MHEFT QCD] MHEFT^2<=2
\end{verbatim}
where schematically this would correspond to $|\mathcal{M}|^2 \sim \bullet 1 + \sqbullet c_i + \star c_i c_j$. The classes of diagrams of fig.~\ref{fig:eft2insert} (appendix~\ref{app:feyndiags}) appear in addition.

\begin{figure}[!htp]
\centering
\setlength{\unitlength}{0.6cm}
\subfigure{
\begin{fmffile}{box}
    \begin{fmfgraph*}(7,4)
        \fmfleft{i1,i2}
        \fmfright{o1,o2}
        \fmf{gluon}{i1,v1}
        \fmflabel{$g$}{i1}
        \fmflabel{$g$}{i2}
        \fmflabel{$h$}{o2}
        \fmflabel{$h$}{o1}
        \fmf{gluon}{i2,v2}
        \fmf{fermion,label=$q$}{v2,v3}
        \fmf{fermion}{v4,v1}
        \fmf{fermion}{v1,v2}
        \fmf{fermion}{v3,v4}
        \fmf{dashes}{v4,o1}
        \fmf{dashes}{v3,o2}
    \end{fmfgraph*}
\end{fmffile}
}
\subfigure{
\begin{fmffile}{triangle}
    \begin{fmfgraph*}(7,4)
        \fmfleft{i1,i2}
        \fmfright{o1,o2}
        \fmflabel{$g$}{i1}
        \fmflabel{$g$}{i2}
        \fmflabel{$h$}{o2}
        \fmflabel{$h$}{o1}
        \fmf{gluon,tension=2}{i1,v1}
        \fmf{gluon,tension=2}{i2,v2}
        \fmf{fermion,label=$q$}{v2,v3}
        \fmf{fermion}{v3,v1}
        \fmf{fermion}{v1,v2}
        \fmf{dashes,tension=3}{v3,v4}
        \fmf{dashes}{v4,o1}
        \fmf{dashes}{v4,o2}
    \end{fmfgraph*}
\end{fmffile}
}

\caption{Example Feynman diagrams for leading-order gluon-fusion Higgs boson pair production in the Standard Model.}
\label{fig:hh_sm}
\end{figure}
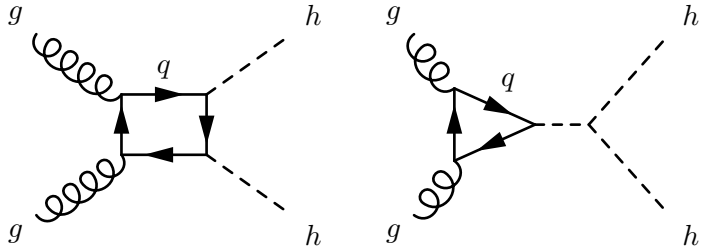

\subsection{Triple Higgs Boson Production}

For the investigation of triple Higgs boson production, which constitutes the focus of this article, we included the interference terms of one or two insertions with the SM and the squares of one-insertion terms through:
\begin{verbatim}
generate g g > h h h [noborn=MHEFT QCD] MHEFT^2<=2
\end{verbatim}
The results with up to three powers of the anomalous operators, more appropriate within the framework of the electroweak chiral Lagrangian (appendix~\ref{app:chiral}), are presented in appendix~\ref{app:triplelimits}. 

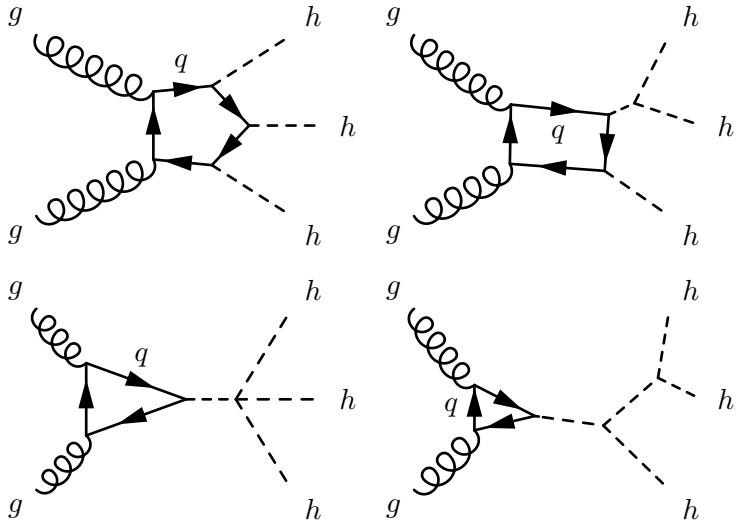
\begin{figure}[!htp]
\centering
\setlength{\unitlength}{0.6cm}
\subfigure{
\begin{fmffile}{pentagon}
    \begin{fmfgraph*}(7,4)
        \fmfleft{i1,i2}
        \fmfright{o1,o2,o3}
        \fmf{gluon}{i1,v1}
        \fmflabel{$g$}{i1}
        \fmflabel{$g$}{i2}
        \fmflabel{$h$}{o3}
        \fmflabel{$h$}{o2}
        \fmflabel{$h$}{o1}
        \fmf{gluon}{i2,v2}
    \fmf{fermion,label=$q$,tension=2}{v2,v3}
        \fmf{fermion}{v3,v4}
        \fmf{fermion}{v4,v5}
        \fmf{fermion,tension=2}{v5,v1}  
        \fmf{fermion}{v1,v2}
        \fmf{dashes}{v5,o1}
        \fmf{dashes}{v4,o2}
        \fmf{dashes}{v3,o3}
    \end{fmfgraph*}
\end{fmffile}
}
\subfigure{
\begin{fmffile}{box_hhh}
    \begin{fmfgraph*}(7,4)
        \fmfleft{i1,i2}
        \fmfright{o1,o2,o3}
        \fmf{gluon}{i1,v1}
        \fmflabel{$g$}{i1}
        \fmflabel{$g$}{i2}
        \fmflabel{$h$}{o2}
        \fmflabel{$h$}{o1}
        \fmflabel{$h$}{o3}
        \fmf{gluon}{i2,v2}
        \fmf{fermion,label=$q$}{v2,v3}
        \fmf{fermion}{v4,v1}
        \fmf{fermion}{v1,v2}
        \fmf{fermion}{v3,v4}
        \fmf{dashes,tension=1.4}{v4,o1}
        \fmf{dashes,tension=4}{v3,v5}
        \fmf{dashes}{v5,o2}
        \fmf{dashes}{v5,o3}
    \end{fmfgraph*}
\end{fmffile}
}
\par\bigskip
\par\bigskip
\subfigure{
\begin{fmffile}{triangle_hhh}
    \begin{fmfgraph*}(7,4)
        \fmfleft{i1,i2}
        \fmfright{o1,o2,o3}
        \fmflabel{$g$}{i1}
        \fmflabel{$g$}{i2}
        \fmflabel{$h$}{o3}
        \fmflabel{$h$}{o2}
        \fmflabel{$h$}{o1}
        \fmf{gluon,tension=2}{i1,v1}
        \fmf{gluon,tension=2}{i2,v2}
        \fmf{fermion,label=$q$}{v2,v3}
        \fmf{fermion}{v3,v1}
        \fmf{fermion}{v1,v2}
        \fmf{dashes,tension=4}{v3,v4}
        \fmf{dashes}{v4,o1}
        \fmf{dashes}{v4,o2}
        \fmf{dashes}{v4,o3}
    \end{fmfgraph*}
\end{fmffile}
}
\subfigure{
\begin{fmffile}{triangle_hhh2}
    \begin{fmfgraph*}(7,4)
        \fmfleft{i1,i2}
        \fmfright{o1,o2,o3}
        \fmflabel{$g$}{i1}
        \fmflabel{$g$}{i2}
        \fmflabel{$h$}{o3}
        \fmflabel{$h$}{o2}
        \fmflabel{$h$}{o1}
        \fmf{gluon,tension=1}{i1,v1}
        \fmf{gluon,tension=1}{i2,v2}
        \fmf{fermion}{v2,v3}
        \fmf{fermion}{v3,v1}
        \fmf{fermion,label=$q$}{v1,v2}
        \fmf{dashes,tension=15}{v3,v4}
        \fmf{dashes,tension=2}{v4,v5}
         \fmf{dashes}{v5,v6}
        \fmf{dashes}{v5,o1}
        \fmf{dashes}{v6,o2}
        \fmf{dashes}{v6,o3}
    \end{fmfgraph*}
\end{fmffile}
}

\caption{Example Feynman diagrams for leading-order gluon-fusion Higgs boson triple production in the Standard Model.}
\label{fig:hhh_sm}
\end{figure}

As in the case of Higgs boson pair production, our current treatment would schematically correspond to $|\mathcal{M}|^2 \sim \bullet 1 + \sqbullet c_i + \star c_i c_j$. The classes of the various diagrams that appear are shown in fig.~\ref{fig:hhh_sm} for the SM, and in figs.~\ref{fig:eft1insert_hhh} for one insertion, and~\ref{fig:eft2insert_hhh} for two insertions, in appendix~\ref{app:feyndiags}.\footnote{We note here that a different number of insertions can be achieved by modifying the \verb!MHEFT^2<=! part of the command. The maximum number of possible insertions depends on the process considered.}

\section{Phenomenology of Triple Higgs Boson Production}\label{sec:pheno}

\subsection{Cross Section Fits}

\begin{figure}[htp]
\begin{center}
  \includegraphics[width=1.0\columnwidth]{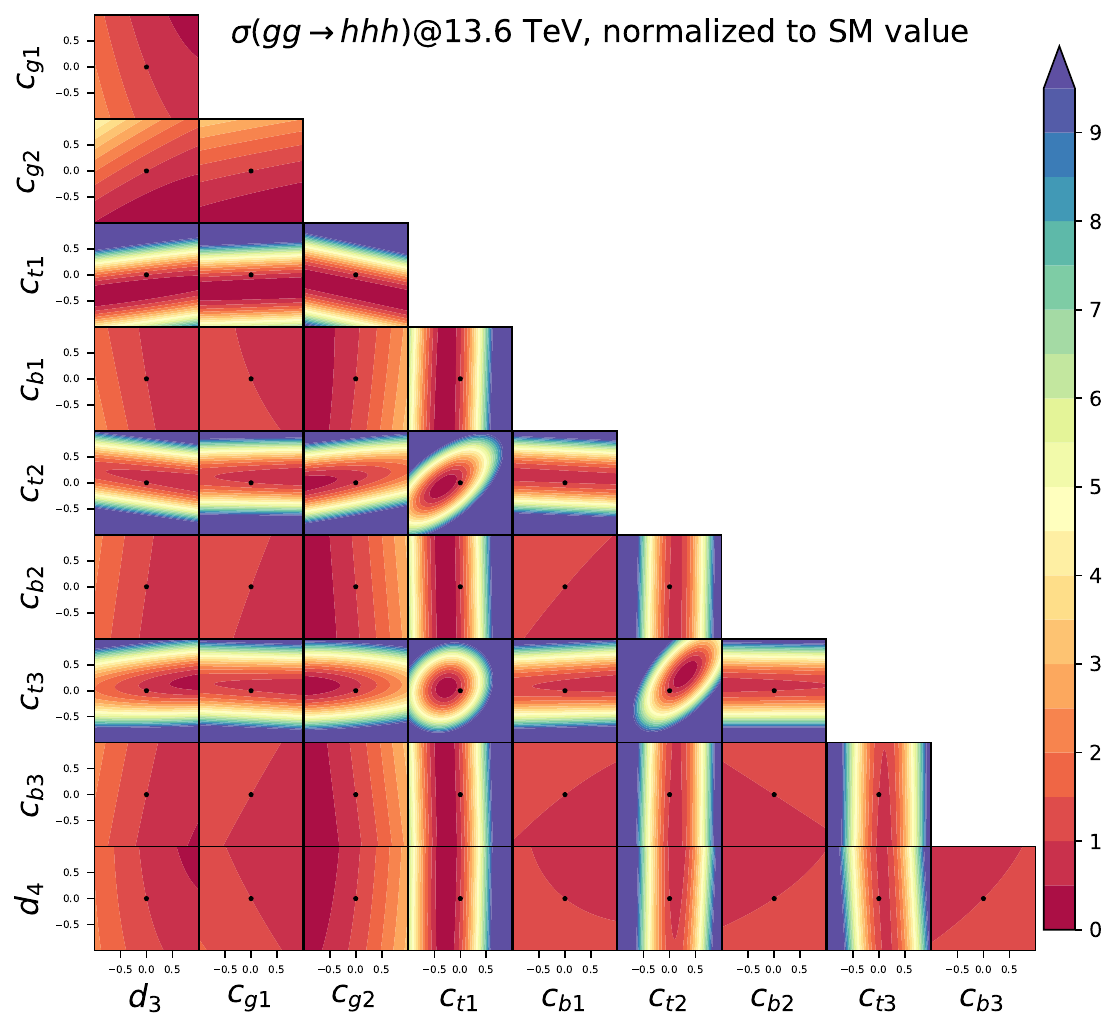}
\caption{\label{fig:hhhfit} Fit of the cross section for triple Higgs boson production at 13.6~TeV, normalized to the SM value. For each combination of couplings, the other couplings have been set to zero for simplicity. Each change of color in the contours represents a shift of a factor of 0.5$\times$ the SM value.}
\end{center}
\end{figure}

By employing our Monte Carlo implementation of eq.~\ref{eq:LgghhPhenoExp}, We have performed cross section fits at a proton-proton collider at 13, 13.6, 14, 27, and 100~TeV, over all coefficients relevant to either Higgs boson pair or triple production. Two-dimensional projections of the fits for $E_\mathrm{CM}=13.6$~TeV appear in fig.~\ref{fig:hhhfit} for triple production. In this figure, for each corresponding plot, all other coefficients have been set to zero, for the sake of simplicity. 

\begin{table}[htp]
    \centering
    \scriptsize
\begin{NiceTabular}{cccccccccccc}[hvlines,corners=NE] 
$d_3$ & -0.750 & 0.292 & & & & \\ 
$d_4$ & -0.158 & -0.0703 & 0.0340 & & & \\ 
$c_{g1}$ & -0.278 & 0.0426 & 0.0484 & 0.0256 & & & \\ 
$c_{g2}$ & 1.39 & -0.704 & -0.0312 & -0.156 & 0.538 & & \\ 
$c_{t1}$ & 6.94 & -3.17 & -0.309 & -0.850 & 5.16 & 12.6 & \\ 
$c_{t2}$ & -3.61 & 4.05 & -0.872 & -0.0482 & -4.15 & -17.6 & 15.3  \\ 
$c_{t3}$ & -2.72 & -1.57 & 1.33 & 0.906 & -0.316 & -4.64 & -18.2 & 13.0  \\ 
$c_{b1}$ & -0.125 & 0.177 & -0.0457 & -0.00903 & -0.166 & -0.675 & 1.38 & -0.941 & 0.0317  \\ 
$c_{b2}$ & 0.106 & -0.0752 & 0.00692 & -0.00740 & 0.0949 & 0.433 & -0.509 & 0.162 & -0.0219 & 0.00489  \\ 
$c_{b3}$ & 0.161 & -0.0809 & -0.00396 & -0.0182 & 0.124 & 0.598 & -0.474 & -0.0434 & -0.0189 & 0.0109 & 0.00719  \\ 
& 1 & $d_3$ & $d_4$ & $c_{g1} $ & $c_{g2}$ & $c_{t1}$ & $c_{t2}$ & $c_{t3}$ & $c_{b1}$ & $c_{b2}$ & $c_{b3}$ \\ 
\end{NiceTabular}
    \caption{Polynomial coefficients, $A_i$ (second column only) and $B_{ij}$, relevant for the determination of the cross section for leading-order Higgs boson triple production, in the form $\sigma/\sigma_\mathrm{SM} - 1 = \sum_i A_i c_i + \sum_{i,j} B_{ij} c_i c_j$, where $c_i \in \left\{ d_3, c_{g1}, c_{g2}, c_{t1}, c_{t2}, c_{b1}, c_{b2}, \right\}$, at $E_\mathrm{CM} = 13.6~\mathrm{TeV}$.}
    \label{tab:hhhcoeffs}
\end{table}

The numerical values of the polynomial coefficients, yielding the cross section in the form $\sigma/\sigma_\mathrm{SM} - 1 = \sum_i A_i c_i + \sum_{i,j} B_{ij} c_i c_j$, where $c_i \in \left\{ d_3, d_4, c_{g1}, c_{g2}, c_{t1}, c_{t2}, c_{t3}, c_{b1}, c_{b2}, c_{b3} \right\}$, are shown for $E_\mathrm{CM}=13.6$~TeV in table~\ref{tab:hhhcoeffs} for triple production. For Higgs boson pair production at $E_\mathrm{CM} = 13.6~\mathrm{TeV}$, see table~\ref{tab:hhcoeffs}, in appendix~\ref{app:fits}. The coefficients for both processes, at $E_\mathrm{CM} = 13,~14,~27$ and $100$~TeV can be found in appendix~\ref{app:fits}. The table should be read as follows: each value shown corresponds to the coefficient multiplying the product of the couplings of the corresponding first column and last row, respectively. As an example, with reference to table~\ref{tab:hhhcoeffs}, the coefficient of the term proportional to $d_4 d_3$ is $-0.0703$, i.e.\ read off the second row, third column. Using these coefficients, one can construct the cross section for any given value of the anomalous couplings. 

All the fits for the signal processes, and subsequent simulations, have been performed using the \verb|MSHT20nlo_as118| PDF set~\cite{Bailey:2020ooq} and the default dynamical scale choice (option 3) in \texttt{MG5\_aMC}, which corresponds to the sum of the transverse mass divided by 2.

We note here that the expected contribution of bottom-quark loops in the SM, both at LHC energies and at 100 TeV, is expected to be $\mathcal{O}(0.1\%)$. Therefore, anomalous couplings of the Higgs boson to the bottom quark are included merely for completeness, and furthermore, this ensures that we can safely neglect charm quark contributions in our analysis.

\subsection{Other Constraints on Anomalous Couplings}

The majority of the anomalous couplings that appear in the phenomenological Lagrangian of eq.~\ref{eq:LgghhPhenoExp} are already tightly constrained by other processes that involve the interactions of gluons, top and bottom quarks with the Higgs boson. The two exceptions that are not presently constrained are the anomalous interactions of three Higgs bosons and two top or bottom quarks, with relevant coefficients $c_{t3}$ and $c_{b3}$, as well as the anomalous modification to the Higgs boson's quartic interaction, related to the $d_4$ coefficient. While it is beyond the scope of the present study to perform a full fit, involving several processes and constraints, with their associated correlations, it is important to provide order-of-magnitude estimates for the two scenarios that we examine: the 13.6~TeV high-luminosity LHC, and the 100~TeV FCC-hh at the end of its lifetime. 

\begin{table}[htp]
    \centering
\begin{NiceTabular}{cccc}[hvlines,corners=NE] 
 \hline
 \multicolumn{4}{|c|}{\bf {Percentage uncertainties}}\\
 \hline
& HL-LHC & FCC-hh & Ref. \\
$\delta (d_3)$ & 50 & 5 & \cite{deBlas:2019rxi} (table 12) \\
$\delta (c_{g1})$ & 2.3 & 0.49 & \cite{deBlas:2019rxi} (table 3)\\ 
$\delta (c_{g2})$ & 5 & 1 & \cite{Azatov:2015oxa} (Figure 12, right)\\
$\delta (c_{t1})$ & 3.3 & 1.0 &\cite{deBlas:2019rxi} (table 3) \\
$\delta (c_{t2})$ & 30 & 10 &\cite{Azatov:2015oxa} (Figure 12, right) \\
$\delta (c_{b1})$ & 3.6 & 0.43 & \cite{deBlas:2019rxi} (table 3)\\
$\delta (c_{b2})$ & 30 & 10 & assumed same as $c_{t2}$\\
\end{NiceTabular}
    \caption{Estimates of percentage uncertainties (\%) obtained on the subset of anomalous couplings that appear in other processes at the HL-LHC and FCC-hh. The last column provides the source for these numbers. }
    \label{tab:OtherConstraints}
\end{table}

We consider the constraints on $c_{t1}$, $c_{b1}$ and $c_{g1}$ that would arise within the ``kappa-0'' scenario, as they are defined in~\cite{deBlas:2019rxi} (table 3). For the HL-LHC we consider those labeled ``HL-LHC", and for the 100~TeV FCC-hh, we consider those projected after the combined results of ``FCC-ee240+FCC-ee365'', ``FCC-eh'', and ``FCC-hh'' for the FCC-hh, collectively labeled ``FCC-ee/eh/hh''. For the modifications to the Higgs boson triple self-coupling, we use the values quoted for the ``di-Higgs exclusive'' results for the ``HL-LHC'', shown in table 12 of~\cite{deBlas:2019rxi}. For $c_{g2}$ and $c_{t2}$, we assume approximate constraints derived from the results of the analysis of~\cite{Azatov:2015oxa}. Due to the lack of concrete studies in the literature for $c_{b2}$, to the best of our knowledge, we have assumed that the constraint is identical to that of $c_{t2}$, obtained from~\cite{Azatov:2015oxa}. For all the coefficients, we assume that the central value of these future measurements will be at zero. The assumed constraints are summarized as percentage uncertainties in table~\ref{tab:OtherConstraints}, along with the corresponding references, for convenience. 

\subsection{Statistical Analysis}\label{sec:statanal}
The goal of the present phenomenological analysis is two-fold: on the one hand, we wish to determine if triple Higgs boson production itself can be observed at the end of the HL-LHC or FCC-hh lifetime, and on the other, we wish to calculate the constraints that would be obtained on the anomalous couplings, given the assumption the SM is the true underlying theory. The first goal corresponds to the null hypothesis being the absence of triple Higgs boson production, whereas the latter corresponds to the null hypothesis being SM triple Higgs boson production, i.e.\ triple production with all the anomalous coupling coefficients set to zero. 

For the first of these goals, i.e.\ to determine the significance, $\Sigma$, of observing triple Higgs boson production for a certain anomalous coupling parameter-space point, we optimize the signal versus background discrimination according to our phenomenological analysis (described in the following section, \ref{sec:phenoanal}), to obtain the maximum significance over the set of cuts for SM triple Higgs boson production. 
Since we expect the number of signal events to be much lower than that for the background, we employ the following formula for the significance: 
\begin{equation}
\Sigma = \frac{S}{\sqrt{B + (\alpha B)^2}}\;,
\end{equation}
where $S$ and $B$ are the expected number of signal and background events, respectively, at a given integrated luminosity $\mathcal{L}$, and $\alpha$ is a factor that we employ to model the systematic uncertainty present in the background estimation. During our search for the maximum SM triple Higgs boson production significance, we used the value $\alpha =0$, i.e.\ we optimized the quantity $S/\sqrt{B}$. Subsequently, all results are presented with this optimal set of cuts, obtained separately for each collider, while setting the systematic uncertainty factor $\alpha = 0.05$. 

We have derived the expected ``evidence'' (3$\sigma$) and ``discovery'' (5$\sigma$) limits on the observation of triple Higgs boson production on the $(c_{t3}, d_4)$-plane, where the novel information coming from triple Higgs boson production is expected to arise. Furthermore, we have derived the 68\% confidence level (C.L.) (1$\sigma$) and 95\% C.L. (2$\sigma$) expected limits on the same plane, given that the SM hypothesis is true (i.e.\ SM triple Higgs boson production). While doing so, we allow the coefficients $d_3$ and $c_{t2}$ to vary. These will be constrained primarily via Higgs boson pair production, and since this is a challenging process in itself, will not be substantially constrained compared to the gluon-Higgs anomalous interactions, or the top-anti-top-Higgs anomalous coupling, $c_{t1}$, see table~\ref{tab:OtherConstraints} for a quantification of this statement. In addition, we do not expect triple Higgs boson production to be competitive to Higgs boson pair production in terms of constraints on $d_3$ and $c_{t2}$, so we ``marginalize'' over these two. Note that, due to the fact that we are assuming minimal flavour violation, as discussed in appendix~\ref{app:d6eftL}, as well as for the sake of simplicity, we do not expect the effect of the $c_{b3}$ coefficient to be as significant as that of $c_{t3}$, and therefore we have set $c_{b3}$ to zero in the phenomenological analysis. 

To derive the evidence and discovery limits for triple Higgs boson production, we assume that the number of signal and background events follow gaussian distributions, and define the uncertainty on the background number of events as:
\begin{equation}
    \delta_\mathrm{B} = \sqrt{B + (\alpha B)^2}\;.
\end{equation}
Then, the probability value ($p$-value) to obtain a given number of signal events, $S(\{c_i\})$, corresponding to the set of anomalous coupling coefficients $\{c_i\}$, is given by: 
\begin{equation}
P(\{c_i\}) = \frac{1}{\sqrt{2 \pi} \delta_\mathrm{B}} \exp \left[ -\frac{ S(\{c_i\})^2}{ 2 \delta_\mathrm{B}^2} \right]\;.
\end{equation}
To take into account the approximate constraints of table~\ref{tab:OtherConstraints}, that would be present, ignoring possible correlations that will exist between constraints, we multiply the $p$-values by a gaussian prior factor:
\begin{equation}\label{eq:pconstrained}
P_C(c_i) = \frac{1}{\sqrt{2 \pi} \delta(c_i)} \exp \left[ -\frac{c_i^2}{ 2 \delta(c_i)^2} \right]\;,
\end{equation}
at a value $c_i$ of each of the anomalous coefficients $d_3$ and $c_{t2}$. We then marginalize by integrating over the $d_3$ and $c_{t2}$ directions. Since this is done over a grid, the integral is represented by a sum over the grid points:  
\begin{equation}\label{eq:marginalization}
    P_M(c_{t3},d_4) = \sum_{(d_3, c_{t2})} \Delta d_3 \Delta c_{t2}  P_C(d_3) P_C(c_{t2}) P(\{c_i\})\;,
\end{equation}
where $\Delta d_3$ and $\Delta c_{t2}$ are the bin widths in the $d_3$ and $c_{t2}$ directions, respectively. The probability is then converted to $\chi^2$ via the \texttt{Python} $\chi^2$ inverse survival function for two degrees of freedom (\texttt{scipy.stats.chi2.isf}), and the 3$\sigma$ and 5$\sigma$ limits are found by drawing the iso-contours corresponding to $\chi^2 - \chi^2_\mathrm{min} = (11.8,28.7)$, respectively. The results are shown in Fig.~\ref{fig:2ddiscovery}, in section~\ref{sec:results}.

For the determination of the constraints obtained for the null hypothesis being SM triple Higgs boson production, we folow a similar procedure. In this scenario, we assume that the number of events for both signal and background follow a gaussian distribution, as we did before, and hence the uncertainty on the number of total expected SM events is given by:
\begin{equation}
    \delta_\mathrm{SM+B} = \sqrt{S_{\mathrm{SM}} + B + (\alpha B)^2}\;,
\end{equation}
where $S_\mathrm{SM}$ represents the expected number of events for SM triple Higgs boson production. Then the $p$-value to obtain a given number of events corresponding to an anomalous coupling parameter-space point, $S(\{c_i\})$, given that the SM is the truth, is given by:
\begin{equation}
\bar{P}(\{c_i\}) = \frac{1}{\sqrt{2 \pi} \delta_\mathrm{SM+B}} \exp \left[ -\frac{ (S_\mathrm{SM} - S(\{c_i\}))^2}{ 2 \delta_\mathrm{SM+B}^2} \right]\;.
\end{equation}
This $p$-value is then multiplied by the prior factors of eq.~\ref{eq:pconstrained}, and summed as before:
\begin{equation}\label{eq:marginalization}
    \bar{P}_M(c_{t3},d_4) = \sum_{(d_3, c_{t2})} \Delta d_3 \Delta c_{t2}  P_C(d_3) P_C(c_{t2}) \bar{P}(\{c_i\})\;,
\end{equation}
followed by a conversion to the $\chi^2$ value as  before. The 68\% (1$\sigma$) and 95\% (2$\sigma$) confidence level (C.L.) intervals are found by drawing the iso-contours corresponding to $\chi^2 - \chi^2_\mathrm{min} = (2.28, 5.99)$, respectively. The results of this analysis are shown in Fig.~\ref{fig:2dconstraints}, in section~\ref{sec:results}. 

To obtain the one-dimensional limits on either of the $d_4$ or $c_{t3}$ couplings alone, we take the procedure one step further: we sum over the additional marginalized coupling as in eq.~\ref{eq:marginalization}, and convert the $p$-value into a $\chi^2$ value. For the triple Higgs boson searches, we calculate the $3\sigma$ and $5\sigma$ evidence and discovery limits by finding the points that correspond to $\chi^2 - \chi^2_\mathrm{min} = (9.00, 25.0)$, respectively. For the expected constraints on the $d_4$ and $c_{t3}$ coefficients, given that the SM is true,  we calculate the $1\sigma$ and $2\sigma$ intervals by finding the points that correspond to $\chi^2 - \chi^2_\mathrm{min} = (0.99, 3.84)$, respectively. The results for the evidence and discovery points, and for the expected limits given that the SM is true, are shown in tables~\ref{tab:disclimits} and~\ref{tab:limits}, respectively, in section~\ref{sec:results}.

In regard to the use of the Gaussian approximation in our analysis, we note here that, while the expected number of events for the case of the SM signal is low and, strictly speaking, Poisson statistics should be employed in that case, we would like to emphasize that this is not the case at the $1,2,3$ and $5\sigma$ exclusion/discovery boundaries that we have deduced. To be concrete, for the HL-LHC case, to obtain a 1$\sigma$ significance, we would need, given our estimates for the expected number of background events (table~\ref{tab:backgrounds}, top) $B=419.6$, the following number of signal events: 

\begin{equation}
S \simeq 1\times \sqrt{B + (\alpha B)^2} \simeq 29.32\;.
\end{equation}

For this expected number of signal events, the Gaussian approximation would yield a $68\%$ confidence-level interval of $29.32 \pm 5.41$, whereas the Poisson approximation would yield $29.32^{+4.48}_{-6.48}$. Given the other uncertainties that enter our phenomenological analysis, it is therefore reasonable to assume that events are Gaussian-distributed at the boundaries of the exclusion/discovery regions. 

\subsection{Phenomenological Analysis}\label{sec:phenoanal}

To obtain constraints on anomalous triple Higgs boson production at proton colliders, following the statistical procedure outlined above, we have performed a hadron-level phenomenological analysis of the 6 $b$-jet final-state originating from the decays of all three Higgs bosons to $b\bar{b}$ quark pairs. We closely follow the analysis of Refs.~\cite{Papaefstathiou:2019ofh,Papaefstathiou:2020lyp}. Parton-level events have been generated using the \texttt{MG5\_aMC} anomalous couplings implementation constructed in the present article, with showering, hadronization, and simulation of the underlying event, performed via the general-purpose \texttt{HERWIG 7} Monte Carlo event generator~\cite{Bahr:2008pv, Gieseke:2011na, Arnold:2012fq, Bellm:2013hwb, Bellm:2015jjp, Bellm:2017bvx, Bellm:2019zci,Bewick:2023tfi}. The event analysis was performed via the \texttt{HwSim} framework addon to \texttt{HERWIG 7}~\cite{hwsim}. No smearing due to the detector resolution or identification efficiencies have been applied to the final objects used in the analysis, apart from a $b$-jet identification efficiency, discussed below. We expect such effects to produce differences in binned jet observables of $\mathcal{O}(10\%)$ at the LHC, see for example~\cite{ATLAS:2020cli,CMS-DP-2021-033}, and anticipate substantial improvement to this at the FCC-hh. 

The branching ratio of $h\rightarrow b\bar{b}$ will be modified primarily due to $c_{t1}$, $c_{g1}$, indirectly through modifications to the $h\rightarrow gg$ and $h \rightarrow \gamma \gamma$ branching ratios, and directly through $c_{b1}$. To take this effect into account, we employed the \texttt{eHDECAY} code~\cite{Contino:2014aaa}. The program \texttt{eHDECAY} includes QCD radiative corrections, and next-to-leading order electroweak corrections are only applied to the SM contributions. For further details, see Ref.~\cite{Contino:2014aaa}. We have performed a fit of the $\texttt{eHDECAY}$ branching ratio $h\rightarrow b\bar{b}$, and we have subsequently normalized this to the latest branching ratio provided by the Higgs Cross Section Working Group's Yellow Report~\cite{yr4page,LHCHiggsCrossSectionWorkingGroup:2016ypw}, $\mathrm{BR}(h\rightarrow b\bar{b})=0.5824$. The fit is then used to rescale the final cross section of $pp \rightarrow hhh \rightarrow (b\bar{b})(b\bar{b})(b\bar{b})$. The background processes containing Higgs bosons turned out to be subdominant with respect to the dominant QCD 6 $b$-jet and $Z$+jets backgrounds, and therefore we did not modify these when deriving the final cross sections. 

For the generation of the backgrounds involving $b$-quarks \textit{not} originating from either a $Z$  or Higgs boson, we imposed the following generation-level cuts for the 100~TeV proton collider: $p_{T,b} >30~\mathrm{GeV}$, $|\eta _j|< 5.0$, and $\Delta R_{b,b} >0.2$. The transverse momentum cut was lowered to $p_{T,b} >20~\mathrm{GeV}$ for 13.6 TeV, except for the QCD 6 $b$-jet background, for which we produced the events inclusively, without any generation cuts.\footnote{In general, the simulation of the QCD induced process $p p\rightarrow (b\bar{b}) (b\bar{b}) (b\bar{b})$ is one of the most challenging aspects of the phenomenological study. The samples are produced in parallel using OMNI cluster at the university of Siegen using the ``gridpack'' option available in \texttt{MG5\_aMC}.} The selection analysis was optimized considering as a main backgrounds the QCD-induced process $p p\rightarrow  (b\bar{b}) (b\bar{b}) (b\bar{b})$, and the $Z$+jets process (represented by $Z+(b\bar{b}) (b\bar{b})$), which we found to be significant at LHC energies. 

The event selection procedure for our analyses proceeds as follows: as in \cite{Papaefstathiou:2019ofh}, an event is considered if it contains at least six $b$-tagged jets, of which only the six ones with the highest $p_{T}$ are taken into account. A universal minimal threshold for the transverse momentum, $p_{T,b}$, of any of the selected $b$-tagged jets is imposed. In addition a universal cut on their maximum pseudo-rapidity, $|\eta_b|$, is also applied. We subsequently make use of the observable:
\begin{eqnarray}
\chi^{2, (6)}=\sum_{qr \in I} (m_{qr}-m_h)^2\;,  
\label{eq:chi2}
\end{eqnarray}
where $I=\{jb_{1}jb_{2},jb_{3}jb_{4},jb_{5}jb_{6}\}$ is the set of all possible 15 pairings of 6-$b$ tagged jets. Out of all the possible combinations we pick the one with the smallest value $\chi_{\rm min}^{2, (6)}$. The pairings of $b$-jets defining $\chi_{\rm min}^{2, (6)}$ constitute our best candidates for the reconstruction of the three Higgs bosons, $h$. Our studies have demonstrated that $\chi_{\rm min}^{2, (6)}$ is one of the most powerful observables to employ in signal versus background discrimination. 

We further refine the discrimination power of the $\chi_{\rm min}^{2, (6)}$ variable by using the individual mass differences $\Delta m=|m_{q r} -m_h|$ in eq.~(\ref{eq:chi2}), sorting them out according to $\Delta m_{\rm min}< \Delta m_{\rm med} < \Delta m_{\rm max}$, and imposing independent cuts on each of them. We also consider the transverse momentum $p_{T}(h^i)$ of each reconstructed Higgs boson candidate. These reconstructed particles are also sorted based on the value of $p_{T}(h^i)$, on which we then impose a cut. Besides the universal minimal threshold on $p_{T,b}$, introduced at the beginning of this section, we impose cuts on the three $b$-jets with the highest transverse momentum $p_{T, b_i}$, for $i=1,2,3$. The set of cuts $p_{T, b_3}<p_{T, b_2}<p_{T, b_1}$ is the second most powerful discriminating observable in our list. Finally, we also considered two additional geometrical observables. The first of them is the distance between $b$-jets in each reconstructed Higgs boson $\Delta R_{bb}(h^i)$. The second one is the distance between the reconstructed Higgs bosons $\Delta R(h^i, h^j)$ themselves. 

Our optimization process then proceeds as in \cite{Papaefstathiou:2019ofh,Papaefstathiou:2020lyp}: we sequentially try different combination of cuts over the observables introduced above on our signal and background samples until we achieve a significance above $2$ or when our number of Monte Carlo events is reduced so drastically that no meaningful statistical conclusions can be derived if this number becomes smaller (this happens for instance when for a given combination of cuts, we are left with less than 10 Monte Carlo events of signal or background). 

\begin{table}
\begin{center}
\begin{tabular}{ |l|l|l| } 
 \hline
 \multicolumn{3}{|c|}{\bf {Optimized cuts}}\\
 \hline
Observable & $13.6$ TeV & $100$ TeV \\ 
  \hline
 $p_{T,b}>$ & $25.95$ GeV &   $35.00$ GeV\\ 
 $|\eta_{b}|<$ & $2.3$ &  $3.3$ \\ 
  $\Delta R_{bb}>$&$0.3$&$0.3$\\
$p_{T,b_i}>$ &$[25.95, 25.95, 25.95]$ GeV $i=1,2,3$&$[170.00, 135.00, 35.00]$ GeV\\
$\chi^{2,(6)} <$ &$27.0$ GeV & $26.0$ GeV\\
$\Delta m_{\rm{ min, med, max}}<$&$[100, 200, 300]$ GeV& $[8, 8, 8]$ GeV\\
$\Delta R_{bb}(h^i)<$ & $[3.5, 3.5, 3.5]$&$[3.5, 3.5, 3.5]$\\
 $\Delta R(h^i,h^j)<$ & $[3.5, 3.5, 3.5]$ & $[3.5, 3.5, 3.5]$\\
 $p_{T}(h^i)>$ & $[0.0, 0.0, 0.0]$ GeV&$[200.0, 190.0, 20.0]$ GeV\\
 \hline
\end{tabular}
\caption{Optimized cuts determined for the phenomenological analysis. The indices $i,j$ can take the values $i,j=1,2,3$. For the cut $\Delta R(h^i, h^j)$ the three pairings correspond to $(1,2), (1,3), (2,3)$. The indexed elements should be read from left to right in increasing order. The last two rows refer to cuts over light jets.}
\label{tab:cuts}
\end{center}
\end{table}

\begin{table}
\begin{center}
\begin{tabular}{ |l|l|l|l| } 
 \hline
 \multicolumn{4}{|c|}{\bf{LHC Analysis $(13.6 ~\rm{TeV})$}}\\
 \hline
 &&&\\
 Process & $\sigma_{\rm{NLO}} (6~b\mathrm{-jet})~[\rm{fb}]$  & $\varepsilon_{\rm{analysis}}$&$N^{\rm{cuts}}_{3 \times 10^{3}~ \rm{fb}^{-1}}$\\
 &&&\\
 \hline
 $ h h h (\rm{SM})$& $1.97 \times 10^{-2}$& $0.12$& $2.77$\\
 \hline
 QCD $(b \bar{b}) (b \bar{b}) (b \bar{b})$& 6136.12  & $1.00 \times 10^{-5}$& 69.67 \\ \hline
   $pp \rightarrow Z (b \bar{b}) (b \bar{b})$ & 61.80  & 0.0045 & 318.17 \\\hline
   $pp \rightarrow Z Z (b \bar{b})$ &  2.16   & 0.0059  & 14.3 \\\hline
   $pp \rightarrow h Z (b \bar{b})$ &  0.45   & 0.0159 & 8.1 \\\hline
   $pp \rightarrow h h Z$ &  0.0374   & 0.034  & 1.45 \\\hline
   $pp \rightarrow h h (b \bar{b})$ & 0.0036& 0.028 & 0.11 \\\hline
   LI $gg \rightarrow h Z Z$ & 0.143   &0.022 & 3.62 \\\hline
   LI $gg \rightarrow Z Z Z$ & 0.124   &0.013 & 1.76 \\\hline
   LI $gg \rightarrow h h Z$ & 0.0458  &0.047 & 2.42\\\hline
\multicolumn{4}{c}{}\\
\hline
\multicolumn{4}{|c|}{\bf{FCC-hh Analysis $(100 ~\rm{TeV})$}}\\
 \hline
 &&&\\
 Process & $\sigma_{\rm{NLO}} (6~b\mathrm{-jet})~[\rm{fb}]$ & $\varepsilon_{\rm{analysis}}$&$N^{\rm{cuts}}_{20 ~ \rm{ab}^{-1}}$\\
 &&&\\
 \hline
 $ h h h (\rm{SM})$& $1.14$ & $0.0115$& $98.90$\\
 \hline
 QCD $(b \bar{b}) (b \bar{b}) (b \bar{b})$&  $56.66\times 10^{3}$  & $1.12 \times 10^{-5} $&4777.71\\\hline
    $pp \rightarrow Z (b \bar{b}) (b \bar{b})$ & 1285.37  & $3.04\times 10^{-5}$ & 294.63 \\\hline
   $pp \rightarrow Z Z (b \bar{b})$ & 49.01 & $2.02\times10^{-5}$  & 7.48\\\hline
   $pp \rightarrow h Z (b \bar{b})$ & 9.87  & $3.04\times 10^{-5}$& 2.26 \\\hline
   $pp \rightarrow h h Z$ &  0.601  & $5.95\times 10^{-4}$& 2.70 \\\hline
   $pp \rightarrow h h (b \bar{b})$ & 0.096  & $8.095\times 10^{-5}$& $\ll 1$ \\\hline
   LI $gg \rightarrow h Z Z$ &   8.28  &  $1.62\times 10^{-4}$& $10.12$ \\\hline
   LI $gg \rightarrow Z Z Z$ &  6.63  & $4.05\times 10^{-5}$ & $2.03$ \\\hline
   LI $gg \rightarrow h h Z$ &   2.65   & $2.54\times10^{-4}$& 5.07\\\hline 
 \end{tabular}
 \caption{The lists of processes considered during our phenomenological analysis, along with their respective cross sections to the 6 $b$-jet final state. The efficiencies $\varepsilon_{\rm analysis}$ and number of events $N_{\mathcal{L}}^{\rm cuts}$, correspond to those obtained after applying the set of cuts given in table~\ref{tab:cuts}. A $b$-jet identification efficiency of $0.85$ (for each $b$-jet) has also been applied to obtain the number of events. For the HL-LHC we considered an integrated luminosity of $\mathcal{L}=3000~\hbox{fb}^{-1}$, and for  for the FCC-hh a luminosity of $\mathcal{L}=20~\hbox{ab}^{-1}$. To approximate higher-order corrections, a $K$-factor $K=2$ has been included for all processes, with respect to the leading-order cross section. The background processes marked with ``LI" represent loop-induced contributions that have been generated separately.}
 \label{tab:backgrounds}
 \end{center}
\end{table}

For $100~\rm{TeV}$, the observables are optimized in the following order, (i) $p_{T,b}$, (ii) $|\eta_b|$, (iii) $\chi^{2, (6)}$, (iv) $\Delta m_{\rm min}$, $\Delta m_{\rm med}$, $\Delta m_{\rm max}$, (v) $p_{T, b_1}$ and (vi) $p_{T, b_2}$. By testing different cut combinations over the variables above we reach a SM triple Higgs boson significance of $S/\sqrt{B}\simeq 2.33\sigma$ without $b$-jet tagging or $S/\sqrt{B}\simeq 1.43 \sigma$ including the $b$-tagging factor (0.85). We note that this is an improvement over the previous result of~\cite{Papaefstathiou:2019ofh}, where the corresponding significance was around $1.7\sigma$ without $b$-tagging. For $13.6~\rm{TeV}$ the optimization of the cuts follows a similar path, however after applying the cut on $\chi^{2, (6)}$ we are left with $\mathcal{O}(10)$ Monte Carlo events for background, and therefore the full procedure stops at a significance of only $0.23\sigma$ without $b$-jet tagging or $0.14\sigma$ with $b$-jet tagging ($0.85$). The concrete values of the cuts applied on the different kinematic variables depend on the center of mass energy of the collider under study, and are summarized on table~\ref{tab:cuts}. We note that the optimal set of cuts for the HL-LHC requires more central $b$-jets than that at the FCC-hh. The complete list of backgrounds, their initial cross sections, efficiency of cuts and expected number of events at each of the full collider luminosities, are given in table~\ref{tab:backgrounds}. It is interesting to note that the $Z+(b\bar{b})(b\bar{b})$ is in fact the dominant one at the HL-LHC, whereas the reverse is true for the FCC-hh. It will be important to validate this result in a future detailed experimental study that consider the full effects of the detectors.

To employ the results of the SM analysis over the whole of the parameter space we are considering, we have performed a polynomial fit of the efficiency of the analysis on the signal, $\varepsilon_\mathrm{analysis}(hhh)$, at various, randomly-chosen, combinations of anomalous coefficient values. In combination with the fits of the cross section, and the fit of the branching ratio of the Higgs boson to $(b\bar{b})$, we can estimate the number of events at a given luminosity, for a given collider for any parameter-space point within the anomalous coupling picture, which we dubbed $S(\{c_i\})$ in our discussion of the statistical approach we take, in section~\ref{sec:statanal}. 

\subsection{Results}\label{sec:results}

\begin{figure}[htp]
\begin{center}
    \includegraphics[width=0.495\columnwidth]{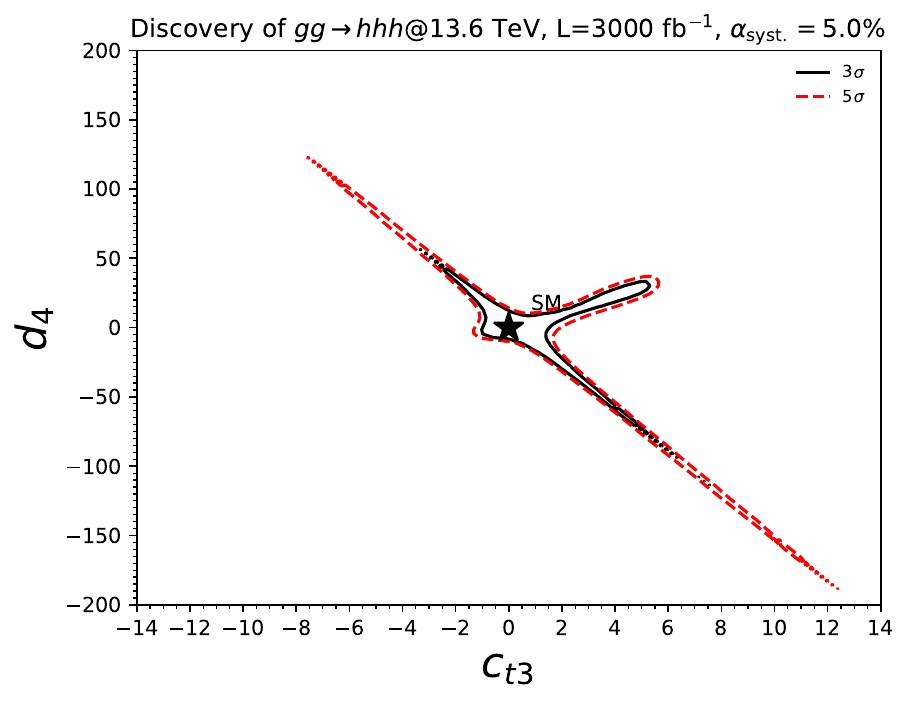}
  \includegraphics[width=0.495\columnwidth]{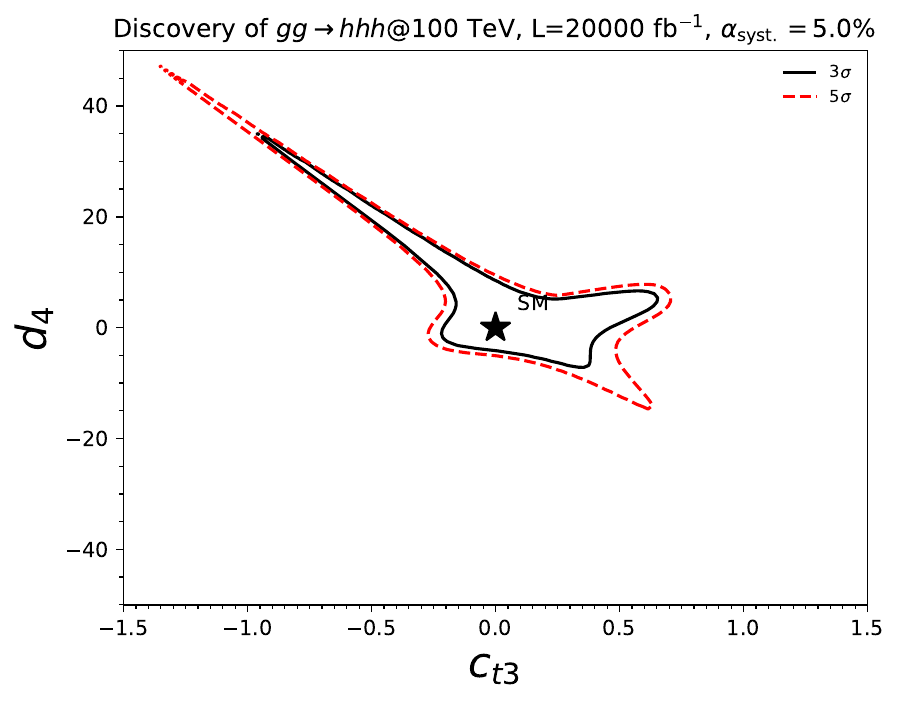}
\caption{\label{fig:2ddiscovery} The 3$\sigma$ evidence (black solid) and 5$\sigma$ discovery (red dashed) curves on the $(c_{t3},d_4)$-plane for triple Higgs boson production at 13~TeV/3000~fb$^{-1}$ (left), and 100~TeV/20~ab$^{-1}$ (right), marginalized over the $c_{t2}$ and $d_3$ anomalous couplings. Note the differences in the axes ranges at each collider.}
\end{center}
\end{figure}

\begin{figure}[htp]
\begin{center}
    \includegraphics[width=0.495\columnwidth]{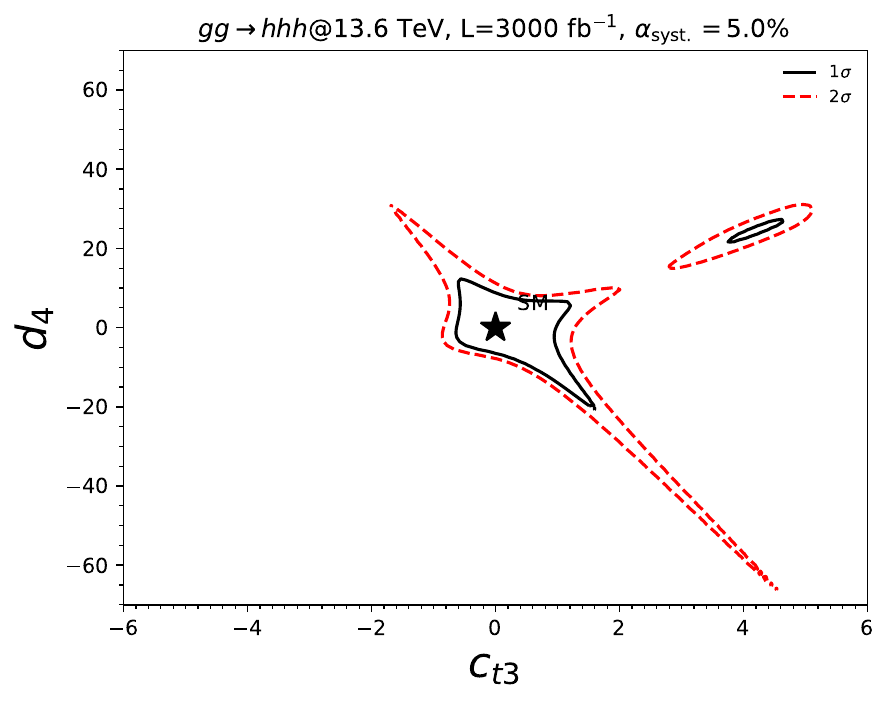}
  \includegraphics[width=0.495\columnwidth]{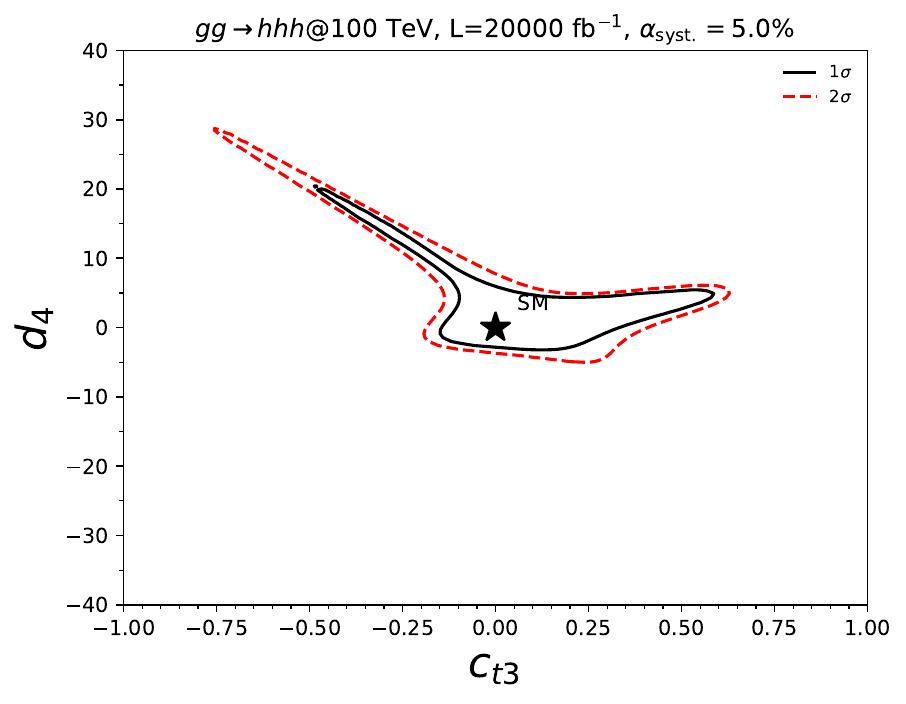}
\caption{\label{fig:2dconstraints} The 68\% C.L. (1$\sigma$, black solid) and 95\% C.L (2$\sigma$, red dashed) limit on the $(c_{t3},d_4)$-plane for triple Higgs boson production at 13~TeV/3000~fb$^{-1}$ (left), and 100~TeV/20~ab$^{-1}$ (right), marginalized over the $c_{t2}$ and $d_3$ anomalous couplings. Note the differences in the axes ranges at each collider. }
\end{center}
\end{figure}

\begin{table}[htp]
    \centering
\begin{NiceTabular}{ccc||cc}[hvlines,corners=NE] 
& HL-LHC 3$\sigma$ & HL-LHC 5$\sigma$ & FCC-hh 3$\sigma$ & FCC-hh 5$\sigma$ \\
$d_4$ & $[-28.0, 41.7]$  & $[-99.5, 152.9]$  & $[-24.9, 20.8]$ & $ [-40.8, 23.1]$     \\
$c_{t3}$ & $[-2.1,5.5]$ & $[-7.1, 11.3]$ & $[-0.8,0.6]$ & $[-1.2, 0.7]$  \\
\end{NiceTabular}
   \caption{ The 3$\sigma$ evidence and 5$\sigma$ discovery limits on for triple Higgs boson production, for the $c_{t3}$ and $d_4$ coefficients at 13~TeV/3000~fb$^{-1}$, and 100~TeV/20~ab$^{-1}$, marginalized over $c_{t2}$, $d_3$ and either $d_4$, or $c_{t3}$. }\label{tab:disclimits}
\end{table}

\begin{table}[htp]
    \centering
\begin{NiceTabular}{ccc||cc}[hvlines,corners=NE] 
& HL-LHC 68\% & HL-LHC 95\% & FCC-hh 68\% & FCC-hh 95\%   \\
$d_4$ & $[-6.6, 12.4]$  & $[-10.0, 21.3]$  & $[-3.9, 10.5]$ & $ [-10.6, 18.8]$     \\
$c_{t3}$ & $[-0.6, 1.1]$ & $[-0.9, 3.6]$ & $[-0.1, 0.3]$ & $[-0.4,  0.6]$  \\
\end{NiceTabular}
   \caption{ The 68\% C.L. (1$\sigma$) and 95\% C.L (2$\sigma$) limits on $c_{t3}$ and $d_4$ for triple Higgs boson production at 13~TeV/3000~fb$^{-1}$, and 100~TeV/20~ab$^{-1}$, marginalized over $c_{t2}$, $d_3$ and either $d_4$, or $c_{t3}$. }\label{tab:limits}
\end{table}

The main results of our two-dimensional analysis over the $(c_{t3},d_4)$-plane are shown in figs.~\ref{fig:2ddiscovery} and~\ref{fig:2dconstraints}. In particular, fig.~\ref{fig:2ddiscovery} shows the potential ``evidence" and ``discovery'' regions (3$\sigma$ and 5$\sigma$, respectively) for triple Higgs boson production at the high-luminosity LHC on the left (13.6 TeV with 3000~fb$^{-1}$ of integrated luminosity), and at a FCC-hh (100 TeV, with 20~ab$^{-1}$) on the right. Evidently, very large modifications to the quartic self-coupling are necessary for discovery of triple Higgs boson production at the HL-LHC, ranging from $d_4\sim 125$ for $c_{t3}~\sim-8$, to $d_4\sim \pm 40$ for $c_{t3}\sim 0$ and then down to $d_4\sim -200$ for $c_{t3}\sim 12$. The situation is greatly improved, as expected, at the FCC-hh, where the range of $d_4$ is reduced to $d_4~\sim 40$ for $c_{t3}\sim-1.5$, and to $d_4\sim -20$ for $c_{t3}\sim 1.0$. It is interesting to note that the whole of the parameter space with $c_{t3} \gtrsim 1.0$, or with $c_{t3} \lesssim -1.5$ is discoverable, at the FCC-hh at 5$\sigma$. For the potential 68\% (1$\sigma$) and 95\% C.L. (2$\sigma$) constraints of fig.~\ref{fig:2dconstraints}, the situation is slightly more encouraging for the HL-LHC, with the whole region of $d_4 \gtrsim 40$ or $d_4 \lesssim -60$ excluded at 95\% C.L.. The corresponding region at 68\% C.L. is $d_4 \gtrsim 20$ and $d_4 \lesssim -30$. For $c_{t3}$, it is evident that all the region $c_{t3} \lesssim -2$ and $c_{t3} \gtrsim 5$ will be excluded at 95\% C.L. and $c_{t3} \lesssim -1$, $c_{t3} \gtrsim 4$ at 68\% C.L.. On the other hand, the FCC-hh will almost be able to exclude the whole positive region of $d_4$ for any value of $c_{t3}$ at 68\% C.L.. This will potentially be achievable if combined with other Higgs boson triple production final states. For the $c_{t3}$ coupling, both the constraints reach the $\mathcal{O}(\mathrm{few}~10\%)$ level for any value of $d_4$. 

The one-dimensional analysis' results, presented in tables~\ref{tab:disclimits} and~\ref{tab:limits}, for the ``evidence'' and ``discovery'' potential, and exclusion limits, respectively, reflect the above conclusions. For instance, it is clear by examining table~\ref{tab:disclimits}, that the HL-LHC will only see evidence of triple Higgs boson production in the 6 $b$-jet final state only if $d_4$ has modifications of $|d_4| \sim \mathcal{O}(\mathrm{few}~10)$, and will only discover it if $|d_4| \sim \mathcal{O}(100)$. On the other hand, there could be evidence or discovery of Higgs boson triple production if $|c_{t3}|\sim \mathcal{O}(\mathrm{1-10})$. The 1$\sigma$ and 2$\sigma$ exclusion regions are much tighter, as expected, with $|d_4| \sim \mathcal{O}(10)$ at 1$\sigma$ or 2$\sigma$ at the HL-LHC, improving somewhat at the FCC-hh, and $|c_{t3}| \sim \mathcal{O}(0.1-1)$, both at the HL-LHC and FCC-hh.

\section{Conclusions} \label{sec:conclusions}

We have conducted a phenomenological analysis of the triple Higgs boson production at the high-luminosity LHC (13.6 TeV with 3000~fb$^{-1}$) and a proposed future circular collider colliding protons at 100 TeV (20~ab$^{-1}$). For this analysis, we have constructed a model to be used with the \texttt{MadGraph5\_aMC@NLO} event generator, which is publicly available at~\cite{gitlabrepo}. By examining the 6 $b$-jet final state, within a signal versus background analysis, we have concluded that interesting constraints can be obtained on the most relevant coefficients, $d_4$ and $c_{t3}$, representing modifications to the Higgs quartic self-coupling, and additional top-Higgs interactions, respectively. These are presented, for the HL-LHC and the FCC-hh, over the $(c_{t3},d_4)$-plane in figs.~\ref{fig:2ddiscovery} and~\ref{fig:2dconstraints}, showing the evidence/discovery potential of the triple Higgs boson production process itself, marginalized over the $d_3$ and $c_{t2}$ coefficients, and the 1/2$\sigma$ exclusion regions arising from the process on that plane. The one-dimensional evidence/discovery regions over either $d_4$, or $c_{t3}$ at both colliders are given in table~\ref{tab:disclimits}, and the possible constraints extracted in table~\ref{tab:limits}. 

The results of our study demonstrate the importance of including additional contributions, beyond the modifications to the self-couplings, when examining multi-Higgs boson production processes, and in particular triple Higgs boson production. We are looking forward to a more detailed study for the HL-LHC, conducted by the ATLAS and CMS collaborations, including detector simulation effects, and the full correlation between other channels. From the phenomenological point of view, improvements will arise by including additional final states, e.g.\ targetting the process $pp \rightarrow hhh \rightarrow (b\bar{b})(b\bar{b})(\tau^+\tau^-)$, or by performing an analysis that leverages machine learning techniques to maximize significance.\footnote{This approach was taken in~\cite{Stylianou:2023xit} at the HL-LHC for modifications of the Higgs boson's self-couplings.} In summary, we believe that the triple Higgs boson production process should constitute part of a full multi-dimensional fit, within the anomalous couplings picture. 

\acknowledgments

We extend our thanks to Alberto Tonero for his valuable early contributions to this project. 
A.P. acknowledges support by the National Science Foundation under Grant No.\ PHY 2210161. G.T.X. received support for this project from the European Union’s Horizon
2020 research and innovation programme under the Marie Sklodowska-Curie grant agreement No 860881-HIDDeN [S.D.G.] and the Marie Sklodowska-Curie grant agreement No
945422. This research was supported by the Deutsche Forschungsgemeinschaft (DFG, German Research Foundation) under grant 396021762 - TRR 257. We thank Wolfgang Kilian for interesting discussions. We would also like to thank the organizers of the HHH workshop that took place in July 2023 at the Inter University Center, in Dubrovnik, Croatia (IUC), for facilitating discussions on the subject matter presented here. 

\appendix

\section{$D=6$ Effective Field Theory}\label{app:d6eftL}

The phenomenological Lagrangian employed in this study can be motivated by considering the $D=6$ effective field theory (EFT) Lagrangian of ref.~\cite{Goertz:2014qta}. The discussion that follows has been adapted for the purposes of the present article. 

New Physics associated to a new scale $\Lambda \gg v$ can be described in a model-independent way by augmenting the Lagrangian of the SM with all possible gauge-invariant operators of mass dimension $D>4$, where the leading effects arise from $D=6$ operators  (neglecting lepton-number violating operators, irrelevant to our study). Working at this level, the extension of the SM that is relevant for multi-Higgs boson production reads:
\begin{equation}
  \label{eq:Lfinal}
  \begin{split}
	{\cal L} = {\cal L}_{\rm SM} &+ \frac{c_H}{2\Lambda^2}(\partial^\mu |H|^2)^2 - \frac{c_6}{\Lambda^2} \lambda_\mathrm{SM} |H|^6
	\\ 
	&- \left( \frac{c_t}{\Lambda^2}y_t |H|^2 \bar Q_L H^c t_R + \frac{c_b}{\Lambda^2}y_b |H|^2 \bar Q_L H b_R + \frac{c_\tau}{\Lambda^2}y_\tau |H|^2 \bar L_L H \tau_R + \text{h.c.} \right)
	\\
	&+ \frac{\alpha_s c_g}{4 \pi \Lambda^2} |H|^2 G_{\mu\nu}^a G^{\mu\nu}_a
	+ \frac{\alpha^{\prime}\,  c_\gamma}{ 4 \pi \Lambda^2} |H|^2 B_{\mu\nu} B^{\mu\nu}\\
	&+ \frac{i g\,  c_{HW} }{16 \pi^2 \Lambda^2 }(D^\mu H)^\dagger\sigma_k (D^\nu H) W_{\mu\nu}^k
	+ \frac{i g^\prime\,  c_{HB}}{ 16 \pi^2 \Lambda^2 }(D^\mu H)^\dagger (D^\nu H) B_{\mu\nu}\\
	&+ \frac{i g\, c_{W}}{2 \Lambda^2} (H^\dagger \sigma_k \overleftrightarrow D^\mu H)
	 D^\nu W_{\mu\nu}^k
	+ \frac{i g^\prime\,  c_B}{2 \Lambda^2} (H^\dagger \overleftrightarrow D^\mu H)
	 \partial^\nu B_{\mu\nu}\\
	& + {\cal L}_{\text{CP}}
	 + {\cal L}_{\text{4f}}\, ,
  \end{split}
\end{equation}
where $\alpha_s$ is the strong coupling constant and $\alpha^{\prime} \equiv g^{\prime \, 2} / 4 \pi$.

The full set of $D = 6$ operators that can be formed out of the SM field content was first obtained in~\cite{Buchmuller:1985jz} and reduced to a non-redundant minimal set in~\cite{Grzadkowski:2010es}. Here, we employed equations of motion to move to the basis used in~\cite{Elias-Miro:2013mua,Pomarol:2013zra}  and then imposed constraints from precision tests to neglect a class of operators whose effect is already constrained
to be at most 1\% with respect to the SM, following~\cite{Dumont:2013wma,Corbett:2012dm,Corbett:2012ja,Corbett:2013pja,Contino:2013kra}. Including these operators would have a negligible numerical impact on the analysis, given the experimental and theoretical errors.

Precision measurements also lead to the approximate restrictions~\cite{Elias-Miro:2013mua}
\begin{equation}
	\label{eq:res}
	\frac{c_{HB}}{16 \pi^2} = - \frac{c_{HW}}{16 \pi^2}	= - c_B = c_W \, ,
\end{equation}
which we will employ in the following.
Thus our setup corresponds to a restricted strongly-interacting light Higgs (SILH) Lagrangian~\cite{Giudice:2007fh} where $c_T$ has been set to zero and the relations (\ref{eq:res}) are used.

We note here that we do not include the effects of the chromomagnetic dipole operator, which has been shown be significant in processes such as $pp \rightarrow t\bar{t}$, $pp \rightarrow h$, $pp\rightarrow t\bar{t} h$,~\cite{Maltoni:2016yxb,Deutschmann:2017qum,DiNoi:2023ygk} as well as Higgs boson pair production~\cite{Heinrich:2023rsd}. Nevertheless, the chromomagnetic operator is subleading from the point of view of weakly interacting UV dynamics~\cite{Grober:2843280}, and we therefore leave it to future work.

The terms contributing to multi-Higgs production via gluon fusion at the LHC at $D=6$ EFT are:
\begin{equation}\label{eq:d6lagr}
\begin{split}
\mathcal{L}_{h^n} = &- \mu^2 |H|^2 - \lambda |H|^4 -  \left( y_t \bar Q_L H^c t_R + y_b \bar Q_L H b_R + \text{h.c.} \right)
\\
&+ \frac{c_H}{2\Lambda^2}(\partial^\mu |H|^2)^2 - \frac{c_6}{\Lambda^2} \lambda_\mathrm{SM} |H|^6  + \frac{\alpha_s c_g}{4\pi\Lambda^2} |H|^2 G_{\mu\nu}^a G^{\mu\nu}_a
\\
&- \left( \frac{c_t}{\Lambda^2} y_t |H|^2 \bar Q_L H^c t_R	+ \frac{c_b}{\Lambda^2}y_b |H|^2 \bar Q_L H b_R + \text{h.c.} \right) , 
\end{split}
\end{equation}
where the first line includes the relevant Standard Model terms that will receive corrections from $D=6$ operators. For the purposes of the current study, we set $C_H = 0$ in what follows. We also assume minimal flavour violation~\cite{DAmbrosio:2002vsn}, which leads to the coefficients of the Yukawa-like terms in the last row of eq.~\ref{eq:d6lagr} being proportional to the (SM-like) Yukawa couplings. 

Examining the relevant terms in the scalar potential and expanding about the electroweak minimum, we get:
\begin{eqnarray}
\mathcal{L}_\mathrm{self} = &-&\mu^2 \frac{ (v+h)^2 } { 2} - \lambda \frac{ (v+h)^4 } { 4} - \frac{ c_6 \lambda_\mathrm{SM} } { \Lambda^2 }  \frac{ (v+h)^6 } { 8 } \nonumber \\
= &-& \frac{ \mu^2 } { 2 } ( v^2 +  2 h v + h^2 ) - \frac{ \lambda} { 4} ( v^4 + 4 h v^3 + 6 h^2 v^2 + 4 h^3 v + h^4 ) \nonumber \\
&-& \frac{ c_6 \lambda_\mathrm{SM} } { 8 \Lambda^2 } ( v^6 + 6 v^5 h + 15 v^4 h^2 + 20 h^3 v^3 + 15 v^2 h^4 + 6 h^5 v + h^6 ) \;.
\end{eqnarray}
Omitting terms with $h^n$, $n>4$, and constant terms we arrive at
\begin{eqnarray}\label{eq:selftriple}
\mathcal{L}_\mathrm{self} =  &-& \frac{ \mu^2 } { 2 } ( 2 h v + h^2 ) - \frac{ \lambda} { 4} ( 4 h v^3 + 6 h^2 v^2 + 4 h^3 v + h^4 ) \nonumber \\
&-& \frac{ c_6 \lambda_\mathrm{SM} } { 8 \Lambda^2 } (  6 h v^5 + 15 h^2 v^4 + 20 h^3 v^3 + 15 h^4 v^2 ) + ~... \;.
\end{eqnarray}

Finally, we focus on the fermion-Higgs boson interactions that receive contributions from
\begin{eqnarray}
\mathcal{L}_{hf} = - \frac{ y_f } { \sqrt{2} } \bar{f}_L ( h + v ) f_R - \frac { c_f y_f } { \Lambda^2 } \frac{ ( v+h )^2 } { 2} \bar{f}_L  \frac{ ( v+h ) } { \sqrt{2} } f_R\, + \text{h.c.}\;,
\end{eqnarray}
where $f = t, b, ~...$, with $f_{L,R}$ the left- and right-handed fields, and the first term comes from the SM whereas the second term is a dimension-6 contribution. Expanding, we obtain
\begin{eqnarray}\label{eq:hf}
\mathcal{L}_{hf} = &-& \frac{ y_f v} { \sqrt{2} }  \left( 1 + \frac{c_t v^2 } { 2 \Lambda^2 } \right)\bar{f}_L f_R \nonumber \\
&-& \frac{ y_f } { \sqrt{2} } \left( 1 + \frac{ 3 c_f v^2 } { 2 \Lambda^2 } \right) \bar{f}_L f_R h \nonumber \\
&-& \frac{ y_f } { \sqrt{2} } \left( \frac{ 3 c_f v } { 2 \Lambda^2 } \right) \bar{f}_L f_R h^2 \nonumber \\
&-& \frac{ y_f } { \sqrt{2} } \left( \frac{ c_f  } { 2 \Lambda^2 } \right) \bar{f}_L f_R h^3 \, + \text{h.c.} \;.
\end{eqnarray}
The first line gives the expression for the modified fermion mass,
\begin{equation}
m_f = \frac{ y_f v} { \sqrt{2} } \left( 1 + \frac{c_t v^2 } { 2 \Lambda^2 } \right) \,,
\end{equation}
and we can re-express eq.~(\ref{eq:hf}) in terms of this:
\begin{eqnarray}\label{eq:hfmass}
\mathcal{L}_{hf} = &-& m_f \bar{f}_L f_R \nonumber \\
&-&   \frac{m_f}{v}  \left( 1 + \frac{  c_f v^2 } {  \Lambda^2 } \right) \bar{f}_L f_R h \nonumber \\
&-&  \frac{m_f}{v} \left( \frac{ 3c_f v } {  2\Lambda^2 } \right) \bar{f}_L f_R h^2 \nonumber \\
&-&   \frac{m_f}{v} \left( \frac{ c_f } {  2\Lambda^2 } \right) \bar{f}_L f_R h^3 
\, + \text{h.c.} \;.
\end{eqnarray}
The final term that we need to consider is
\begin{eqnarray}\label{eq:hg}
\mathcal{L}_{hg} &=& \frac{ \alpha_s c_g } { 4\pi \Lambda^2} |H|^2 G_{\mu\nu}^a G^{\mu\nu}_a \nonumber \\
&=& \frac{ \alpha_s c_g } { 4\pi \Lambda^2}  \frac{(h + v)^2}{2} G_{\mu\nu}^a G^{\mu\nu}_a \nonumber \\
&=& \frac{ \alpha_s c_g } { 4 \pi \Lambda^2} ( h v + \frac{h^2}{2} ) G_{\mu\nu}^a G^{\mu\nu}_a + ~...\;,
\end{eqnarray}
where the omitted constant term can be absorbed into an unobservable re-definition of the gluon wave function. 

We thus obtain the following interactions in terms of the Higgs boson scalar $h$, relevant to Higgs boson multi-production:
\begin{equation}\label{eq:Lgghmulti}
\begin{split}
\mathcal{L}_\mathrm{D=6} = &- \frac{ m_h^2 } { 2 v } \left( 1 + c_6 \right) h^3 - \frac{ m_h^2 } { 8 v^2 }  \left( 1  + 6 c_6 \right) h^4
\\
&+ \frac{\alpha_s c_g} {4 \pi } \left( \frac{h}{v} + \frac{h^2} {2v^2} \right) G_{\mu\nu}^a G^{\mu\nu}_a
\\
&- \left[ \frac{m_t}{v} \left( 1 + c_t   \right) \bar{t}_L t_R h +  \frac{m_b}{v} \left( 1 + c_b   \right) \bar{b}_L b_R h + \text{h.c.} \right]
\\
&- \left[ \frac{m_t}{v^2}\left( \frac{3 c_t}{2} \right) \bar{t}_L t_R h^2 +  \frac{m_b}{v^2}\left( \frac{3 c_b}{2}  \right) \bar{b}_L b_R h^2 +  \text{h.c.} \right] 
\\
&- \left[ \frac{m_t}{v^3}\left( \frac{c_t}{2} \right) \bar{t}_L t_R h^3 +  \frac{m_b}{v^3}\left( \frac{c_b}{2}  \right) \bar{b}_L b_R h^3 +  \text{h.c.} \right],
\end{split}
\end{equation}
where we have explicitly written down the contributing components of the $Q_L$ doublets.

\section{The Electroweak Chiral Lagrangian Including a Light Higgs Boson}\label{app:chiral}

The $D=6$ Lagrangian of eq.~\ref{eq:d6lagr} represents a ``linear'' realization of an effective field theory, including the leading terms in terms of ``canonical dimensions'' $D$, and assuming baryon- and lepton-number conservation. This is also sometimes referred to as ``SMEFT''. An alternative way to organize the EFT is by chiral dimensions, which yields a ``nonlinear'' realization. 

In this generalization of the SM, the gauge interactions are unchanged (at leading order), but general anomalous couplings are introduced for the physical Higgs boson. To do this in a consistent, gauge-invariant way, the scalar fields of the theory have to be decomposed into the three Goldstone fields $\varphi^a$, described by:
\begin{equation}
U = \exp ( 2 i \varphi^a T^a / v) = \exp(2 i \Phi/v)\;,
\end{equation}
where 
\begin{equation}
\Phi = \varphi^a T^a = \frac{1}{\sqrt{2}}\left(\begin{matrix}
\varphi^0/\sqrt{2} & \varphi^+ \\
\varphi^- & -\varphi^0/\sqrt{2}
\end{matrix}\right)\;,
\end{equation}
and where $T^a$ are the generators of $\mathrm{SU}(2)$. The polar decomposition of the Higgs doublet $H$ is then given by:
\begin{equation}
    H = \frac{v+h}{\sqrt{2}} U \left(\begin{matrix} 0 \\ 1 \end{matrix}\right)\;,
\end{equation}
where $h$ is the physical Higgs boson. 

Under electroweak gauge transformations $\mathrm{SU}(2)_L \times \mathrm{U}(1)_Y$:
\begin{equation}
U \rightarrow g_L U g_Y^\dagger\;,\;\;\;\; h\rightarrow h\;,\;\;\;\; g_{L,R} \in \mathrm{SU}(2)_{L,R}\;,
\end{equation}
such that the $h$ is invariant, and its couplings can be consistently modified. The transofmrations $g_L$, and the $\mathrm{U}(1)_Y$ subgroup of $g_R$ are gauged, so that the covariant derivatives are given by:\footnote{The term ``nonlinear'' describing this EFT comes from the fact that the scalar sector of the SM possesses a larger symmetry $\mathrm{SU}_L \times \mathrm{SU}_R$, known as the chiral electroweak symmetry, under which the Goldstone bosons $\varphi^a$ transform nonlinearly, in contrast to the usual Higgs doublet field, which transforms linearly~\cite{LHCHiggsCrossSectionWorkingGroup:2016ypw}.}
\begin{equation}
\mathrm{D}_\mu U = \partial_\mu U + ig W_\mu U - i g' B_\mu U T_3\;,\;\;\;\; \mathrm{D}_\mu h = \partial_\mu h\;.
\end{equation}

New dynamics may appear in the Higgs sector at a new physics scale $f \sim \mathcal{O}(\mathrm{TeV})$. This could be, e.g.\ composite, pseudo-Goldstone Higgs particles~\cite{Agashe:2004rs,Contino:2006qr,Contino:2010rs,Falkowski:2007hz,Carena:2014ria}, but also by other models with a modified Higgs sector with either weak or strong couplings. We can then parametrize deviations of the SM couplings by a quantity $\xi = v^2/f^2$, where $v=246$~GeV is the electroweak scale. Anomalous contributions, with respect to the SM, of order $\xi$ in the Higgs couplings will generically lead to a cut-off $\Lambda = 4\pi f$ in the effective description of the new Higgs dynamics. This picture might be supplemented by TeV-scale (order $f$) new degrees of freedom (non-standard fermions, extra pseudo-Goldstone bosons), integrated out in the EFT at the electroweak scale $v$. The EFT can then be organized in full generality~\cite{Buchalla:2014eca} as a double expansion in $\xi = v^2/f^2$ and $f^2/\Lambda^2 = 1/16\pi^2$, which are the two dimensionless parameters that can be formed out of the three relevant scales $v$, $f$ and $\Lambda$. They are both small under the condition $v \ll f \ll \Lambda$. The expansion in $\xi$ amounts to an expansion of the Lagrangian in operators of increasing canonical dimension. The expansion in $f^2/\Lambda^2$, corresponds to a loop expansion or, equivalently, to an expansion in terms of increasing chiral dimension ($\chi$). To leading order in chiral dimensions, and to first order in $\xi$, the SM effective Lagrangian can be written in nonlinear notation as 
\begin{eqnarray}\label{eq:L2chiral}
{\cal L}_2 &=& -\frac{1}{2} \langle G_{\mu\nu}G^{\mu\nu}\rangle
-\frac{1}{2}\langle W_{\mu\nu}W^{\mu\nu}\rangle 
-\frac{1}{4} B_{\mu\nu}B^{\mu\nu}
+\bar q i\!\not\!\! Dq +\bar l i\!\not\!\! Dl
 +\bar u i\!\not\!\! Du +\bar d i\!\not\!\! Dd +\bar e i\!\not\!\! De 
\nonumber\\
&& +\frac{v^2}{4}\ \l L_\mu L^\mu \r\, \left( 1+F_U(h)\right)
+\frac{1}{2} \partial_\mu h \partial^\mu h - V(h) \nonumber\\
&& - v \left[ \bar q \left( Y_u +
       \sum^3_{n=1} Y^{(n)}_u \left(\frac{h}{v}\right)^n \right) U P_+r 
+ \bar q \left( Y_d + 
     \sum^3_{n=1} Y^{(n)}_d \left(\frac{h}{v}\right)^n \right) U P_-r
  \right. \nonumber\\ 
&& \quad\quad\left. + \bar l \left( Y_e +
   \sum^3_{n=1} Y^{(n)}_e \left(\frac{h}{v}\right)^n \right) U P_-\eta 
+ {\rm h.c.}\right]\;,
\end{eqnarray}
with $L_\mu=i UD_\mu U^\dagger$, $P_\pm = 1/2\pm T_3$, and
\begin{align}
F_U &=(2-a_2)\frac{h}{v} +(1-2a_2)\left(\frac{h}{v}\right)^2 -\frac{4}{3} a_2 
\left(\frac{h}{v}\right)^3-\frac{1}{3} a_2 \left(\frac{h}{v}\right)^4\;,
\end{align}
\begin{align}
V&=\frac{m^2_h}{2} h^2 + \frac{m^2_h v^2}{2}\left[ 
\left( 1+\frac{4}{3}a_1-\frac{3}{2}a_2\right)\left(\frac{h}{v}\right)^3
+\left(\frac{1}{4}+ 2 a_1 - \frac{25}{12}a_2\right)
\left(\frac{h}{v}\right)^4 \right. \nonumber\\ 
&\left. \hspace*{3cm} +(a_1-a_2)\left(\frac{h}{v}\right)^5+
\frac{a_1-a_2}{6}\left(\frac{h}{v}\right)^6\right]\;,
\end{align}
\begin{align}
Y^{(1)}_f&=\left(1-\frac{a_2}{2}\right)Y_f+ 2\bar Y_f,\qquad
Y^{(2)}_f=3 Y^{(3)}_f = -\frac{a_2}{2} Y_f + 3 \bar Y_f,\qquad f=u,d,e\;.
\end{align}
For generality, we have included generic flavor matrices $\bar Y_f$ arising 
at NLO. In scenarios with minimal flavor violation \cite{DAmbrosio:2002vsn}, 
we can assume $\bar Y_f \propto Y_f$, as we did for the $D=6$ EFT (i.e.\ linear) scenario. 

When applying the chiral Lagrangian formalism to processes that arise only at one-loop level in the SM, such as $h\rightarrow gg$, $h\rightarrow \gamma \gamma$ or $h\rightarrow Z\gamma$, terms at NLO (chiral dimension 4, or loop order 1) become relevant, in addition to loops with modified couplings from eq.~\ref{eq:L2chiral}. For the complete list of NLO operators we refer the reader to~\cite{Buchalla:2013rka}. The relevant terms here are $\sim e^2 F_{\mu\nu} F^{\mu\nu} h$, $\sim eg' F_{\mu\nu}Z^{\mu\nu}h$ and $\sim g_s^2 \left< G_{\mu\nu} G^{\mu\nu}\right> h$. 

In summary, the terms from the electroweak chiral Lagrangian relevant to multi-Higgs boson production, up to $gg \rightarrow hhh$, are given by~\cite{LHCHiggsCrossSectionWorkingGroup:2016ypw,Buchalla:2018yce}:
\begin{equation}\label{eq:LgghhPhenoExpChiral}
\begin{split}
\mathcal{L}_{\chi} \supset &- c_{hhh}\lambda_\mathrm{SM} v  h^3 - c_{hhhh} \frac{ \lambda_\mathrm{SM} } { 4 } h^4
\\
&+ \frac{\alpha_s } {8 \pi}  \left( c_{ggh} \frac{h}{v} + c_{gghh} \frac{h^2} {v^2} \right) G_{\mu\nu}^a G^{\mu\nu}_a
\\
&- \left[ \frac{m_t}{v} c_{t} \bar{t}_L t_R h +  \frac{m_b}{v} c_{b} \bar{b}_L b_R h + \text{h.c.} \right]
\\
&- \left[ \frac{m_t}{v^2} c_{tt} \bar{t}_L t_R h^2 +  \frac{m_b}{v^2} c_{bb} \bar{b}_L b_R h^2 +  \text{h.c.} \right] 
\\
&- \left[ \frac{m_t}{v^3}\left( c_{ttt} \right) \bar{t}_L t_R h^3 +  \frac{m_b}{v^3}\left( c_{bbb}  \right) \bar{b}_L b_R h^3 +  \text{h.c.} \right],
\end{split}
\end{equation}
where we have again taken $\lambda_\mathrm{SM} \equiv m_h^2 / 2v^2$. The identifications that yield the phenomenological Lagrangian employed in this article, eq.~\ref{eq:LgghhPhenoExp}, are: $c_{hhh} \rightarrow (1+c_3)$, $c_{hhhh} \rightarrow (1+d_4)$, $c_{ggh} \rightarrow 2c_{g1}/3$, $c_{gghh}\rightarrow c_{g2}/3$, $c_f \rightarrow (1+c_{f1})$, $c_{ff} \rightarrow c_{f2}$ and $c_{fff} \rightarrow c_{f3}/2$.

\section{Validation of the Monte Carlo Implementation}\label{app:validation}

\begin{figure}[htp]
\centering
\setlength{\unitlength}{0.6cm}
\subfigure{
\begin{fmffile}{ttHhhh}
    \begin{fmfgraph*}(7,4)
        \fmfleft{i1,i2}
        \fmfright{o1,o2,o3}
        \fmflabel{$\bar{t}$}{i1}
        \fmflabel{$t$}{i2}
        \fmflabel{$h$}{o3}
        \fmflabel{$h$}{o2}
        \fmflabel{$h$}{o1}
        \fmf{fermion}{i2,v1}
        \fmf{fermion}{v1,i1}  
        \fmf{dashes, label=$H$}{v1,v2}
        \fmf{dashes}{v2,o1}
        \fmf{dashes}{v2,o2}
        \fmf{dashes}{v2,o3}
    \end{fmfgraph*}
\end{fmffile}
}
\subfigure{
\begin{fmffile}{tthhhEFT}
    \begin{fmfgraph*}(7,4)
        \fmfleft{i1,i2}
        \fmfright{o1,o2,o3}
        \fmflabel{$\bar{t}$}{i1}
        \fmflabel{$t$}{i2}
        \fmflabel{$h$}{o3}
        \fmflabel{$h$}{o2}
        \fmflabel{$h$}{o1}
        \fmf{fermion}{i2,v1}
        \fmf{fermion}{v1,i1}  
        \fmfblob{10}{v1}      
        \fmf{dashes}{v1,o1}
        \fmf{dashes}{v1,o2}
        \fmf{dashes}{v1,o3}
    \end{fmfgraph*}
\end{fmffile}
}
\caption{The $t\bar{t} \rightarrow hhh$ process used to validate the implementation of the $t\bar{t}hhh$ vertex. The process is shown in the model with a new heavy scalar ($H$) and in the limit of $M_H \gg \sqrt{\hat{s}}$.}
\label{fig:tthhhEFT}
\end{figure}
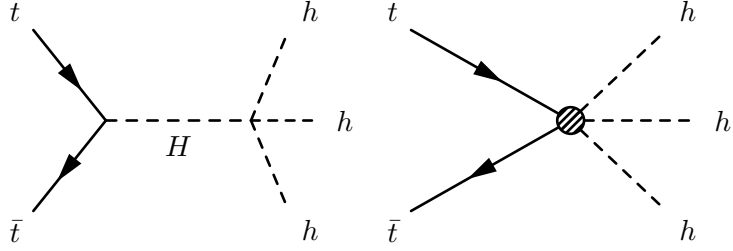

\begin{figure}[htp]
\begin{center}
  \includegraphics[width=0.6\columnwidth]{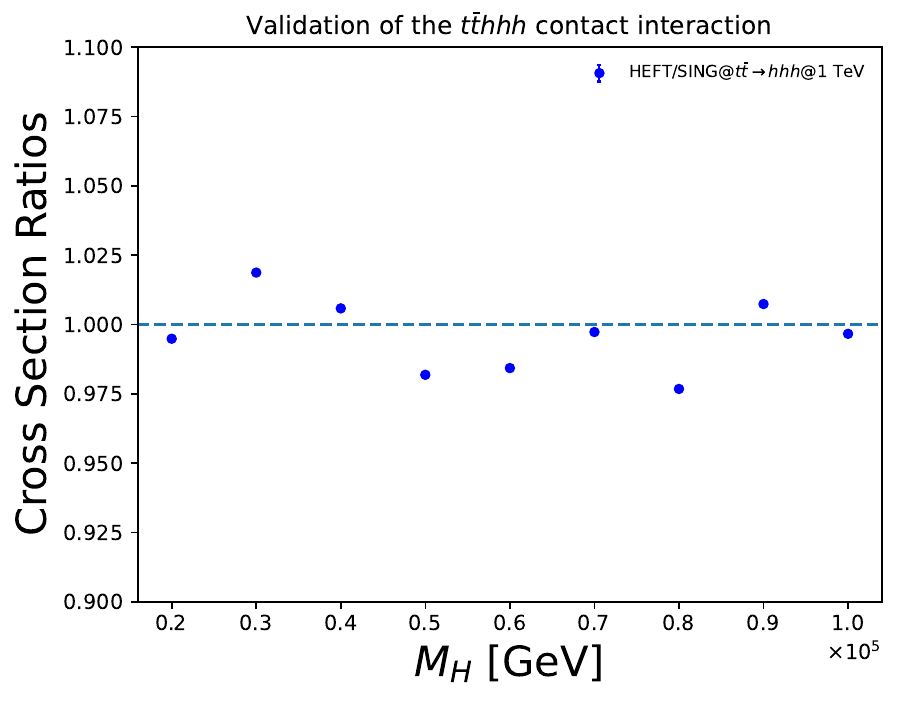}
\caption{\label{fig:tthhh} The ratio of cross sections between for $t\bar{t} \rightarrow hhh$ between the anomalous interaction (HEFT) and the heavy scalar ($H$) descriptions. See main text for further details.}
\end{center}
\end{figure}

Almost all vertices relevant in triple Higgs boson production are present in pair production. Therefore, these were validated using the \texttt{HiggsPair} model in \texttt{HERWIG 7} Monte Carlo event generator, with minor modifications, in the $gg \rightarrow hh$ process. 
Validation of the new $t\bar{t}hhh$ contact interaction was performed by considering the processes $t\bar{t} \rightarrow hhh$ at 1~TeV in the $t\bar{t}$ center-of-mass frame without a PDF involved. The processes were calculated both in the HEFT and in a model with a heavy scalar ($H$) that couples to $t\bar{t}$ and $hhh$ only. This implies taking the limit of the effective field theory directly and checking whether the effective vertex functions as expected. The matching of the coefficient of eq.~\ref{eq:LgghhPhenoExp} with the singlet model, e.g. of~\cite{Papaefstathiou:2020iag}, implies that $c_{t3} = 2 v^2/M_H^2$, when the the quartic coupling between the heavy scalar and the three Higgs bosons is set to $\lambda_{1112} = 1$ and the mixing angle $\theta = \pi/2$ such that the SM Higgs boson is decoupled. Figure~\ref{fig:tthhh} shows the ratio of the anomalous $t\bar{t}hhh$ interaction cross section over the heavy scalar cross section for various masses of the heavy scalar, chosen to be much higher than the center-of-mass energy.

\section{Feynman Diagrams}\label{app:feyndiags}

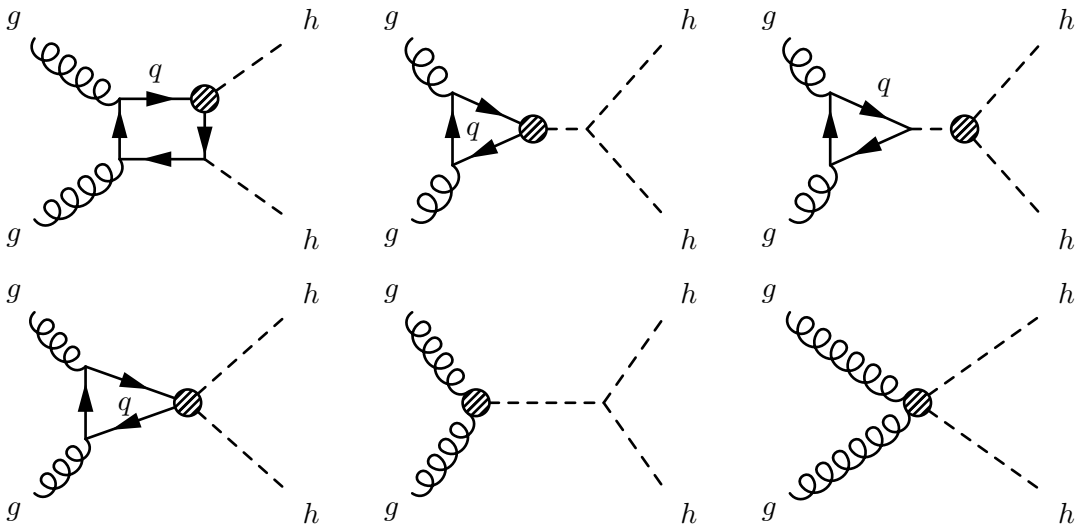
\begin{figure}[htp]
    \centering
    \setlength{\unitlength}{0.6cm}
\subfigure{    
\begin{fmffile}{box1i}
    \begin{fmfgraph*}(7,4)
        \fmfleft{i1,i2}
        \fmfright{o1,o2}
        \fmf{gluon}{i1,v1}
        \fmflabel{$g$}{i1}
        \fmflabel{$g$}{i2}
        \fmflabel{$h$}{o2}
        \fmflabel{$h$}{o1}
        \fmf{gluon}{i2,v2}
        \fmfblob{10}{v3}    
        \fmf{fermion,label=$q$}{v2,v3}
        \fmf{fermion}{v4,v1}
        \fmf{fermion}{v1,v2}
        \fmf{fermion}{v3,v4}
        \fmf{dashes}{v4,o1}
        \fmf{dashes}{v3,o2}
    \end{fmfgraph*}
\end{fmffile}
}
\subfigure{
\begin{fmffile}{triangle1i}
    \begin{fmfgraph*}(7,4)
        \fmfleft{i1,i2}
        \fmfright{o1,o2}
        \fmflabel{$g$}{i1}
        \fmflabel{$g$}{i2}
        \fmflabel{$h$}{o2}
        \fmflabel{$h$}{o1}
        \fmf{gluon,tension=2}{i1,v1}
        \fmf{gluon,tension=2}{i2,v2}
        \fmf{fermion,label=$q$}{v2,v3}
        \fmf{fermion}{v3,v1}
        \fmf{fermion}{v1,v2}
        \fmf{dashes,tension=3}{v3,v4}
        \fmfblob{10}{v3}
        \fmf{dashes}{v4,o1}
        \fmf{dashes}{v4,o2}
    \end{fmfgraph*}
\end{fmffile}
}
\subfigure{
\begin{fmffile}{triangle1i2}
    \begin{fmfgraph*}(7,4)
        \fmfleft{i1,i2}
        \fmfright{o1,o2}
        \fmflabel{$g$}{i1}
        \fmflabel{$g$}{i2}
        \fmflabel{$h$}{o2}
        \fmflabel{$h$}{o1}
        \fmf{gluon,tension=2}{i1,v1}
        \fmf{gluon,tension=2}{i2,v2}
        \fmf{fermion,label=$q$}{v2,v3}
        \fmf{fermion}{v3,v1}
        \fmf{fermion}{v1,v2}
        \fmf{dashes,tension=3}{v3,v4}
        \fmfblob{10}{v4}
        \fmf{dashes}{v4,o1}
        \fmf{dashes}{v4,o2}
    \end{fmfgraph*}
\end{fmffile}
}
\par\bigskip 
\par\bigskip
\subfigure{
\begin{fmffile}{triangle1i3}
    \begin{fmfgraph*}(7,4)
        \fmfleft{i1,i2}
        \fmfright{o1,o2}
        \fmflabel{$g$}{i1}
        \fmflabel{$g$}{i2}
        \fmflabel{$h$}{o2}
        \fmflabel{$h$}{o1}
        \fmf{gluon,tension=2}{i1,v1}
        \fmf{gluon,tension=2}{i2,v2}
        \fmf{fermion,label=$q$}{v2,v3}
        \fmf{fermion}{v3,v1}
        \fmf{fermion}{v1,v2}
        \fmf{dashes,tension=3}{v3,v4}
        \fmfblob{10}{v3}
        \fmf{dashes}{v3,o1}
        \fmf{dashes}{v3,o2}
    \end{fmfgraph*}
\end{fmffile}
}
\subfigure{
\begin{fmffile}{gghhi1}
    \begin{fmfgraph*}(7,4)
        \fmfleft{i1,i2}
        \fmfright{o1,o2}
        \fmflabel{$g$}{i1}
        \fmflabel{$g$}{i2}
        \fmflabel{$h$}{o2}
        \fmflabel{$h$}{o1}
        \fmf{gluon}{i1,v1}
        \fmf{gluon}{i2,v1}
        \fmf{dashes}{v1,v2}
        \fmfblob{10}{v1}
        \fmf{dashes}{v2,o1}
        \fmf{dashes}{v2,o2}
    \end{fmfgraph*}
\end{fmffile}
}
\subfigure{
\begin{fmffile}{gghh}
    \begin{fmfgraph*}(7,4)
        \fmfleft{i1,i2}
        \fmfright{o1,o2}
        \fmflabel{$g$}{i1}
        \fmflabel{$g$}{i2}
        \fmflabel{$h$}{o2}
        \fmflabel{$h$}{o1}
        \fmf{gluon}{i1,v1}
        \fmf{gluon}{i2,v1}
        \fmfblob{10}{v1}
        \fmf{dashes}{v1,o1}
        \fmf{dashes}{v1,o2}
    \end{fmfgraph*}
\end{fmffile}
}
\par\bigskip
    \caption{Example Feynman diagrams with \textit{one} EFT operator insertion contributing to Higgs boson pair production.}
    \label{fig:eft1insert}
\end{figure}

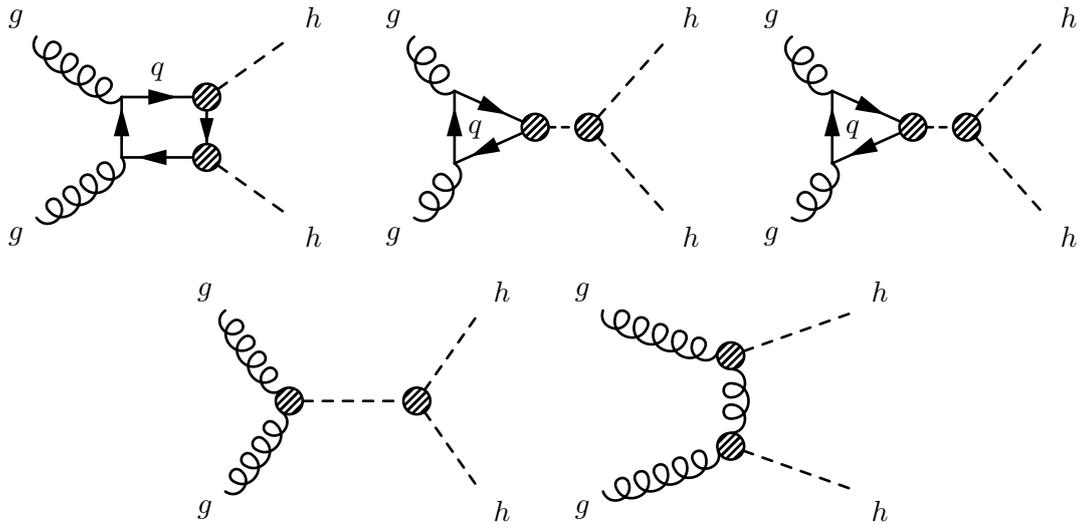
\begin{figure}[htp]
    \centering
    \setlength{\unitlength}{0.6cm}
\subfigure{    
\begin{fmffile}{box2i}
    \begin{fmfgraph*}(7,4)
        \fmfleft{i1,i2}
        \fmfright{o1,o2}
        \fmf{gluon}{i1,v1}
        \fmflabel{$g$}{i1}
        \fmflabel{$g$}{i2}
        \fmflabel{$h$}{o2}
        \fmflabel{$h$}{o1}
        \fmf{gluon}{i2,v2}
        \fmfblob{10}{v3}   
        \fmfblob{10}{v4} 
        \fmf{fermion,label=$q$}{v2,v3}
        \fmf{fermion}{v4,v1}
        \fmf{fermion}{v1,v2}
        \fmf{fermion}{v3,v4}
        \fmf{dashes}{v4,o1}
        \fmf{dashes}{v3,o2}
    \end{fmfgraph*}
\end{fmffile}
}
\subfigure{
\begin{fmffile}{triangle2i}
    \begin{fmfgraph*}(7,4)
        \fmfleft{i1,i2}
        \fmfright{o1,o2}
        \fmflabel{$g$}{i1}
        \fmflabel{$g$}{i2}
        \fmflabel{$h$}{o2}
        \fmflabel{$h$}{o1}
        \fmf{gluon,tension=2}{i1,v1}
        \fmf{gluon,tension=2}{i2,v2}
        \fmf{fermion,label=$q$}{v2,v3}
        \fmf{fermion}{v3,v1}
        \fmf{fermion}{v1,v2}
        \fmf{dashes,tension=3}{v3,v4}
        \fmfblob{10}{v3}
        \fmfblob{10}{v4}
        \fmf{dashes}{v4,o1}
        \fmf{dashes}{v4,o2}
    \end{fmfgraph*}
\end{fmffile}
}
\subfigure{
\begin{fmffile}{triangle2i2}
    \begin{fmfgraph*}(7,4)
        \fmfleft{i1,i2}
        \fmfright{o1,o2}
        \fmflabel{$g$}{i1}
        \fmflabel{$g$}{i2}
        \fmflabel{$h$}{o2}
        \fmflabel{$h$}{o1}
        \fmf{gluon,tension=2}{i1,v1}
        \fmf{gluon,tension=2}{i2,v2}
        \fmf{fermion,label=$q$}{v2,v3}
        \fmf{fermion}{v3,v1}
        \fmf{fermion}{v1,v2}
        \fmf{dashes,tension=3}{v3,v4}
        \fmfblob{10}{v4}
        \fmfblob{10}{v3}
        \fmf{dashes}{v4,o1}
        \fmf{dashes}{v4,o2}
    \end{fmfgraph*}
\end{fmffile}
}
\par\bigskip 
\par\bigskip
\subfigure{
\begin{fmffile}{gghhi2}
    \begin{fmfgraph*}(7,4)
        \fmfleft{i1,i2}
        \fmfright{o1,o2}
        \fmflabel{$g$}{i1}
        \fmflabel{$g$}{i2}
        \fmflabel{$h$}{o2}
        \fmflabel{$h$}{o1}
        \fmf{gluon}{i1,v1}
        \fmf{gluon}{i2,v1}
        \fmf{dashes}{v1,v2}
        \fmfblob{10}{v1}
        \fmfblob{10}{v2}
        \fmf{dashes}{v2,o1}
        \fmf{dashes}{v2,o2}
    \end{fmfgraph*}
\end{fmffile}
}
\subfigure{
\begin{fmffile}{gghhtchan}
    \begin{fmfgraph*}(7,4)
        \fmfleft{i1,i2}
        \fmfright{o1,o2}
        \fmflabel{$g$}{i1}
        \fmflabel{$g$}{i2}
        \fmflabel{$h$}{o2}
        \fmflabel{$h$}{o1}
        \fmf{gluon}{i1,v1}
        \fmf{gluon}{i2,v2}
        \fmf{gluon}{v1,v2}
        \fmfblob{10}{v1}
        \fmfblob{10}{v2}
        \fmf{dashes}{v1,o1}
        \fmf{dashes}{v2,o2}
    \end{fmfgraph*}
\end{fmffile}
}
\par\bigskip
    \caption{Example Feynman diagrams with \textit{two} EFT operator insertions contributing to Higgs boson pair production.}
    \label{fig:eft2insert}
\end{figure}

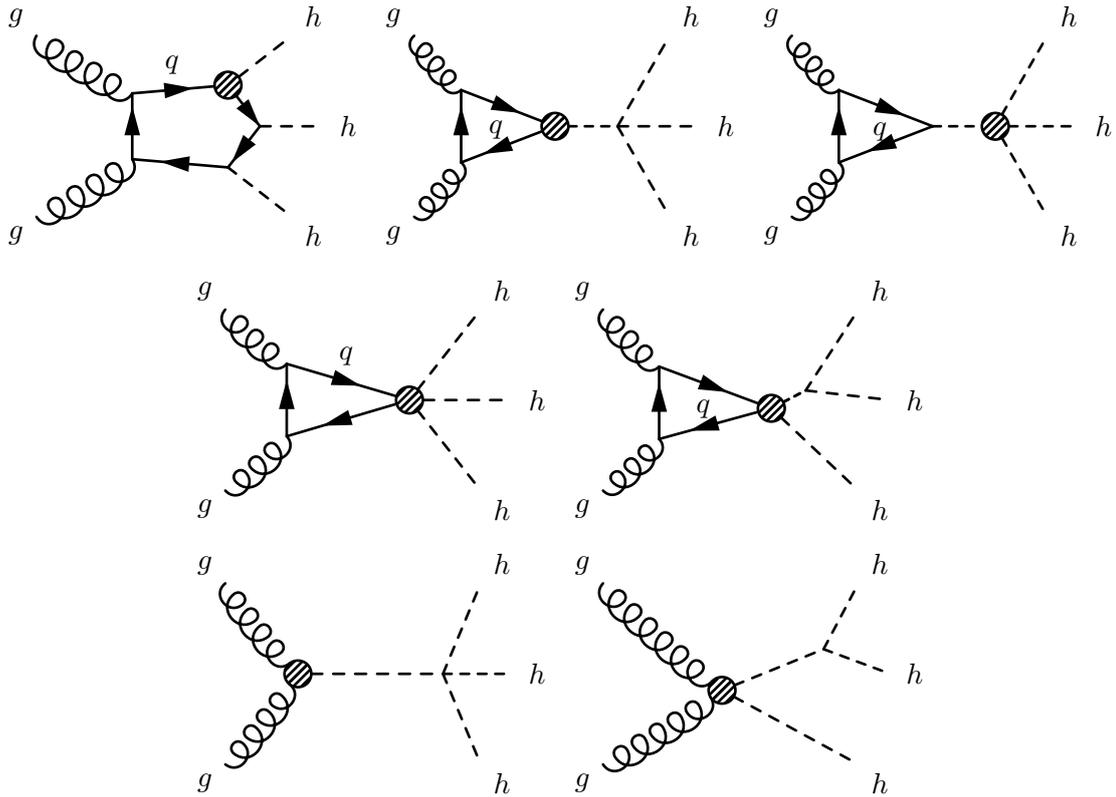
\begin{figure}[htp]
    \centering
    \setlength{\unitlength}{0.6cm}
\subfigure{    
\begin{fmffile}{box1ihhh}
    \begin{fmfgraph*}(7,4)
        \fmfleft{i1,i2}
        \fmfright{o1,o2,o3}
        \fmf{gluon}{i1,v1}
        \fmflabel{$g$}{i1}
        \fmflabel{$g$}{i2}
        \fmflabel{$h$}{o2}
        \fmflabel{$h$}{o3}
        \fmflabel{$h$}{o1}
        \fmf{gluon}{i2,v2}
        \fmfblob{10}{v3}    
        \fmf{fermion,label=$q$}{v2,v3}
        \fmf{fermion}{v3,v4}
        \fmf{fermion}{v4,v5}
        \fmf{fermion}{v5,v1}
        \fmf{fermion}{v1,v2}
        \fmf{dashes}{v3,o3}
        \fmf{dashes}{v4,o2}
        \fmf{dashes}{v5,o1}
    \end{fmfgraph*}
\end{fmffile}
}
\subfigure{
\begin{fmffile}{triangle1ihhh}
    \begin{fmfgraph*}(7,4)
        \fmfleft{i1,i2}
        \fmfright{o1,o2,o3}
        \fmflabel{$g$}{i1}
        \fmflabel{$g$}{i2}
        \fmflabel{$h$}{o2}
        \fmflabel{$h$}{o1}
        \fmflabel{$h$}{o3}
        \fmf{gluon,tension=2}{i1,v1}
        \fmf{gluon,tension=2}{i2,v2}
        \fmf{fermion,label=$q$}{v2,v3}
        \fmf{fermion}{v3,v1}
        \fmf{fermion}{v1,v2}
        \fmf{dashes,tension=3}{v3,v4}
        \fmfblob{10}{v3}
        \fmf{dashes}{v4,o1}
        \fmf{dashes}{v4,o2}
        \fmf{dashes}{v4,o3}
    \end{fmfgraph*}
\end{fmffile}
}
\subfigure{
\begin{fmffile}{triangle1i2hhh}
    \begin{fmfgraph*}(7,4)
        \fmfleft{i1,i2}
        \fmfright{o1,o2,o3}
        \fmflabel{$g$}{i1}
        \fmflabel{$g$}{i2}
        \fmflabel{$h$}{o2}
        \fmflabel{$h$}{o1}
        \fmflabel{$h$}{o3}
        \fmf{gluon,tension=2}{i1,v1}
        \fmf{gluon,tension=2}{i2,v2}
        \fmf{fermion,label=$q$}{v2,v3}
        \fmf{fermion}{v3,v1}
        \fmf{fermion}{v1,v2}
        \fmf{dashes,tension=3}{v3,v4}
        \fmfblob{10}{v4}
        \fmf{dashes}{v4,o1}
        \fmf{dashes}{v4,o2}
        \fmf{dashes}{v4,o3}
    \end{fmfgraph*}
\end{fmffile}
}
\par\bigskip 
\par\bigskip
\subfigure{
\begin{fmffile}{triangle1i3hhh}
    \begin{fmfgraph*}(7,4)
        \fmfleft{i1,i2}
        \fmfright{o1,o2,o3}
        \fmflabel{$g$}{i1}
        \fmflabel{$g$}{i2}
        \fmflabel{$h$}{o2}
        \fmflabel{$h$}{o1}
        \fmflabel{$h$}{o3}
        \fmf{gluon,tension=2}{i1,v1}
        \fmf{gluon,tension=2}{i2,v2}
        \fmf{fermion,label=$q$}{v2,v3}
        \fmf{fermion}{v3,v1}
        \fmf{fermion}{v1,v2}
        \fmf{dashes,tension=3}{v3,v4}
        \fmfblob{10}{v3}
        \fmf{dashes}{v3,o1}
        \fmf{dashes}{v3,o2}
        \fmf{dashes}{v3,o3}
    \end{fmfgraph*}
\end{fmffile}
}
\subfigure{
\begin{fmffile}{triangle1i4hhh}
    \begin{fmfgraph*}(7,4)
        \fmfleft{i1,i2}
        \fmfright{o1,o2,o3}
        \fmflabel{$g$}{i1}
        \fmflabel{$g$}{i2}
        \fmflabel{$h$}{o2}
        \fmflabel{$h$}{o1}
        \fmflabel{$h$}{o3}
        \fmf{gluon,tension=2}{i1,v1}
        \fmf{gluon,tension=2}{i2,v2}
        \fmf{fermion,label=$q$}{v2,v3}
        \fmf{fermion}{v3,v1}
        \fmf{fermion}{v1,v2}
        \fmf{dashes,tension=3}{v3,v4}
        \fmfblob{10}{v3}
        \fmf{dashes}{v3,o1}
        \fmf{dashes}{v3,v4}
        \fmf{dashes}{v4,o2}
        \fmf{dashes}{v4,o3}
    \end{fmfgraph*}
\end{fmffile}
}
\par\bigskip 
\par\bigskip
\subfigure{
\begin{fmffile}{gghhhi1}
    \begin{fmfgraph*}(7,4)
        \fmfleft{i1,i2}
        \fmfright{o1,o2,o3}
        \fmflabel{$g$}{i1}
        \fmflabel{$g$}{i2}
        \fmflabel{$h$}{o3}
        \fmflabel{$h$}{o2}
        \fmflabel{$h$}{o1}
        \fmf{gluon}{i1,v1}
        \fmf{gluon}{i2,v1}
        \fmf{dashes}{v1,v2}
        \fmfblob{10}{v1}
        \fmf{dashes}{v2,o1}
        \fmf{dashes}{v2,o2}
        \fmf{dashes}{v2,o3}
    \end{fmfgraph*}
\end{fmffile}
}
\subfigure{
\begin{fmffile}{gghhh}
    \begin{fmfgraph*}(7,4)
        \fmfleft{i1,i2}
        \fmfright{o1,o2,o3}
        \fmflabel{$g$}{i1}
        \fmflabel{$g$}{i2}
        \fmflabel{$h$}{o2}
        \fmflabel{$h$}{o3}
        \fmflabel{$h$}{o1}
        \fmf{gluon}{i1,v1}
        \fmf{gluon}{i2,v1}
        \fmfblob{10}{v1}
        \fmf{dashes}{v1,o1}
        \fmf{dashes}{v1,v2}
        \fmf{dashes}{v2,o2}
         \fmf{dashes}{v2,o3}
    \end{fmfgraph*}
\end{fmffile}
}
\par\bigskip
    \caption{Example Feynman diagrams with \textit{one} EFT operator insertion contributing to Higgs boson triple production.}
    \label{fig:eft1insert_hhh}
\end{figure}

\begin{figure}[htp]
    \centering
    \setlength{\unitlength}{0.6cm}
\subfigure{    
\begin{fmffile}{box2ihhh}
    \begin{fmfgraph*}(7,4)
        \fmfleft{i1,i2}
        \fmfright{o1,o2,o3}
        \fmf{gluon}{i1,v1}
        \fmflabel{$g$}{i1}
        \fmflabel{$g$}{i2}
        \fmflabel{$h$}{o2}
        \fmflabel{$h$}{o3}
        \fmflabel{$h$}{o1}
        \fmf{gluon}{i2,v2}
        \fmfblob{10}{v3}   
        \fmfblob{10}{v4} 
        \fmf{fermion,label=$q$}{v2,v3}
        \fmf{fermion}{v3,v4}  
        \fmf{fermion}{v4,v5}
        \fmf{fermion}{v5,v1}
        \fmf{fermion}{v1,v2}
        \fmf{dashes}{v3,o3}
        \fmf{dashes}{v4,o2}
        \fmf{dashes}{v5,o1}
    \end{fmfgraph*}
\end{fmffile}
}
\subfigure{
\begin{fmffile}{triangle2i2hhh}
    \begin{fmfgraph*}(7,4)
        \fmfleft{i1,i2}
        \fmfright{o1,o2,o3}
        \fmflabel{$g$}{i1}
        \fmflabel{$g$}{i2}
        \fmflabel{$h$}{o2}
        \fmflabel{$h$}{o1}
        \fmflabel{$h$}{o3}
        \fmf{gluon,tension=2}{i1,v1}
        \fmf{gluon,tension=2}{i2,v2}
        \fmf{fermion,label=$q$}{v2,v3}
        \fmf{fermion}{v3,v1}
        \fmf{fermion}{v1,v2}
        \fmf{dashes,tension=4}{v3,v4}
        \fmfblob{10}{v4}
        \fmfblob{10}{v3}
        \fmf{dashes}{v4,o1}
        \fmf{dashes}{v4,v5}
        \fmf{dashes}{v5,o2}
        \fmf{dashes}{v5,o3}
    \end{fmfgraph*}
\end{fmffile}
}
\subfigure{
\begin{fmffile}{triangle2i2hhh2}
    \begin{fmfgraph*}(7,4)
        \fmfleft{i1,i2}
        \fmfright{o1,o2,o3}
        \fmflabel{$g$}{i1}
        \fmflabel{$g$}{i2}
        \fmflabel{$h$}{o2}
        \fmflabel{$h$}{o3}
        \fmflabel{$h$}{o1}
        \fmf{gluon,tension=2}{i1,v1}
        \fmf{gluon,tension=2}{i2,v2}
        \fmf{fermion,label=$q$}{v2,v3}
        \fmf{fermion}{v3,v1}
        \fmf{fermion}{v1,v2}
        \fmf{dashes,tension=3}{v3,v4}
        \fmfblob{10}{v4}
        \fmfblob{10}{v3}
        \fmf{dashes}{v4,o1}
        \fmf{dashes}{v4,o2}
        \fmf{dashes}{v4,o3}
    \end{fmfgraph*}
\end{fmffile}
}
\par\bigskip 
\par\bigskip
\subfigure{
\begin{fmffile}{gghhhi2}
    \begin{fmfgraph*}(7,4)
        \fmfleft{i1,i2}
        \fmfright{o1,o2,o3}
        \fmflabel{$g$}{i1}
        \fmflabel{$g$}{i2}
        \fmflabel{$h$}{o2}
        \fmflabel{$h$}{o3}
        \fmflabel{$h$}{o1}
        \fmf{gluon}{i1,v1}
        \fmf{gluon}{i2,v1}
        \fmf{dashes}{v1,v2}
        \fmfblob{10}{v1}
        \fmfblob{10}{v2}
        \fmf{dashes}{v2,o1}
        \fmf{dashes}{v2,o2}
        \fmf{dashes}{v2,o3}
    \end{fmfgraph*}
\end{fmffile}
}
\subfigure{
\begin{fmffile}{gghhhi22}
    \begin{fmfgraph*}(7,4)
        \fmfleft{i1,i2}
        \fmfright{o1,o2,o3}
        \fmflabel{$g$}{i1}
        \fmflabel{$g$}{i2}
        \fmflabel{$h$}{o2}
        \fmflabel{$h$}{o3}
        \fmflabel{$h$}{o1}
        \fmf{gluon}{i1,v1}
        \fmf{gluon}{i2,v1}
        \fmf{dashes}{v1,v2}
        \fmfblob{10}{v1}
        \fmfblob{10}{v2}
        \fmf{dashes}{v2,o1}
        \fmf{dashes}{v2,v3}
        \fmf{dashes}{v3,o2}
        \fmf{dashes}{v3,o3}
    \end{fmfgraph*}
\end{fmffile}
}
\subfigure{
\begin{fmffile}{gghhhtchan}
    \begin{fmfgraph*}(7,4)
        \fmfleft{i1,i2}
        \fmfright{o1,o2,o3}
        \fmflabel{$g$}{i1}
        \fmflabel{$g$}{i2}
        \fmflabel{$h$}{o2}
        \fmflabel{$h$}{o3}
        \fmflabel{$h$}{o1}
        \fmf{gluon}{i1,v1}
        \fmf{gluon}{i2,v2}
        \fmf{gluon}{v1,v2}
        \fmfblob{10}{v1}
        \fmfblob{10}{v2}
        \fmf{dashes}{v1,o1}
        \fmf{dashes}{v2,o2}
        \fmf{dashes}{v2,o3}
    \end{fmfgraph*}
    \end{fmffile}
}
\par\bigskip
    \caption{Example Feynman diagrams with \textit{two} EFT operator insertions contributing to Higgs boson triple production.}
    \label{fig:eft2insert_hhh}
\end{figure}
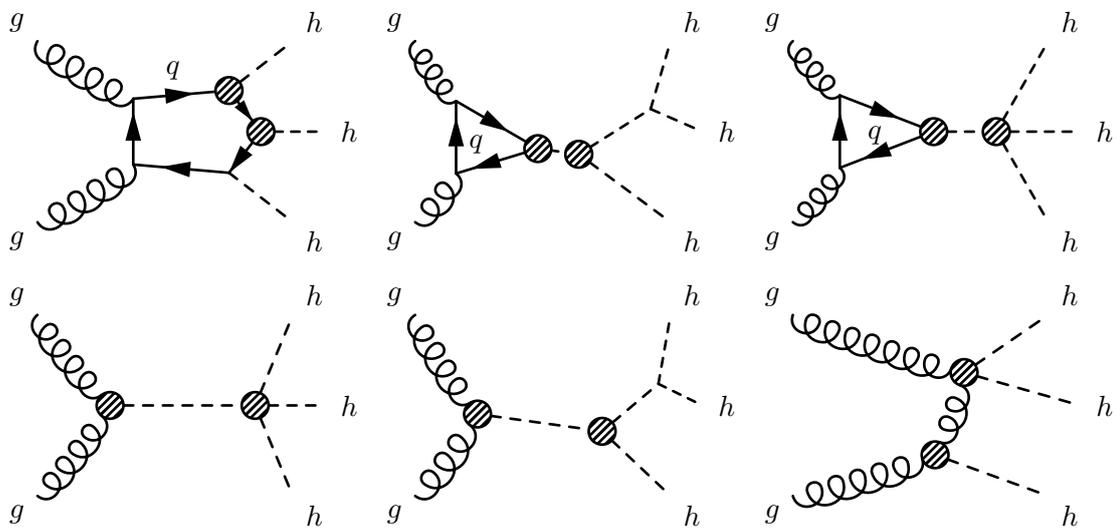

Figures~\ref{fig:eft1insert} and~\ref{fig:eft2insert} represent the Feynman diagrams for either one or two insertions of the operators used in the present article in Higgs boson pair production. Figures~\ref{fig:eft1insert_hhh} and~\ref{fig:eft2insert_hhh} represent the Feynman diagrams for either one or two insertions in the context of Higgs boson triple production. 
\clearpage

\section{Discovery and Limits with Up to Three Powers of Anomalous Operators}\label{app:triplelimits}

\begin{figure}[htp]
\begin{center}
    \includegraphics[width=0.495\columnwidth]{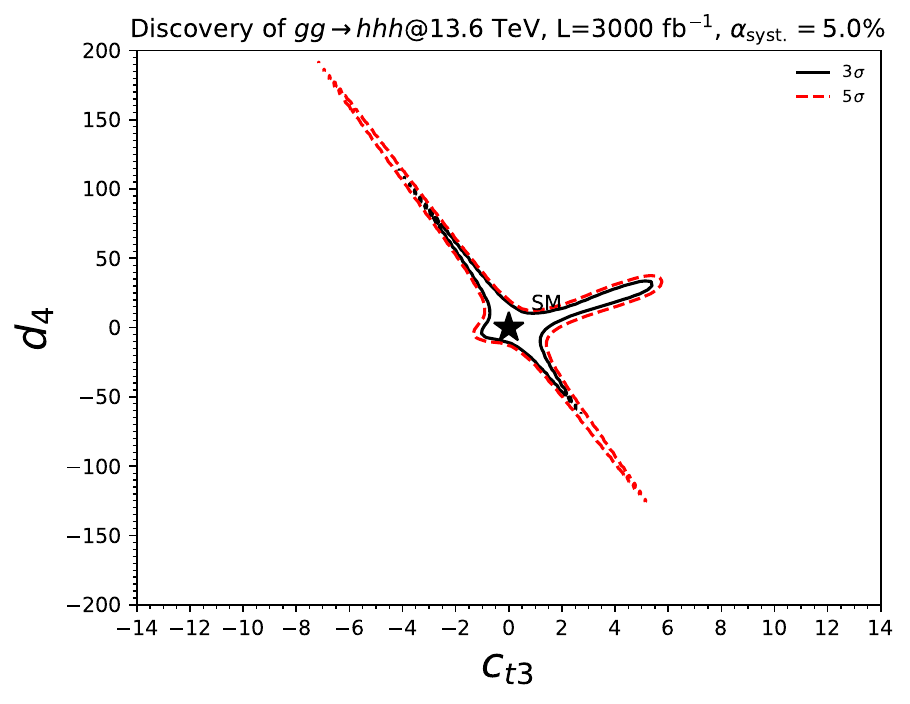}
  \includegraphics[width=0.495\columnwidth]{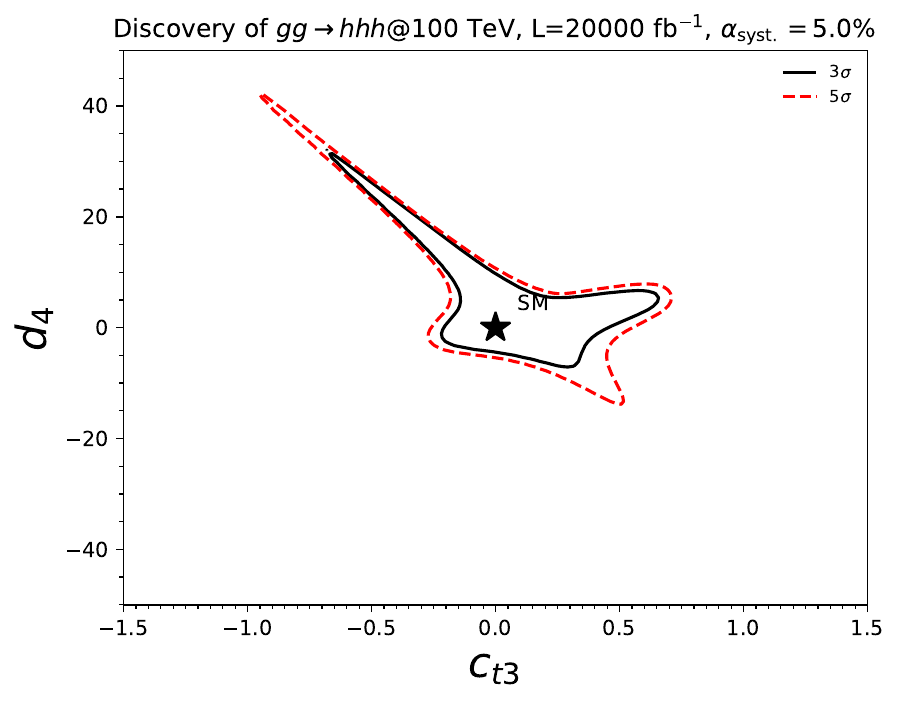}
\caption{\label{fig:2ddiscovery_tripleins} The 3$\sigma$ evidence (black solid) and 5$\sigma$ discovery (red dashed) curves on the $(c_{t3},d_4)$-plane for triple Higgs boson production with up to \textit{triple powers} of anomalous operators at the matrix element-squared level, at 13~TeV/3000~fb$^{-1}$ (left), and 100~TeV/20~ab$^{-1}$ (right), marginalized over the $c_{t2}$ and $d_3$ anomalous couplings. Note the differences in the axes ranges at each collider.}
\end{center}
\end{figure}

\begin{figure}[htp]
\begin{center}
    \includegraphics[width=0.495\columnwidth]{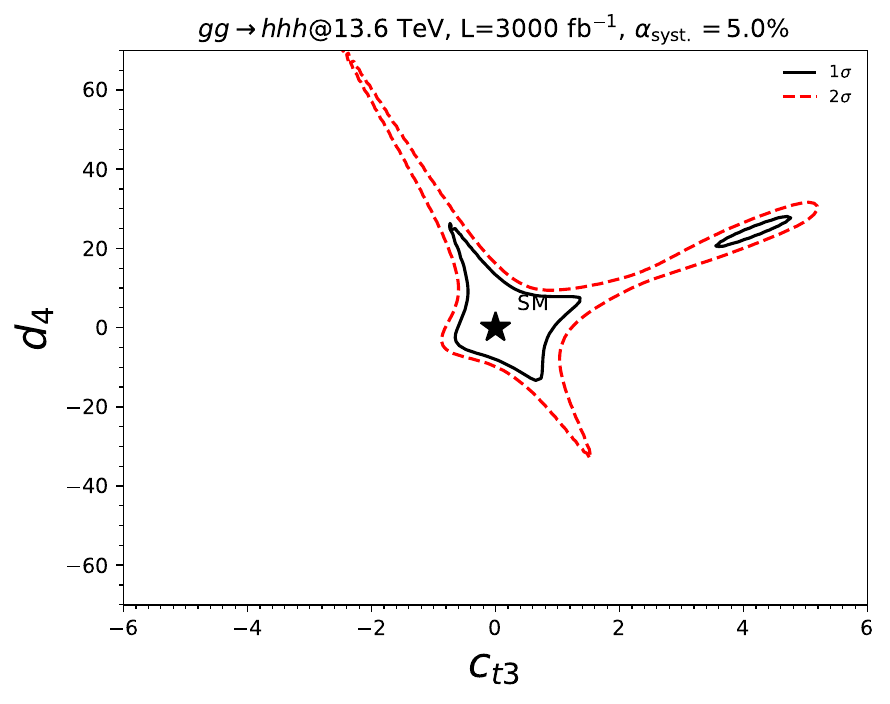}
  \includegraphics[width=0.495\columnwidth]{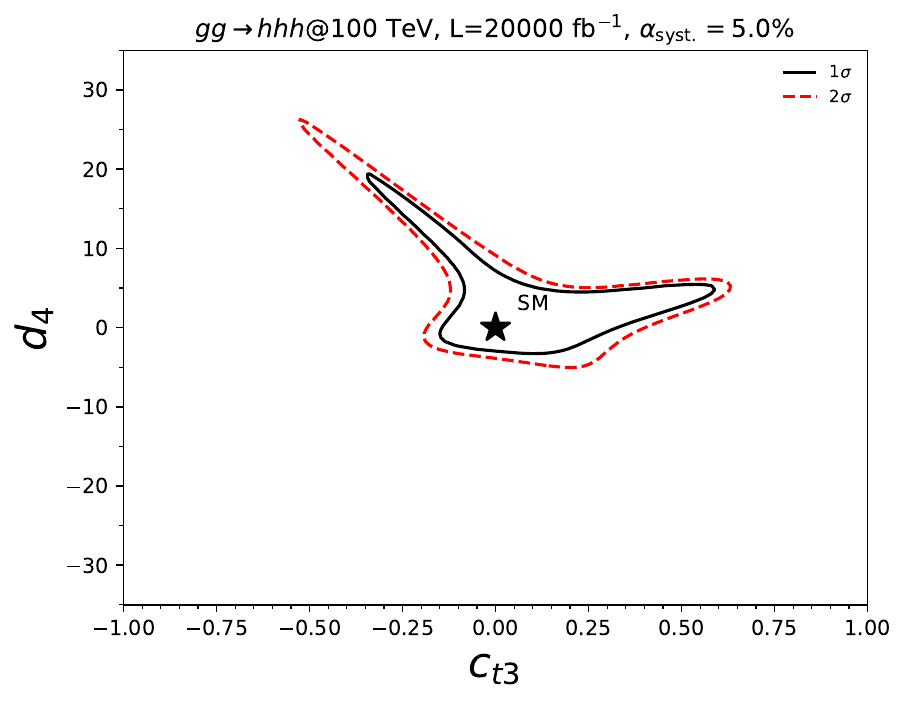}
\caption{\label{fig:2dconstraints_tripleins} The 68\% C.L. (1$\sigma$, black solid) and 95\% C.L (2$\sigma$, red dashed) limit on the $(c_{t3},d_4)$-plane for triple Higgs boson production with up to \textit{triple powers} of anomalous operators at the matrix element-squared level, at 13~TeV/3000~fb$^{-1}$ (left), and 100~TeV/20~ab$^{-1}$ (right), marginalized over the $c_{t2}$ and $d_3$ anomalous couplings. Note the differences in the axes ranges at each collider. }
\end{center}
\end{figure}

\begin{table}[htp]
    \centering
\begin{NiceTabular}{ccc||cc}[hvlines,corners=NE] 
& HL-LHC 3$\sigma$ & HL-LHC 5$\sigma$ & FCC-hh 3$\sigma$ & FCC-hh 5$\sigma$ \\
$d_4$ & $[-38.4, 36.0]$  & $[-90.3, 81.0]$  & $[-17.8, 20.8]$ & $ [-28.9, 23.2]$     \\
$c_{t3}$ & $[-2.8, 2.5]$ & $[-6.4, 5.7]$ & $[-0.5,0.6]$ & $[-0.9, 0.7]$  \\
\end{NiceTabular}
   \caption{ The 3$\sigma$ evidence and 5$\sigma$ discovery limits on for triple Higgs boson production with up to \textit{triple powers} of anomalous operators at the matrix element-squared level, for the $c_{t3}$ and $d_4$ coefficients at 13~TeV/3000~fb$^{-1}$, and 100~TeV/20~ab$^{-1}$, marginalized over $c_{t2}$, $d_3$ and either $d_4$, or $c_{t3}$. }\label{tab:disclimits_tripleins}
\end{table}

\begin{table}[htp]
    \centering
\begin{NiceTabular}{ccc||cc}[hvlines,corners=NE] 
& HL-LHC 68\% & HL-LHC 95\% & FCC-hh 68\% & FCC-hh 95\%   \\
$d_4$ & $[-6.1, 9.7]$  & $[-10.4, 16.1]$  & $[-4.0, 9.6]$ & $ [-10.1, 19.6]$     \\
$c_{t3}$ & $[-0.5, 0.9]$ & $[-1.0, 3.9]$ & $[-0.1, 0.3]$ & $[-0.7,  0.3]$  \\
\end{NiceTabular}
   \caption{ The 68\% C.L. (1$\sigma$) and 95\% C.L (2$\sigma$) limits on $c_{t3}$ and $d_4$ for triple Higgs boson production with up to \textit{triple powers} of anomalous operators at the matrix element-squared level, at 13~TeV/3000~fb$^{-1}$, and 100~TeV/20~ab$^{-1}$, marginalized over $c_{t2}$, $d_3$ and either $d_4$, or $c_{t3}$. }\label{tab:limits_tripleins}
\end{table}

We present here for completeness, the evidence/discovery boundaries and the 1$\sigma$ and 2$\sigma$ limits for the HL-LHC and FCC-hh, obtained with up to three powers of the anomalous operators at the matrix element-squared level. Since we are only considering the set $\{d_4, c_{t3}, d_3, c_{t2}\}$ in our phenomenological analysis, there can only be diagrams with up to two insertions, and the third power only comes from interference terms of these diagrams with single-insertion operators. This can be generated within \texttt{MG5\_aMC}, by:
\begin{verbatim}
generate g g > h h h [noborn=MHEFT QCD] MHEFT^2<=3
\end{verbatim}
Figures~\ref{fig:2ddiscovery_tripleins} and~\ref{fig:2dconstraints_tripleins} are to be compared to figs.~\ref{fig:2ddiscovery} and~\ref{fig:2dconstraints}, respectively, and tables~\ref{tab:disclimits_tripleins} and~\ref{tab:limits_tripleins} to tables~\ref{tab:disclimits} and~\ref{tab:limits}, respectively. As can be clearly observed by examining the plots, as well as by the one-dimensional quantitative limits in the tables, this treatment does not substantially alter either the expected evidence/discovery boundaries, or the limits. 

\section{Cross Section Fits for Triple and Double Higgs Boson Production at Various Energies}\label{app:fits}

\begin{table}[htp]
    \centering
\begin{NiceTabular}{ccccccccc}[hvlines, corners = NE] 
$d_3$ & -0.786 & 0.181 & & & & \\ 
$c_{g1}$ & -0.386 & 0.0412 & 0.150 & & & \\ 
$c_{g2}$ & 0.971 & -0.123 & -0.715 & 0.853 & & \\ 
$c_{t1}$ & 4.86 & -1.87 & -1.02 & 2.56 & 5.91 & \\ 
$c_{t2}$ & -5.57 & 1.70 & 2.08 & -5.06 & -13.9 & 10.0  \\ 
$c_{b1}$ & -0.0900 & -0.0656 & 0.224 & -0.526 & -0.298 & 1.17 & 0.0964  \\ 
$c_{b2}$ & 0.0629 & 0.0668 & -0.199 & 0.468 & 0.224 & -1.01 & -0.174 & 0.0786  \\ 
& 1 & $d_3$ & $c_{g1} $ & $c_{g2}$ & $c_{t1}$ & $c_{t2}$ & $c_{b1}$ & $c_{b2}$ \\ 

\end{NiceTabular}
    \caption{Polynomial coefficients, $A_i$ (second column only) and $B_{ij}$, relevant for the determination of the cross section for leading-order Higgs boson pair production, in the form $\sigma/\sigma_\mathrm{SM} - 1 = \sum_i A_i c_i + \sum_{i,j} B_{ij} c_i c_j$, where $c_i \in \left\{ d_3, c_{g1}, c_{g2}, c_{t1}, c_{t2}, c_{b1}, c_{b2}, \right\}$, at $E_\mathrm{CM} = 13.6~\mathrm{TeV}$.}
    \label{tab:hhcoeffs}
\end{table}

\begin{table}[htp]
    \centering
\begin{NiceTabular}{ccccccccc}[hvlines, corners = NE] 
$d_3$ & -0.846 & 0.179 & & & & \\
$c_{g1}$ & -0.385 & 0.164 & 0.406 & & & \\
$c_{g2}$ & 1.03 & -0.438 & -1.07 & 0.777 & & \\
$c_{t1}$ & 4.92 & -2.08 & -1.63 & 3.34 & 6.37 & \\
$c_{t2}$ & -5.85 & 2.48 & 3.09 & -5.32 & -16.2 & 11.2  \\
$c_{b1}$ & -0.00554 & 0.00287 & 0.244 & -0.289 & -0.238 & 0.662 & 0.0400  \\
$c_{b2}$ & -0.0117 & 0.00464 & -0.138 & 0.159 & 0.101 & -0.339 & -0.0461 & 0.0133  \\
& 1 & $d_3$ & $c_{g1} $ & $c_{g2}$ & $c_{t1}$ & $c_{t2}$ & $c_{b1}$ & $c_{b2}$ \\
\end{NiceTabular}
    \caption{Polynomial coefficients, $A_i$ (second column only) and $B_{ij}$, relevant for the determination of the cross section for leading-order Higgs boson pair production, in the form $\sigma/\sigma_\mathrm{SM} - 1 = \sum_i A_i c_i + \sum_{i,j} B_{ij} c_i c_j$, where $c_i \in \left\{ d_3, c_{g1}, c_{g2}, c_{t1}, c_{t2}, c_{b1}, c_{b2}, \right\}$, at $E_\mathrm{CM} = 13~\mathrm{TeV}$.}
    \label{tab:hhcoeffs13}
\end{table}

\begin{table}[htp]
    \centering
    \scriptsize
\begin{NiceTabular}{cccccccccccc}[hvlines,corners=NE] 
$d_3$ & -0.824 & 0.225 & & & & \\
$d_4$ & -0.147 & -0.0640 & 0.0755 & & & \\
$c_{g1}$ & -0.229 & 0.123 & -0.0151 & 0.0168 & & & \\
$c_{g2}$ & 1.35 & -0.609 & -0.0411 & -0.168 & 0.470 & & \\
$c_{t1}$ & 6.76 & -3.22 & -0.00692 & -0.888 & 4.78 & 12.3 & \\
$c_{t2}$ & -3.82 & 3.24 & -1.59 & 0.864 & -3.37 & -19.5 & 16.1  \\
$c_{t3}$ & -2.76 & -0.384 & 1.92 & -0.0730 & -1.16 & -3.36 & -17.6 & 12.4  \\
$c_{b1}$ & -0.121 & 0.167 & -0.123 & 0.0440 & -0.137 & -0.871 & 2.00 & -1.45 & 0.0661  \\
$c_{b2}$ & 0.0382 & -0.0607 & 0.0478 & -0.0159 & 0.0470 & 0.306 & -0.748 & 0.566 & -0.0501 & 0.00950  \\
$c_{b3}$ & 0.0676 & -0.0445 & 0.0137 & -0.0120 & 0.0535 & 0.294 & -0.378 & 0.135 & -0.0217 & 0.00804 & 0.00239  \\
& 1 & $d_3$ & $d_4$ & $c_{g1} $ & $c_{g2}$ & $c_{t1}$ & $c_{t2}$ & $c_{t3}$ & $c_{b1}$ & $c_{b2}$ & $c_{b3}$ \\
\end{NiceTabular}
    \caption{Polynomial coefficients, $A_i$ (second column only) and $B_{ij}$, relevant for the determination of the cross section for leading-order Higgs boson triple production, in the form $\sigma/\sigma_\mathrm{SM} - 1 = \sum_i A_i c_i + \sum_{i,j} B_{ij} c_i c_j$, where $c_i \in \left\{ d_3, c_{g1}, c_{g2}, c_{t1}, c_{t2}, c_{b1}, c_{b2}, \right\}$, at $E_\mathrm{CM} = 13~\mathrm{TeV}$.}
    \label{tab:hhhcoeffs13}
\end{table}

\begin{table}[htp]
    \centering
\begin{NiceTabular}{ccccccccc}[hvlines, corners = NE] 
$d_3$ & -0.857 & 0.282 & & & & \\
$c_{g1}$ & -0.386 & 0.291 & 0.0772 & & & \\
$c_{g2}$ & 0.957 & -1.08 & -0.613 & 1.38 & & \\
$c_{t1}$ & 5.00 & -2.38 & -1.12 & 3.20 & 6.39 & \\
$c_{t2}$ & -5.86 & 3.50 & 1.76 & -6.17 & -15.8 & 11.0  \\
$c_{b1}$ & -0.0500 & 0.165 & 0.101 & -0.514 & -0.298 & 0.865 & 0.0529  \\
$c_{b2}$ & -0.00429 & -0.0655 & -0.0420 & 0.228 & 0.0704 & -0.324 & -0.0489 & 0.0115  \\
& 1 & $d_3$ & $c_{g1} $ & $c_{g2}$ & $c_{t1}$ & $c_{t2}$ & $c_{b1}$ & $c_{b2}$ \\
\end{NiceTabular}
    \caption{Polynomial coefficients, $A_i$ (second column only) and $B_{ij}$, relevant for the determination of the cross section for leading-order Higgs boson pair production, in the form $\sigma/\sigma_\mathrm{SM} - 1 = \sum_i A_i c_i + \sum_{i,j} B_{ij} c_i c_j$, where $c_i \in \left\{ d_3, c_{g1}, c_{g2}, c_{t1}, c_{t2}, c_{b1}, c_{b2}, \right\}$, at $E_\mathrm{CM} = 14~\mathrm{TeV}$.}
    \label{tab:hhcoeffs14}
\end{table}

\begin{table}[htp]
    \centering
    \scriptsize
\begin{NiceTabular}{cccccccccccc}[hvlines,corners=NE] 
$d_3$ & -0.827 & 0.414 & & & & \\
$d_4$ & -0.211 & -0.110 & 0.0511 & & & \\
$c_{g1}$ & -0.208 & 0.210 & -0.0282 & 0.0265 & & & \\
$c_{g2}$ & 1.28 & -0.700 & -0.0655 & -0.177 & 0.440 & & \\
$c_{t1}$ & 6.67 & -3.57 & -0.372 & -0.901 & 4.55 & 11.8 & \\
$c_{t2}$ & -3.73 & 4.96 & -0.993 & 1.26 & -3.59 & -18.2 & 15.5  \\
$c_{t3}$ & -2.67 & -2.19 & 1.61 & -0.558 & -0.551 & -3.38 & -18.1 & 12.9  \\
$c_{b1}$ & -0.0693 & 0.187 & -0.0567 & 0.0473 & -0.0998 & -0.496 & 1.24 & -0.974 & 0.0268  \\
$c_{b2}$ & 0.0853 & 0.136 & -0.0783 & 0.0346 & -0.00548 & -0.00213 & 1.04 & -1.27 & 0.0525 & 0.0319  \\
$c_{b3}$ & 0.0827 & 0.00135 & -0.0231 & 0.000427 & 0.0404 & 0.216 & 0.0953 & -0.350 & 0.00865 & 0.0160 & 0.00300  \\
& 1 & $d_3$ & $d_4$ & $c_{g1} $ & $c_{g2}$ & $c_{t1}$ & $c_{t2}$ & $c_{t3}$ & $c_{b1}$ & $c_{b2}$ & $c_{b3}$ \\
\end{NiceTabular}
    \caption{Polynomial coefficients, $A_i$ (second column only) and $B_{ij}$, relevant for the determination of the cross section for leading-order Higgs boson triple production, in the form $\sigma/\sigma_\mathrm{SM} - 1 = \sum_i A_i c_i + \sum_{i,j} B_{ij} c_i c_j$, where $c_i \in \left\{ d_3, c_{g1}, c_{g2}, c_{t1}, c_{t2}, c_{b1}, c_{b2}, \right\}$, at $E_\mathrm{CM} = 14~\mathrm{TeV}$.}
    \label{tab:hhhcoeffs14}
\end{table}

\begin{table}[htp]
    \centering
\begin{NiceTabular}{ccccccccc}[hvlines, corners = NE] 
$d_3$ & -0.690 & 0.643 & & & & \\
$c_{g1}$ & -0.393 & -0.0942 & 0.0638 & & & \\
$c_{g2}$ & 0.828 & -1.23 & 0.0452 & 0.600 & & \\
$c_{t1}$ & 4.62 & -3.54 & -0.482 & 3.67 & 7.15 & \\
$c_{t2}$ & -5.66 & 3.00 & 0.882 & -3.29 & -15.0 & 8.52  \\
$c_{b1}$ & 0.0110 & 0.186 & -0.0437 & -0.167 & -0.327 & 0.159 & 0.0172  \\
$c_{b2}$ & -0.0117 & -0.126 & 0.0307 & 0.113 & 0.214 & -0.0967 & -0.0236 & 0.00807  \\
& 1 & $d_3$ & $c_{g1} $ & $c_{g2}$ & $c_{t1}$ & $c_{t2}$ & $c_{b1}$ & $c_{b2}$ \\
\end{NiceTabular}
    \caption{Polynomial coefficients, $A_i$ (second column only) and $B_{ij}$, relevant for the determination of the cross section for leading-order Higgs boson pair production, in the form $\sigma/\sigma_\mathrm{SM} - 1 = \sum_i A_i c_i + \sum_{i,j} B_{ij} c_i c_j$, where $c_i \in \left\{ d_3, c_{g1}, c_{g2}, c_{t1}, c_{t2}, c_{b1}, c_{b2}, \right\}$, at $E_\mathrm{CM} = 27~\mathrm{TeV}$.}
    \label{tab:hhcoeffs27}
\end{table}

\begin{table}[htp]
    \centering
    \scriptsize
\begin{NiceTabular}{cccccccccccc}[hvlines,corners=NE] 
$d_3$ & -0.675 & 0.193 & & & & \\
$d_4$ & -0.124 & -0.113 & 0.0795 & & & \\
$c_{g1}$ & -0.388 & 0.198 & -0.0413 & 0.0516 & & & \\
$c_{g2}$ & 1.25 & -0.696 & 0.191 & -0.358 & 0.628 & & \\
$c_{t1}$ & 7.00 & -2.82 & 0.0138 & -1.55 & 5.17 & 12.9 & \\
$c_{t2}$ & -3.88 & 3.35 & -1.75 & 1.61 & -5.96 & -19.5 & 16.9  \\
$c_{t3}$ & -3.62 & -0.745 & 2.15 & -0.129 & 1.15 & -7.00 & -18.4 & 15.5  \\
$c_{b1}$ & 0.0800 & 0.0750 & -0.105 & 0.0275 & -0.127 & -0.0145 & 1.16 & -1.41 & 0.0345  \\
$c_{b2}$ & 0.0412 & -0.0807 & 0.0628 & -0.0362 & 0.142 & 0.337 & -0.941 & 0.756 & -0.0414 & 0.0145  \\
$c_{b3}$ & 0.0562 & -0.0116 & -0.0107 & -0.00778 & 0.0224 & 0.176 & -0.0138 & -0.194 & 0.00701 & -0.00196 & 0.000963  \\
& 1 & $d_3$ & $d_4$ & $c_{g1} $ & $c_{g2}$ & $c_{t1}$ & $c_{t2}$ & $c_{t3}$ & $c_{b1}$ & $c_{b2}$ & $c_{b3}$ \\
\end{NiceTabular}
    \caption{Polynomial coefficients, $A_i$ (second column only) and $B_{ij}$, relevant for the determination of the cross section for leading-order Higgs boson triple production, in the form $\sigma/\sigma_\mathrm{SM} - 1 = \sum_i A_i c_i + \sum_{i,j} B_{ij} c_i c_j$, where $c_i \in \left\{ d_3, c_{g1}, c_{g2}, c_{t1}, c_{t2}, c_{b1}, c_{b2}, \right\}$, at $E_\mathrm{CM} = 27~\mathrm{TeV}$.}
    \label{tab:hhhcoeffs27}
\end{table}

\begin{table}[htp]
    \centering
\begin{NiceTabular}{ccccccccc}[hvlines, corners = NE] 
$d_3$ & -0.644 & 0.105 & & & & \\
$c_{g1}$ & -0.276 & 0.0790 & 0.0322 & & & \\
$c_{g2}$ & 0.644 & -0.0898 & -0.406 & 2.01 & & \\
$c_{t1}$ & 4.60 & -1.44 & -0.753 & 2.91 & 5.55 & \\
$c_{t2}$ & -5.75 & 1.74 & 1.08 & -5.34 & -14.5 & 9.86  \\
$c_{b1}$ & -0.0759 & 0.0174 & 0.0295 & -0.253 & -0.260 & 0.427 & 0.00829  \\
$c_{b2}$ & 0.0391 & 0.0119 & -0.0714 & 0.805 & 0.387 & -0.839 & -0.0490 & 0.0830  \\
& 1 & $d_3$ & $c_{g1} $ & $c_{g2}$ & $c_{t1}$ & $c_{t2}$ & $c_{b1}$ & $c_{b2}$ \\
\end{NiceTabular}
    \caption{Polynomial coefficients, $A_i$ (second column only) and $B_{ij}$, relevant for the determination of the cross section for leading-order Higgs boson pair production, in the form $\sigma/\sigma_\mathrm{SM} - 1 = \sum_i A_i c_i + \sum_{i,j} B_{ij} c_i c_j$, where $c_i \in \left\{ d_3, c_{g1}, c_{g2}, c_{t1}, c_{t2}, c_{b1}, c_{b2}, \right\}$, at $E_\mathrm{CM} = 100~\mathrm{TeV}$.}
    \label{tab:hhcoeffs100}
\end{table}

\begin{table}[htp]
    \centering
    \scriptsize
\begin{NiceTabular}{cccccccccccc}[hvlines,corners=NE] 
$d_3$ & -0.643 & 0.185 & & & & \\
$d_4$ & -0.175 & -0.0386 & 0.0352 & & & \\
$c_{g1}$ & -0.236 & 0.0972 & 0.00817 & 0.0153 & & & \\
$c_{g2}$ & 1.29 & -0.811 & 0.119 & -0.204 & 0.898 & & \\
$c_{t1}$ & 6.43 & -2.46 & -0.331 & -0.810 & 5.10 & 10.8 & \\
$c_{t2}$ & -3.93 & 3.10 & -0.724 & 0.704 & -7.00 & -17.1 & 14.2  \\
$c_{t3}$ & -4.64 & -1.26 & 2.01 & 0.186 & 3.73 & -8.21 & -21.9 & 28.6  \\
$c_{b1}$ & -0.0679 & 0.0514 & -0.0113 & 0.0119 & -0.116 & -0.290 & 0.466 & -0.342 & 0.00383  \\
$c_{b2}$ & 0.0750 & -0.0144 & -0.0122 & -0.00757 & 0.0246 & 0.217 & -0.0383 & -0.338 & -0.000788 & 0.00169  \\
$c_{b3}$ & -0.0157 & 0.0642 & -0.0330 & 0.00962 & -0.154 & -0.195 & 0.697 & -0.962 & 0.0113 & 0.00292 & 0.0108  \\
& 1 & $d_3$ & $d_4$ & $c_{g1} $ & $c_{g2}$ & $c_{t1}$ & $c_{t2}$ & $c_{t3}$ & $c_{b1}$ & $c_{b2}$ & $c_{b3}$ \\
\end{NiceTabular}
    \caption{Polynomial coefficients, $A_i$ (second column only) and $B_{ij}$, relevant for the determination of the cross section for leading-order Higgs boson triple production, in the form $\sigma/\sigma_\mathrm{SM} - 1 = \sum_i A_i c_i + \sum_{i,j} B_{ij} c_i c_j$, where $c_i \in \left\{ d_3, c_{g1}, c_{g2}, c_{t1}, c_{t2}, c_{b1}, c_{b2}, \right\}$, at $E_\mathrm{CM} = 100~\mathrm{TeV}$.}
    \label{tab:hhhcoeffs100}
\end{table}

This appendix provides the coefficients of the polynomial function for the cross section $\sigma/\sigma_\mathrm{SM} - 1 = \sum_i A_i c_i + \sum_{i,j} B_{ij} c_i c_j$, where $c_i \in \left\{ d_3, d_4, c_{g1}, c_{g2}, c_{t1}, c_{t2}, c_{t3}, c_{b1}, c_{b2}, c_{b3} \right\}$, for Higgs boson pair and production triple production. The coefficients for Higgs boson pair production for $E_\mathrm{CM} = 13.6,~13,~14,~27$, and $100$~TeV are shown in tables~\ref{tab:hhcoeffs},~\ref{tab:hhcoeffs13},~\ref{tab:hhcoeffs14},~\ref{tab:hhcoeffs27},~\ref{tab:hhcoeffs100}, respectively, and for triple production in tables~\ref{tab:hhhcoeffs},~\ref{tab:hhhcoeffs13},~\ref{tab:hhhcoeffs14},~\ref{tab:hhhcoeffs27},~\ref{tab:hhhcoeffs100}.

\bibliographystyle{JHEP}
\bibliography{biblio.bib}

\providecommand{\href}[2]{#2}\begingroup\raggedright\begin{thebibliography}{100}

\bibitem{gitlabrepo}
A.~Papaefstathiou, G.~Tetlalmatzi-Xolocotzi and A.~Tonero, ``{Triple Higgs
  Boson Production in a Higgs Effective Theory}.''
  https://gitlab.com/apapaefs/multihiggs\_loop\_sm, 2023.

\bibitem{Higgs:1964pj}
P.W.~Higgs, \emph{{Broken Symmetries and the Masses of Gauge Bosons}},
  \href{https://doi.org/10.1103/PhysRevLett.13.508}{\emph{Phys. Rev. Lett.}
  {\bfseries 13} (1964) 508}.

\bibitem{Englert:1964et}
F.~Englert and R.~Brout, \emph{{Broken Symmetry and the Mass of Gauge Vector
  Mesons}}, \href{https://doi.org/10.1103/PhysRevLett.13.321}{\emph{Phys. Rev.
  Lett.} {\bfseries 13} (1964) 321}.

\bibitem{Guralnik:1964eu}
G.S.~Guralnik, C.R.~Hagen and T.W.B.~Kibble, \emph{{Global Conservation Laws
  and Massless Particles}},
  \href{https://doi.org/10.1103/PhysRevLett.13.585}{\emph{Phys. Rev. Lett.}
  {\bfseries 13} (1964) 585}.

\bibitem{ATLAS:2012yve}
{\scshape ATLAS} collaboration, \emph{{Observation of a new particle in the
  search for the Standard Model Higgs boson with the ATLAS detector at the
  LHC}}, \href{https://doi.org/10.1016/j.physletb.2012.08.020}{\emph{Phys.
  Lett. B} {\bfseries 716} (2012) 1}
  [\href{https://arxiv.org/abs/1207.7214}{{\ttfamily 1207.7214}}].

\bibitem{CMS:2012qbp}
{\scshape CMS} collaboration, \emph{{Observation of a New Boson at a Mass of
  125 GeV with the CMS Experiment at the LHC}},
  \href{https://doi.org/10.1016/j.physletb.2012.08.021}{\emph{Phys. Lett. B}
  {\bfseries 716} (2012) 30} [\href{https://arxiv.org/abs/1207.7235}{{\ttfamily
  1207.7235}}].

\bibitem{Dorsch:2013wja}
G.C.~Dorsch, S.J.~Huber and J.M.~No, \emph{{A strong electroweak phase
  transition in the 2HDM after LHC8}},
  \href{https://doi.org/10.1007/JHEP10(2013)029}{\emph{JHEP} {\bfseries 10}
  (2013) 029} [\href{https://arxiv.org/abs/1305.6610}{{\ttfamily 1305.6610}}].

\bibitem{No:2013wsa}
J.M.~No and M.~Ramsey-Musolf, \emph{{Probing the Higgs Portal at the LHC
  Through Resonant di-Higgs Production}},
  \href{https://doi.org/10.1103/PhysRevD.89.095031}{\emph{Phys. Rev. D}
  {\bfseries 89} (2014) 095031}
  [\href{https://arxiv.org/abs/1310.6035}{{\ttfamily 1310.6035}}].

\bibitem{Curtin:2014jma}
D.~Curtin, P.~Meade and C.-T.~Yu, \emph{{Testing Electroweak Baryogenesis with
  Future Colliders}},
  \href{https://doi.org/10.1007/JHEP11(2014)127}{\emph{JHEP} {\bfseries 11}
  (2014) 127} [\href{https://arxiv.org/abs/1409.0005}{{\ttfamily 1409.0005}}].

\bibitem{Profumo:2014opa}
S.~Profumo, M.J.~Ramsey-Musolf, C.L.~Wainwright and P.~Winslow,
  \emph{{Singlet-catalyzed electroweak phase transitions and precision Higgs
  boson studies}},
  \href{https://doi.org/10.1103/PhysRevD.91.035018}{\emph{Phys. Rev. D}
  {\bfseries 91} (2015) 035018}
  [\href{https://arxiv.org/abs/1407.5342}{{\ttfamily 1407.5342}}].

\bibitem{White:2016nbo}
G.A.~White, \emph{{A Pedagogical Introduction to Electroweak Baryogenesis}}, .

\bibitem{Basler:2016obg}
P.~Basler, M.~Krause, M.~Muhlleitner, J.~Wittbrodt and A.~Wlotzka,
  \emph{{Strong First Order Electroweak Phase Transition in the CP-Conserving
  2HDM Revisited}}, \href{https://doi.org/10.1007/JHEP02(2017)121}{\emph{JHEP}
  {\bfseries 02} (2017) 121}
  [\href{https://arxiv.org/abs/1612.04086}{{\ttfamily 1612.04086}}].

\bibitem{Bernon:2017jgv}
J.~Bernon, L.~Bian and Y.~Jiang, \emph{{A new insight into the phase transition
  in the early Universe with two Higgs doublets}},
  \href{https://doi.org/10.1007/JHEP05(2018)151}{\emph{JHEP} {\bfseries 05}
  (2018) 151} [\href{https://arxiv.org/abs/1712.08430}{{\ttfamily
  1712.08430}}].

\bibitem{Kurup:2017dzf}
G.~Kurup and M.~Perelstein, \emph{{Dynamics of Electroweak Phase Transition In
  Singlet-Scalar Extension of the Standard Model}},
  \href{https://doi.org/10.1103/PhysRevD.96.015036}{\emph{Phys. Rev. D}
  {\bfseries 96} (2017) 015036}
  [\href{https://arxiv.org/abs/1704.03381}{{\ttfamily 1704.03381}}].

\bibitem{Dorsch:2017nza}
G.C.~Dorsch, S.J.~Huber, K.~Mimasu and J.M.~No, \emph{{The Higgs Vacuum
  Uplifted: Revisiting the Electroweak Phase Transition with a Second Higgs
  Doublet}}, \href{https://doi.org/10.1007/JHEP12(2017)086}{\emph{JHEP}
  {\bfseries 12} (2017) 086}
  [\href{https://arxiv.org/abs/1705.09186}{{\ttfamily 1705.09186}}].

\bibitem{Ramsey-Musolf:2019lsf}
M.J.~Ramsey-Musolf, \emph{{The electroweak phase transition: a collider
  target}}, \href{https://doi.org/10.1007/JHEP09(2020)179}{\emph{JHEP}
  {\bfseries 09} (2020) 179}
  [\href{https://arxiv.org/abs/1912.07189}{{\ttfamily 1912.07189}}].

\bibitem{Li:2019tfd}
H.-L.~Li, M.~Ramsey-Musolf and S.~Willocq, \emph{{Probing a scalar
  singlet-catalyzed electroweak phase transition with resonant di-Higgs boson
  production in the $4b$ channel}},
  \href{https://doi.org/10.1103/PhysRevD.100.075035}{\emph{Phys. Rev. D}
  {\bfseries 100} (2019) 075035}
  [\href{https://arxiv.org/abs/1906.05289}{{\ttfamily 1906.05289}}].

\bibitem{Papaefstathiou:2020iag}
A.~Papaefstathiou and G.~White, \emph{{The electro-weak phase transition at
  colliders: confronting theoretical uncertainties and complementary
  channels}}, \href{https://doi.org/10.1007/JHEP05(2021)099}{\emph{JHEP}
  {\bfseries 05} (2021) 099}
  [\href{https://arxiv.org/abs/2010.00597}{{\ttfamily 2010.00597}}].

\bibitem{Papaefstathiou:2021glr}
A.~Papaefstathiou and G.~White, \emph{{The Electro-Weak Phase Transition at
  Colliders: Discovery Post-Mortem}},
  \href{https://doi.org/10.1007/JHEP02(2022)185}{\emph{JHEP} {\bfseries 02}
  (2022) 185} [\href{https://arxiv.org/abs/2108.11394}{{\ttfamily
  2108.11394}}].

\bibitem{Papaefstathiou:2022oyi}
A.~Papaefstathiou, T.~Robens and G.~White, \emph{{Signal strength and W-boson
  mass measurements as a probe of the electro-weak phase transition at
  colliders - Snowmass White Paper}},  in \emph{{Snowmass 2021}}, 5, 2022
  [\href{https://arxiv.org/abs/2205.14379}{{\ttfamily 2205.14379}}].

\bibitem{Zuk:2022qwx}
J.~Zuk, C.~Balazs, A.~Papaefstathiou and G.~White, \emph{{The Effective
  Potential in Fermi Gauges Beyond the Standard Model}},
  \href{https://arxiv.org/abs/2212.04046}{{\ttfamily 2212.04046}}.

\bibitem{Buchmuller:1985jz}
W.~Buchmuller and D.~Wyler, \emph{{Effective Lagrangian Analysis of New
  Interactions and Flavor Conservation}},
  \href{https://doi.org/10.1016/0550-3213(86)90262-2}{\emph{Nucl. Phys. B}
  {\bfseries 268} (1986) 621}.

\bibitem{Grzadkowski:2010es}
B.~Grzadkowski, M.~Iskrzynski, M.~Misiak and J.~Rosiek, \emph{{Dimension-Six
  Terms in the Standard Model Lagrangian}},
  \href{https://doi.org/10.1007/JHEP10(2010)085}{\emph{JHEP} {\bfseries 10}
  (2010) 085} [\href{https://arxiv.org/abs/1008.4884}{{\ttfamily 1008.4884}}].

\bibitem{Elias-Miro:2013mua}
J.~Elias-Miro, J.R.~Espinosa, E.~Masso and A.~Pomarol, \emph{{Higgs windows to
  new physics through d=6 operators: constraints and one-loop anomalous
  dimensions}}, \href{https://doi.org/10.1007/JHEP11(2013)066}{\emph{JHEP}
  {\bfseries 11} (2013) 066} [\href{https://arxiv.org/abs/1308.1879}{{\ttfamily
  1308.1879}}].

\bibitem{Feruglio:1992wf}
F.~Feruglio, \emph{{The Chiral approach to the electroweak interactions}},
  \href{https://doi.org/10.1142/S0217751X93001946}{\emph{Int. J. Mod. Phys. A}
  {\bfseries 8} (1993) 4937}
  [\href{https://arxiv.org/abs/hep-ph/9301281}{{\ttfamily hep-ph/9301281}}].

\bibitem{Bagger:1993zf}
J.~Bagger, V.D.~Barger, K.-m.~Cheung, J.F.~Gunion, T.~Han, G.A.~Ladinsky
  et~al., \emph{{The Strongly interacting W W system: Gold plated modes}},
  \href{https://doi.org/10.1103/PhysRevD.49.1246}{\emph{Phys. Rev. D}
  {\bfseries 49} (1994) 1246}
  [\href{https://arxiv.org/abs/hep-ph/9306256}{{\ttfamily hep-ph/9306256}}].

\bibitem{Koulovassilopoulos:1993pw}
V.~Koulovassilopoulos and R.S.~Chivukula, \emph{{The Phenomenology of a
  nonstandard Higgs boson in W(L) W(L) scattering}},
  \href{https://doi.org/10.1103/PhysRevD.50.3218}{\emph{Phys. Rev. D}
  {\bfseries 50} (1994) 3218}
  [\href{https://arxiv.org/abs/hep-ph/9312317}{{\ttfamily hep-ph/9312317}}].

\bibitem{Burgess:1999ha}
C.P.~Burgess, J.~Matias and M.~Pospelov, \emph{{A Higgs or not a Higgs? What to
  do if you discover a new scalar particle}},
  \href{https://doi.org/10.1142/S0217751X02009813}{\emph{Int. J. Mod. Phys. A}
  {\bfseries 17} (2002) 1841}
  [\href{https://arxiv.org/abs/hep-ph/9912459}{{\ttfamily hep-ph/9912459}}].

\bibitem{Wang:2006im}
L.-M.~Wang and Q.~Wang, \emph{{Nonstandard Higgs in electroweak chiral
  Lagrangian}},  \href{https://arxiv.org/abs/hep-ph/0605104}{{\ttfamily
  hep-ph/0605104}}.

\bibitem{Grinstein:2007iv}
B.~Grinstein and M.~Trott, \emph{{A Higgs-Higgs bound state due to new physics
  at a TeV}}, \href{https://doi.org/10.1103/PhysRevD.76.073002}{\emph{Phys.
  Rev. D} {\bfseries 76} (2007) 073002}
  [\href{https://arxiv.org/abs/0704.1505}{{\ttfamily 0704.1505}}].

\bibitem{Alonso:2012px}
R.~Alonso, M.B.~Gavela, L.~Merlo, S.~Rigolin and J.~Yepes, \emph{{The Effective
  Chiral Lagrangian for a Light Dynamical ''Higgs Particle''}},
  \href{https://doi.org/10.1016/j.physletb.2013.04.037}{\emph{Phys. Lett. B}
  {\bfseries 722} (2013) 330}
  [\href{https://arxiv.org/abs/1212.3305}{{\ttfamily 1212.3305}}].

\bibitem{Buchalla:2013eza}
G.~Buchalla, O.~Cat\'a and C.~Krause, \emph{{On the Power Counting in Effective
  Field Theories}},
  \href{https://doi.org/10.1016/j.physletb.2014.02.015}{\emph{Phys. Lett. B}
  {\bfseries 731} (2014) 80} [\href{https://arxiv.org/abs/1312.5624}{{\ttfamily
  1312.5624}}].

\bibitem{Buchalla:2012qq}
G.~Buchalla and O.~Cata, \emph{{Effective Theory of a Dynamically Broken
  Electroweak Standard Model at NLO}},
  \href{https://doi.org/10.1007/JHEP07(2012)101}{\emph{JHEP} {\bfseries 07}
  (2012) 101} [\href{https://arxiv.org/abs/1203.6510}{{\ttfamily 1203.6510}}].

\bibitem{Buchalla:2013rka}
G.~Buchalla, O.~Cat\`a and C.~Krause, \emph{{Complete Electroweak Chiral
  Lagrangian with a Light Higgs at NLO}},
  \href{https://doi.org/10.1016/j.nuclphysb.2014.01.018}{\emph{Nucl. Phys. B}
  {\bfseries 880} (2014) 552}
  [\href{https://arxiv.org/abs/1307.5017}{{\ttfamily 1307.5017}}].

\bibitem{Buchalla:2017jlu}
G.~Buchalla, O.~Cata, A.~Celis, M.~Knecht and C.~Krause, \emph{{Complete
  One-Loop Renormalization of the Higgs-Electroweak Chiral Lagrangian}},
  \href{https://doi.org/10.1016/j.nuclphysb.2018.01.009}{\emph{Nucl. Phys. B}
  {\bfseries 928} (2018) 93}
  [\href{https://arxiv.org/abs/1710.06412}{{\ttfamily 1710.06412}}].

\bibitem{CMS:2017yfv}
{\scshape CMS} collaboration, \emph{{Search for Higgs boson pair production in
  the $bb\tau\tau$ final state in proton-proton collisions at
  $\sqrt{(}s)=8\text{ }\text{ }\mathrm{TeV}$}},
  \href{https://doi.org/10.1103/PhysRevD.96.072004}{\emph{Phys. Rev. D}
  {\bfseries 96} (2017) 072004}
  [\href{https://arxiv.org/abs/1707.00350}{{\ttfamily 1707.00350}}].

\bibitem{CMS:2017hea}
{\scshape CMS} collaboration, \emph{{Search for Higgs boson pair production in
  events with two bottom quarks and two tau leptons in
  proton\textendash{}proton collisions at $\sqrt s$ =13TeV}},
  \href{https://doi.org/10.1016/j.physletb.2018.01.001}{\emph{Phys. Lett. B}
  {\bfseries 778} (2018) 101}
  [\href{https://arxiv.org/abs/1707.02909}{{\ttfamily 1707.02909}}].

\bibitem{CMS:2017rpp}
{\scshape CMS} collaboration, \emph{{Search for resonant and nonresonant Higgs
  boson pair production in the $ \mathrm{b}\overline{\mathrm{b}}\mathit{\ell
  \nu \ell \nu } $ final state in proton-proton collisions at $ \sqrt{s}=13 $
  TeV}}, \href{https://doi.org/10.1007/JHEP01(2018)054}{\emph{JHEP} {\bfseries
  01} (2018) 054} [\href{https://arxiv.org/abs/1708.04188}{{\ttfamily
  1708.04188}}].

\bibitem{CMS:2018tla}
{\scshape CMS} collaboration, \emph{{Search for Higgs boson pair production in
  the $\gamma\gamma\mathrm{b\overline{b}}$ final state in pp collisions at
  $\sqrt{s}=$ 13 TeV}},
  \href{https://doi.org/10.1016/j.physletb.2018.10.056}{\emph{Phys. Lett. B}
  {\bfseries 788} (2019) 7} [\href{https://arxiv.org/abs/1806.00408}{{\ttfamily
  1806.00408}}].

\bibitem{CMS:2018vjd}
{\scshape CMS} collaboration, \emph{{Search for production of Higgs boson pairs
  in the four b quark final state using large-area jets in proton-proton
  collisions at $\sqrt{s}=$ 13 TeV}},
  \href{https://doi.org/10.1007/JHEP01(2019)040}{\emph{JHEP} {\bfseries 01}
  (2019) 040} [\href{https://arxiv.org/abs/1808.01473}{{\ttfamily
  1808.01473}}].

\bibitem{CMS:2018sxu}
{\scshape CMS} collaboration, \emph{{Search for nonresonant Higgs boson pair
  production in the $\mathrm{b\overline{b}b\overline{b}}$ final state at
  $\sqrt{s} =$ 13 TeV}},
  \href{https://doi.org/10.1007/JHEP04(2019)112}{\emph{JHEP} {\bfseries 04}
  (2019) 112} [\href{https://arxiv.org/abs/1810.11854}{{\ttfamily
  1810.11854}}].

\bibitem{CMS:2018ipl}
{\scshape CMS} collaboration, \emph{{Combination of searches for Higgs boson
  pair production in proton-proton collisions at $\sqrt{s} = $ 13 TeV}},
  \href{https://doi.org/10.1103/PhysRevLett.122.121803}{\emph{Phys. Rev. Lett.}
  {\bfseries 122} (2019) 121803}
  [\href{https://arxiv.org/abs/1811.09689}{{\ttfamily 1811.09689}}].

\bibitem{CMS:2020tkr}
{\scshape CMS} collaboration, \emph{{Search for nonresonant Higgs boson pair
  production in final states with two bottom quarks and two photons in
  proton-proton collisions at $ \sqrt{s} $ = 13 TeV}},
  \href{https://doi.org/10.1007/JHEP03(2021)257}{\emph{JHEP} {\bfseries 03}
  (2021) 257} [\href{https://arxiv.org/abs/2011.12373}{{\ttfamily
  2011.12373}}].

\bibitem{CMS:2022cpr}
{\scshape CMS} collaboration, \emph{{Search for Higgs Boson Pair Production in
  the Four b Quark Final State in Proton-Proton Collisions at s=13\,\,TeV}},
  \href{https://doi.org/10.1103/PhysRevLett.129.081802}{\emph{Phys. Rev. Lett.}
  {\bfseries 129} (2022) 081802}
  [\href{https://arxiv.org/abs/2202.09617}{{\ttfamily 2202.09617}}].

\bibitem{CMS:2022hgz}
{\scshape CMS} collaboration, \emph{{Search for nonresonant Higgs boson pair
  production in final state with two bottom quarks and two tau leptons in
  proton-proton collisions at s=13~TeV}},
  \href{https://doi.org/10.1016/j.physletb.2022.137531}{\emph{Phys. Lett. B}
  {\bfseries 842} (2023) 137531}
  [\href{https://arxiv.org/abs/2206.09401}{{\ttfamily 2206.09401}}].

\bibitem{CMS:2022kdx}
{\scshape CMS} collaboration, \emph{{Search for Higgs boson pairs decaying to
  WW*WW*, WW*$\tau\tau$, and $\tau\tau\tau\tau$ in proton-proton collisions at
  $\sqrt{s}$ = 13 TeV}},
  \href{https://doi.org/10.1007/JHEP07(2023)095}{\emph{JHEP} {\bfseries 07}
  (2023) 095} [\href{https://arxiv.org/abs/2206.10268}{{\ttfamily
  2206.10268}}].

\bibitem{CMS:2022omp}
{\scshape CMS} collaboration, \emph{{Search for nonresonant Higgs boson pair
  production in the four leptons plus twob jets final state in proton-proton
  collisions at $ \sqrt{s} $ = 13 TeV}},
  \href{https://doi.org/10.1007/JHEP06(2023)130}{\emph{JHEP} {\bfseries 06}
  (2023) 130} [\href{https://arxiv.org/abs/2206.10657}{{\ttfamily
  2206.10657}}].

\bibitem{ATLAS:2014pjm}
{\scshape ATLAS} collaboration, \emph{{Search For Higgs Boson Pair Production
  in the $\gamma\gamma b\bar{b}$ Final State using $pp$ Collision Data at
  $\sqrt{s}=8$ TeV from the ATLAS Detector}},
  \href{https://doi.org/10.1103/PhysRevLett.114.081802}{\emph{Phys. Rev. Lett.}
  {\bfseries 114} (2015) 081802}
  [\href{https://arxiv.org/abs/1406.5053}{{\ttfamily 1406.5053}}].

\bibitem{ATLAS:2015zug}
{\scshape ATLAS} collaboration, \emph{{Search for Higgs boson pair production
  in the $b\bar{b}b\bar{b}$ final state from pp collisions at $\sqrt{s} = 8$
  TeVwith the ATLAS detector}},
  \href{https://doi.org/10.1140/epjc/s10052-015-3628-x}{\emph{Eur. Phys. J. C}
  {\bfseries 75} (2015) 412}
  [\href{https://arxiv.org/abs/1506.00285}{{\ttfamily 1506.00285}}].

\bibitem{ATLAS:2015sxd}
{\scshape ATLAS} collaboration, \emph{{Searches for Higgs boson pair production
  in the $hh\to bb\tau\tau, \gamma\gamma WW^*, \gamma\gamma bb, bbbb$ channels
  with the ATLAS detector}},
  \href{https://doi.org/10.1103/PhysRevD.92.092004}{\emph{Phys. Rev. D}
  {\bfseries 92} (2015) 092004}
  [\href{https://arxiv.org/abs/1509.04670}{{\ttfamily 1509.04670}}].

\bibitem{ATLAS:2016paq}
{\scshape ATLAS} collaboration, \emph{{Search for pair production of Higgs
  bosons in the $b\bar{b}b\bar{b}$ final state using proton--proton collisions
  at $\sqrt{s} = 13$ TeV with the ATLAS detector}},
  \href{https://doi.org/10.1103/PhysRevD.94.052002}{\emph{Phys. Rev. D}
  {\bfseries 94} (2016) 052002}
  [\href{https://arxiv.org/abs/1606.04782}{{\ttfamily 1606.04782}}].

\bibitem{ATLAS:2018rnh}
{\scshape ATLAS} collaboration, \emph{{Search for pair production of Higgs
  bosons in the $b\bar{b}b\bar{b}$ final state using proton-proton collisions
  at $\sqrt{s} = 13$ TeV with the ATLAS detector}},
  \href{https://doi.org/10.1007/JHEP01(2019)030}{\emph{JHEP} {\bfseries 01}
  (2019) 030} [\href{https://arxiv.org/abs/1804.06174}{{\ttfamily
  1804.06174}}].

\bibitem{ATLAS:2018dpp}
{\scshape ATLAS} collaboration, \emph{{Search for Higgs boson pair production
  in the $\gamma\gamma b\bar{b}$ final state with 13 TeV $pp$ collision data
  collected by the ATLAS experiment}},
  \href{https://doi.org/10.1007/JHEP11(2018)040}{\emph{JHEP} {\bfseries 11}
  (2018) 040} [\href{https://arxiv.org/abs/1807.04873}{{\ttfamily
  1807.04873}}].

\bibitem{ATLAS:2018hqk}
{\scshape ATLAS} collaboration, \emph{{Search for Higgs boson pair production
  in the $\gamma\gamma WW^{*}$ channel using $pp$ collision data recorded at
  $\sqrt{s} = 13$ TeV with the ATLAS detector}},
  \href{https://doi.org/10.1140/epjc/s10052-018-6457-x}{\emph{Eur. Phys. J. C}
  {\bfseries 78} (2018) 1007}
  [\href{https://arxiv.org/abs/1807.08567}{{\ttfamily 1807.08567}}].

\bibitem{ATLAS:2018uni}
{\scshape ATLAS} collaboration, \emph{{Search for resonant and non-resonant
  Higgs boson pair production in the ${b\bar{b}\tau^+\tau^-}$ decay channel in
  $pp$ collisions at $\sqrt{s}=13$ TeV with the ATLAS detector}},
  \href{https://doi.org/10.1103/PhysRevLett.121.191801}{\emph{Phys. Rev. Lett.}
  {\bfseries 121} (2018) 191801}
  [\href{https://arxiv.org/abs/1808.00336}{{\ttfamily 1808.00336}}].

\bibitem{ATLAS:2018fpd}
{\scshape ATLAS} collaboration, \emph{{Search for Higgs boson pair production
  in the $b\bar{b}WW^{*}$ decay mode at $\sqrt{s}=13$ TeV with the ATLAS
  detector}}, \href{https://doi.org/10.1007/JHEP04(2019)092}{\emph{JHEP}
  {\bfseries 04} (2019) 092}
  [\href{https://arxiv.org/abs/1811.04671}{{\ttfamily 1811.04671}}].

\bibitem{ATLAS:2018ili}
{\scshape ATLAS} collaboration, \emph{{Search for Higgs boson pair production
  in the $WW^{(*)}WW^{(*)}$ decay channel using ATLAS data recorded at
  $\sqrt{s}=13$ TeV}},
  \href{https://doi.org/10.1007/JHEP05(2019)124}{\emph{JHEP} {\bfseries 05}
  (2019) 124} [\href{https://arxiv.org/abs/1811.11028}{{\ttfamily
  1811.11028}}].

\bibitem{ATLAS:2019qdc}
{\scshape ATLAS} collaboration, \emph{{Combination of searches for Higgs boson
  pairs in $pp$ collisions at $\sqrt{s} = $13 TeV with the ATLAS detector}},
  \href{https://doi.org/10.1016/j.physletb.2019.135103}{\emph{Phys. Lett. B}
  {\bfseries 800} (2020) 135103}
  [\href{https://arxiv.org/abs/1906.02025}{{\ttfamily 1906.02025}}].

\bibitem{ATLAS:2019vwv}
{\scshape ATLAS} collaboration, \emph{{Search for non-resonant Higgs boson pair
  production in the $bb\ell\nu\ell\nu$ final state with the ATLAS detector in
  $pp$ collisions at $\sqrt{s} = 13$ TeV}},
  \href{https://doi.org/10.1016/j.physletb.2019.135145}{\emph{Phys. Lett. B}
  {\bfseries 801} (2020) 135145}
  [\href{https://arxiv.org/abs/1908.06765}{{\ttfamily 1908.06765}}].

\bibitem{ATLAS:2020jgy}
{\scshape ATLAS} collaboration, \emph{{Search for the $HH \rightarrow b \bar{b}
  b \bar{b}$ process via vector-boson fusion production using proton-proton
  collisions at $\sqrt{s} = 13$ TeV with the ATLAS detector}},
  \href{https://doi.org/10.1007/JHEP07(2020)108}{\emph{JHEP} {\bfseries 07}
  (2020) 108} [\href{https://arxiv.org/abs/2001.05178}{{\ttfamily
  2001.05178}}].

\bibitem{ATLAS:2021ifb}
{\scshape ATLAS} collaboration, \emph{{Search for Higgs boson pair production
  in the two bottom quarks plus two photons final state in $pp$ collisions at
  $\sqrt{s}=13$ TeV with the ATLAS detector}},
  \href{https://doi.org/10.1103/PhysRevD.106.052001}{\emph{Phys. Rev. D}
  {\bfseries 106} (2022) 052001}
  [\href{https://arxiv.org/abs/2112.11876}{{\ttfamily 2112.11876}}].

\bibitem{ATLAS:2022xzm}
{\scshape ATLAS} collaboration, \emph{{Search for resonant and non-resonant
  Higgs boson pair production in the $ b\overline{b}{\tau}^{+}{\tau}^{-} $
  decay channel using 13 TeV pp collision data from the ATLAS detector}},
  \href{https://doi.org/10.1007/JHEP07(2023)040}{\emph{JHEP} {\bfseries 07}
  (2023) 040} [\href{https://arxiv.org/abs/2209.10910}{{\ttfamily
  2209.10910}}].

\bibitem{ATLAS:2022jtk}
{\scshape ATLAS} collaboration, \emph{{Constraints on the Higgs boson
  self-coupling from single- and double-Higgs production with the ATLAS
  detector using pp collisions at s=13 TeV}},
  \href{https://doi.org/10.1016/j.physletb.2023.137745}{\emph{Phys. Lett. B}
  {\bfseries 843} (2023) 137745}
  [\href{https://arxiv.org/abs/2211.01216}{{\ttfamily 2211.01216}}].

\bibitem{ATLAS:2023qzf}
{\scshape ATLAS} collaboration, \emph{{Search for nonresonant pair production
  of Higgs bosons in the bb\textasciimacron{}bb\textasciimacron{} final state
  in pp collisions at s=13\,\,TeV with the ATLAS detector}},
  \href{https://doi.org/10.1103/PhysRevD.108.052003}{\emph{Phys. Rev. D}
  {\bfseries 108} (2023) 052003}
  [\href{https://arxiv.org/abs/2301.03212}{{\ttfamily 2301.03212}}].

\bibitem{ATLAS:2023gzn}
{\scshape ATLAS} collaboration, \emph{{Studies of new Higgs boson interactions
  through nonresonant $HH$ production in the $b\bar{b}\gamma\gamma$ final state
  in $pp$ collisions at $\sqrt{s}=13$ TeV with the ATLAS detector}},
  \href{https://arxiv.org/abs/2310.12301}{{\ttfamily 2310.12301}}.

\bibitem{Baur:2002rb}
U.~Baur, T.~Plehn and D.L.~Rainwater, \emph{{Measuring the Higgs Boson Self
  Coupling at the LHC and Finite Top Mass Matrix Elements}},
  \href{https://doi.org/10.1103/PhysRevLett.89.151801}{\emph{Phys. Rev. Lett.}
  {\bfseries 89} (2002) 151801}
  [\href{https://arxiv.org/abs/hep-ph/0206024}{{\ttfamily hep-ph/0206024}}].

\bibitem{Baur:2002qd}
U.~Baur, T.~Plehn and D.L.~Rainwater, \emph{{Determining the Higgs Boson
  Selfcoupling at Hadron Colliders}},
  \href{https://doi.org/10.1103/PhysRevD.67.033003}{\emph{Phys. Rev. D}
  {\bfseries 67} (2003) 033003}
  [\href{https://arxiv.org/abs/hep-ph/0211224}{{\ttfamily hep-ph/0211224}}].

\bibitem{Baur:2003gp}
U.~Baur, T.~Plehn and D.L.~Rainwater, \emph{{Probing the Higgs selfcoupling at
  hadron colliders using rare decays}},
  \href{https://doi.org/10.1103/PhysRevD.69.053004}{\emph{Phys. Rev. D}
  {\bfseries 69} (2004) 053004}
  [\href{https://arxiv.org/abs/hep-ph/0310056}{{\ttfamily hep-ph/0310056}}].

\bibitem{Dolan:2012ac}
M.J.~Dolan, C.~Englert and M.~Spannowsky, \emph{{New Physics in LHC Higgs boson
  pair production}},
  \href{https://doi.org/10.1103/PhysRevD.87.055002}{\emph{Phys. Rev. D}
  {\bfseries 87} (2013) 055002}
  [\href{https://arxiv.org/abs/1210.8166}{{\ttfamily 1210.8166}}].

\bibitem{Papaefstathiou:2012qe}
A.~Papaefstathiou, L.L.~Yang and J.~Zurita, \emph{{Higgs boson pair production
  at the LHC in the $b \bar{b} W^+ W^-$ channel}},
  \href{https://doi.org/10.1103/PhysRevD.87.011301}{\emph{Phys. Rev. D}
  {\bfseries 87} (2013) 011301}
  [\href{https://arxiv.org/abs/1209.1489}{{\ttfamily 1209.1489}}].

\bibitem{Cao:2013si}
J.~Cao, Z.~Heng, L.~Shang, P.~Wan and J.M.~Yang, \emph{{Pair Production of a
  125 GeV Higgs Boson in MSSM and NMSSM at the LHC}},
  \href{https://doi.org/10.1007/JHEP04(2013)134}{\emph{JHEP} {\bfseries 04}
  (2013) 134} [\href{https://arxiv.org/abs/1301.6437}{{\ttfamily 1301.6437}}].

\bibitem{Goertz:2013kp}
F.~Goertz, A.~Papaefstathiou, L.L.~Yang and J.~Zurita, \emph{{Higgs Boson
  self-coupling measurements using ratios of cross sections}},
  \href{https://doi.org/10.1007/JHEP06(2013)016}{\emph{JHEP} {\bfseries 06}
  (2013) 016} [\href{https://arxiv.org/abs/1301.3492}{{\ttfamily 1301.3492}}].

\bibitem{Arbey:2013jla}
A.~Arbey, M.~Battaglia and F.~Mahmoudi, \emph{{Supersymmetric Heavy Higgs
  Bosons at the LHC}},
  \href{https://doi.org/10.1103/PhysRevD.88.015007}{\emph{Phys. Rev. D}
  {\bfseries 88} (2013) 015007}
  [\href{https://arxiv.org/abs/1303.7450}{{\ttfamily 1303.7450}}].

\bibitem{deFlorian:2013uza}
D.~de~Florian and J.~Mazzitelli, \emph{{Two-loop virtual corrections to Higgs
  pair production}},
  \href{https://doi.org/10.1016/j.physletb.2013.06.046}{\emph{Phys. Lett. B}
  {\bfseries 724} (2013) 306}
  [\href{https://arxiv.org/abs/1305.5206}{{\ttfamily 1305.5206}}].

\bibitem{Gupta:2013zza}
R.S.~Gupta, H.~Rzehak and J.D.~Wells, \emph{{How well do we need to measure the
  Higgs boson mass and self-coupling?}},
  \href{https://doi.org/10.1103/PhysRevD.88.055024}{\emph{Phys. Rev. D}
  {\bfseries 88} (2013) 055024}
  [\href{https://arxiv.org/abs/1305.6397}{{\ttfamily 1305.6397}}].

\bibitem{Ellwanger:2013ova}
U.~Ellwanger, \emph{{Higgs pair production in the NMSSM at the LHC}},
  \href{https://doi.org/10.1007/JHEP08(2013)077}{\emph{JHEP} {\bfseries 08}
  (2013) 077} [\href{https://arxiv.org/abs/1306.5541}{{\ttfamily 1306.5541}}].

\bibitem{Barr:2013tda}
A.J.~Barr, M.J.~Dolan, C.~Englert and M.~Spannowsky, \emph{{Di-Higgs final
  states augMT2ed -- selecting $hh$ events at the high luminosity LHC}},
  \href{https://doi.org/10.1016/j.physletb.2013.12.011}{\emph{Phys. Lett. B}
  {\bfseries 728} (2014) 308}
  [\href{https://arxiv.org/abs/1309.6318}{{\ttfamily 1309.6318}}].

\bibitem{Maierhofer:2013sha}
P.~Maierh\"ofer and A.~Papaefstathiou, \emph{{Higgs Boson pair production
  merged to one jet}},
  \href{https://doi.org/10.1007/JHEP03(2014)126}{\emph{JHEP} {\bfseries 03}
  (2014) 126} [\href{https://arxiv.org/abs/1401.0007}{{\ttfamily 1401.0007}}].

\bibitem{deFlorian:2013jea}
D.~de~Florian and J.~Mazzitelli, \emph{{Higgs Boson Pair Production at
  Next-to-Next-to-Leading Order in QCD}},
  \href{https://doi.org/10.1103/PhysRevLett.111.201801}{\emph{Phys. Rev. Lett.}
  {\bfseries 111} (2013) 201801}
  [\href{https://arxiv.org/abs/1309.6594}{{\ttfamily 1309.6594}}].

\bibitem{Dolan:2013rja}
M.J.~Dolan, C.~Englert, N.~Greiner and M.~Spannowsky, \emph{{Further on up the
  road: $hhjj$ production at the LHC}},
  \href{https://doi.org/10.1103/PhysRevLett.112.101802}{\emph{Phys. Rev. Lett.}
  {\bfseries 112} (2014) 101802}
  [\href{https://arxiv.org/abs/1310.1084}{{\ttfamily 1310.1084}}].

\bibitem{Goertz:2013eka}
F.~Goertz, A.~Papaefstathiou, L.L.~Yang and J.~Zurita, \emph{{Measuring the
  Higgs boson self-coupling at the LHC using ratios of cross sections}},  in
  \emph{{25th Rencontres de Blois on Particle Physics and Cosmology}}, 9, 2013
  [\href{https://arxiv.org/abs/1309.3805}{{\ttfamily 1309.3805}}].

\bibitem{Frederix:2014hta}
R.~Frederix, S.~Frixione, V.~Hirschi, F.~Maltoni, O.~Mattelaer, P.~Torrielli
  et~al., \emph{{Higgs pair production at the LHC with NLO and parton-shower
  effects}}, \href{https://doi.org/10.1016/j.physletb.2014.03.026}{\emph{Phys.
  Lett. B} {\bfseries 732} (2014) 142}
  [\href{https://arxiv.org/abs/1401.7340}{{\ttfamily 1401.7340}}].

\bibitem{Baglio:2014nea}
J.~Baglio, O.~Eberhardt, U.~Nierste and M.~Wiebusch, \emph{{Benchmarks for
  Higgs Pair Production and Heavy Higgs boson Searches in the Two-Higgs-Doublet
  Model of Type II}},
  \href{https://doi.org/10.1103/PhysRevD.90.015008}{\emph{Phys. Rev. D}
  {\bfseries 90} (2014) 015008}
  [\href{https://arxiv.org/abs/1403.1264}{{\ttfamily 1403.1264}}].

\bibitem{FerreiradeLima:2014qkf}
D.E.~Ferreira~de Lima, A.~Papaefstathiou and M.~Spannowsky, \emph{{Standard
  model Higgs boson pair production in the $(b\bar{b})(b\bar{b})$ final
  state}}, \href{https://doi.org/10.1007/JHEP08(2014)030}{\emph{JHEP}
  {\bfseries 08} (2014) 030} [\href{https://arxiv.org/abs/1404.7139}{{\ttfamily
  1404.7139}}].

\bibitem{deFlorian:2014rta}
D.~de~Florian and J.~Mazzitelli, \emph{{Next-to-Next-to-Leading Order QCD
  Corrections to Higgs Boson Pair Production}},
  \href{https://doi.org/10.22323/1.211.0029}{\emph{PoS} {\bfseries LL2014}
  (2014) 029} [\href{https://arxiv.org/abs/1405.4704}{{\ttfamily 1405.4704}}].

\bibitem{Hespel:2014sla}
B.~Hespel, D.~Lopez-Val and E.~Vryonidou, \emph{{Higgs pair production via
  gluon fusion in the Two-Higgs-Doublet Model}},
  \href{https://doi.org/10.1007/JHEP09(2014)124}{\emph{JHEP} {\bfseries 09}
  (2014) 124} [\href{https://arxiv.org/abs/1407.0281}{{\ttfamily 1407.0281}}].

\bibitem{Barger:2014taa}
V.~Barger, L.L.~Everett, C.B.~Jackson, A.D.~Peterson and G.~Shaughnessy,
  \emph{{New physics in resonant production of Higgs boson pairs}},
  \href{https://doi.org/10.1103/PhysRevLett.114.011801}{\emph{Phys. Rev. Lett.}
  {\bfseries 114} (2015) 011801}
  [\href{https://arxiv.org/abs/1408.0003}{{\ttfamily 1408.0003}}].

\bibitem{Godunov:2014waa}
S.I.~Godunov, M.I.~Vysotsky and E.V.~Zhemchugov, \emph{{Double Higgs production
  at LHC, see-saw type II and Georgi-Machacek model}},
  \href{https://doi.org/10.1134/S1063776115030073}{\emph{J. Exp. Theor. Phys.}
  {\bfseries 120} (2015) 369}
  [\href{https://arxiv.org/abs/1408.0184}{{\ttfamily 1408.0184}}].

\bibitem{Liu:2014rba}
N.~Liu, S.~Hu, B.~Yang and J.~Han, \emph{{Impact of top-Higgs couplings on
  Di-Higgs production at future colliders}},
  \href{https://doi.org/10.1007/JHEP01(2015)008}{\emph{JHEP} {\bfseries 01}
  (2015) 008} [\href{https://arxiv.org/abs/1408.4191}{{\ttfamily 1408.4191}}].

\bibitem{Maltoni:2014eza}
F.~Maltoni, E.~Vryonidou and M.~Zaro, \emph{{Top-quark mass effects in double
  and triple Higgs production in gluon-gluon fusion at NLO}},
  \href{https://doi.org/10.1007/JHEP11(2014)079}{\emph{JHEP} {\bfseries 11}
  (2014) 079} [\href{https://arxiv.org/abs/1408.6542}{{\ttfamily 1408.6542}}].

\bibitem{Chen:2014ask}
C.-Y.~Chen, S.~Dawson and I.M.~Lewis, \emph{{Exploring resonant di-Higgs boson
  production in the Higgs singlet model}},
  \href{https://doi.org/10.1103/PhysRevD.91.035015}{\emph{Phys. Rev. D}
  {\bfseries 91} (2015) 035015}
  [\href{https://arxiv.org/abs/1410.5488}{{\ttfamily 1410.5488}}].

\bibitem{Barr:2014sga}
A.J.~Barr, M.J.~Dolan, C.~Englert, D.E.~Ferreira~de Lima and M.~Spannowsky,
  \emph{{Higgs Self-Coupling Measurements at a 100 TeV Hadron Collider}},
  \href{https://doi.org/10.1007/JHEP02(2015)016}{\emph{JHEP} {\bfseries 02}
  (2015) 016} [\href{https://arxiv.org/abs/1412.7154}{{\ttfamily 1412.7154}}].

\bibitem{MartinLozano:2015vtq}
V.~Mart\'\i{}n~Lozano, J.M.~Moreno and C.B.~Park, \emph{{Resonant Higgs boson
  pair production in the $ hh\to b\overline{b}\ WW\to
  b\overline{b}{\ell}^{+}\nu {\ell}^{-}\overline{\nu} $ decay channel}},
  \href{https://doi.org/10.1007/JHEP08(2015)004}{\emph{JHEP} {\bfseries 08}
  (2015) 004} [\href{https://arxiv.org/abs/1501.03799}{{\ttfamily
  1501.03799}}].

\bibitem{Papaefstathiou:2015iba}
A.~Papaefstathiou, \emph{{Discovering Higgs boson pair production through rare
  final states at a 100 TeV collider}},
  \href{https://doi.org/10.1103/PhysRevD.91.113016}{\emph{Phys. Rev. D}
  {\bfseries 91} (2015) 113016}
  [\href{https://arxiv.org/abs/1504.04621}{{\ttfamily 1504.04621}}].

\bibitem{Dawson:2015oha}
S.~Dawson, A.~Ismail and I.~Low, \emph{{What\textquoteright{}s in the loop? The
  anatomy of double Higgs production}},
  \href{https://doi.org/10.1103/PhysRevD.91.115008}{\emph{Phys. Rev. D}
  {\bfseries 91} (2015) 115008}
  [\href{https://arxiv.org/abs/1504.05596}{{\ttfamily 1504.05596}}].

\bibitem{Kotwal:2015rba}
A.V.~Kotwal, S.~Chekanov and M.~Low, \emph{{Double Higgs Boson Production in
  the 4$\tau$ Channel from Resonances in Longitudinal Vector Boson Scattering
  at a 100 TeV Collider}},
  \href{https://doi.org/10.1103/PhysRevD.91.114018}{\emph{Phys. Rev. D}
  {\bfseries 91} (2015) 114018}
  [\href{https://arxiv.org/abs/1504.08042}{{\ttfamily 1504.08042}}].

\bibitem{Lu:2015jza}
C.-T.~Lu, J.~Chang, K.~Cheung and J.S.~Lee, \emph{{An exploratory study of
  Higgs-boson pair production}},
  \href{https://doi.org/10.1007/JHEP08(2015)133}{\emph{JHEP} {\bfseries 08}
  (2015) 133} [\href{https://arxiv.org/abs/1505.00957}{{\ttfamily
  1505.00957}}].

\bibitem{Carvalho:2015ttv}
A.~Carvalho, M.~Dall'Osso, T.~Dorigo, F.~Goertz, C.A.~Gottardo and M.~Tosi,
  \emph{{Higgs Pair Production: Choosing Benchmarks With Cluster Analysis}},
  \href{https://doi.org/10.1007/JHEP04(2016)126}{\emph{JHEP} {\bfseries 04}
  (2016) 126} [\href{https://arxiv.org/abs/1507.02245}{{\ttfamily
  1507.02245}}].

\bibitem{Cao:2015oxx}
Q.-H.~Cao, Y.~Liu and B.~Yan, \emph{{Measuring trilinear Higgs coupling in WHH
  and ZHH productions at the high-luminosity LHC}},
  \href{https://doi.org/10.1103/PhysRevD.95.073006}{\emph{Phys. Rev. D}
  {\bfseries 95} (2017) 073006}
  [\href{https://arxiv.org/abs/1511.03311}{{\ttfamily 1511.03311}}].

\bibitem{Batell:2015koa}
B.~Batell, M.~McCullough, D.~Stolarski and C.B.~Verhaaren, \emph{{Putting a
  Stop to di-Higgs Modifications}},
  \href{https://doi.org/10.1007/JHEP09(2015)216}{\emph{JHEP} {\bfseries 09}
  (2015) 216} [\href{https://arxiv.org/abs/1508.01208}{{\ttfamily
  1508.01208}}].

\bibitem{Dawson:2015haa}
S.~Dawson and I.M.~Lewis, \emph{{NLO corrections to double Higgs boson
  production in the Higgs singlet model}},
  \href{https://doi.org/10.1103/PhysRevD.92.094023}{\emph{Phys. Rev. D}
  {\bfseries 92} (2015) 094023}
  [\href{https://arxiv.org/abs/1508.05397}{{\ttfamily 1508.05397}}].

\bibitem{Cao:2015oaa}
Q.-H.~Cao, B.~Yan, D.-M.~Zhang and H.~Zhang, \emph{{Resolving the Degeneracy in
  Single Higgs Production with Higgs Pair Production}},
  \href{https://doi.org/10.1016/j.physletb.2015.11.045}{\emph{Phys. Lett. B}
  {\bfseries 752} (2016) 285}
  [\href{https://arxiv.org/abs/1508.06512}{{\ttfamily 1508.06512}}].

\bibitem{Kanemura:2016tan}
S.~Kanemura, K.~Kaneta, N.~Machida, S.~Odori and T.~Shindou, \emph{{Single and
  double production of the Higgs boson at hadron and lepton colliders in
  minimal composite Higgs models}},
  \href{https://doi.org/10.1103/PhysRevD.94.015028}{\emph{Phys. Rev. D}
  {\bfseries 94} (2016) 015028}
  [\href{https://arxiv.org/abs/1603.05588}{{\ttfamily 1603.05588}}].

\bibitem{Contino:2016spe}
R.~Contino et~al., \emph{{Physics at a 100 TeV pp collider: Higgs and EW
  symmetry breaking studies}},
  \href{https://arxiv.org/abs/1606.09408}{{\ttfamily 1606.09408}}.

\bibitem{Cao:2016zob}
Q.-H.~Cao, G.~Li, B.~Yan, D.-M.~Zhang and H.~Zhang, \emph{{Double Higgs
  production at the 14 TeV LHC and a 100 TeV $pp$ collider}},
  \href{https://doi.org/10.1103/PhysRevD.96.095031}{\emph{Phys. Rev. D}
  {\bfseries 96} (2017) 095031}
  [\href{https://arxiv.org/abs/1611.09336}{{\ttfamily 1611.09336}}].

\bibitem{Banerjee:2016nzb}
S.~Banerjee, B.~Batell and M.~Spannowsky, \emph{{Invisible decays in Higgs
  boson pair production}},
  \href{https://doi.org/10.1103/PhysRevD.95.035009}{\emph{Phys. Rev. D}
  {\bfseries 95} (2017) 035009}
  [\href{https://arxiv.org/abs/1608.08601}{{\ttfamily 1608.08601}}].

\bibitem{Huang:2017jws}
T.~Huang, J.M.~No, L.~Perni\'e, M.~Ramsey-Musolf, A.~Safonov, M.~Spannowsky
  et~al., \emph{{Resonant di-Higgs boson production in the $b{\bar b}WW$
  channel: Probing the electroweak phase transition at the LHC}},
  \href{https://doi.org/10.1103/PhysRevD.96.035007}{\emph{Phys. Rev. D}
  {\bfseries 96} (2017) 035007}
  [\href{https://arxiv.org/abs/1701.04442}{{\ttfamily 1701.04442}}].

\bibitem{Nakamura:2017irk}
K.~Nakamura, K.~Nishiwaki, K.-y.~Oda, S.C.~Park and Y.~Yamamoto,
  \emph{{Di-higgs enhancement by neutral scalar as probe of new colored
  sector}}, \href{https://doi.org/10.1140/epjc/s10052-017-4835-4}{\emph{Eur.
  Phys. J. C} {\bfseries 77} (2017) 273}
  [\href{https://arxiv.org/abs/1701.06137}{{\ttfamily 1701.06137}}].

\bibitem{Lewis:2017dme}
I.M.~Lewis and M.~Sullivan, \emph{{Benchmarks for Double Higgs Production in
  the Singlet Extended Standard Model at the LHC}},
  \href{https://doi.org/10.1103/PhysRevD.96.035037}{\emph{Phys. Rev. D}
  {\bfseries 96} (2017) 035037}
  [\href{https://arxiv.org/abs/1701.08774}{{\ttfamily 1701.08774}}].

\bibitem{DiLuzio:2017tfn}
L.~Di~Luzio, R.~Gr\"ober and M.~Spannowsky, \emph{{Maxi-sizing the trilinear
  Higgs self-coupling: how large could it be?}},
  \href{https://doi.org/10.1140/epjc/s10052-017-5361-0}{\emph{Eur. Phys. J. C}
  {\bfseries 77} (2017) 788}
  [\href{https://arxiv.org/abs/1704.02311}{{\ttfamily 1704.02311}}].

\bibitem{Grober:2017gut}
R.~Grober, M.~Muhlleitner and M.~Spira, \emph{{Higgs Pair Production at NLO QCD
  for CP-violating Higgs Sectors}},
  \href{https://doi.org/10.1016/j.nuclphysb.2017.10.002}{\emph{Nucl. Phys. B}
  {\bfseries 925} (2017) 1} [\href{https://arxiv.org/abs/1705.05314}{{\ttfamily
  1705.05314}}].

\bibitem{Zurita:2017sfg}
J.~Zurita, \emph{{Di-Higgs production at the LHC and beyond}},  in \emph{{5th
  Large Hadron Collider Physics Conference}}, 8, 2017
  [\href{https://arxiv.org/abs/1708.00892}{{\ttfamily 1708.00892}}].

\bibitem{Arganda:2017wjh}
E.~Arganda, J.L.~D\'\i{}az-Cruz, N.~Mileo, R.A.~Morales and A.~Szynkman,
  \emph{{Search strategies for pair production of heavy Higgs bosons decaying
  invisibly at the LHC}},
  \href{https://doi.org/10.1016/j.nuclphysb.2018.02.004}{\emph{Nucl. Phys. B}
  {\bfseries 929} (2018) 171}
  [\href{https://arxiv.org/abs/1710.07254}{{\ttfamily 1710.07254}}].

\bibitem{Adhikary:2017jtu}
A.~Adhikary, S.~Banerjee, R.K.~Barman, B.~Bhattacherjee and S.~Niyogi,
  \emph{{Revisiting the non-resonant Higgs pair production at the HL-LHC}},
  \href{https://doi.org/10.1007/JHEP07(2018)116}{\emph{JHEP} {\bfseries 07}
  (2018) 116} [\href{https://arxiv.org/abs/1712.05346}{{\ttfamily
  1712.05346}}].

\bibitem{Bauer:2017cov}
M.~Bauer, M.~Carena and A.~Carmona, \emph{{Higgs Pair Production as a Signal of
  Enhanced Yukawa Couplings}},
  \href{https://doi.org/10.1103/PhysRevLett.121.021801}{\emph{Phys. Rev. Lett.}
  {\bfseries 121} (2018) 021801}
  [\href{https://arxiv.org/abs/1801.00363}{{\ttfamily 1801.00363}}].

\bibitem{Maltoni:2018ttu}
F.~Maltoni, D.~Pagani and X.~Zhao, \emph{{Constraining the Higgs self-couplings
  at e$^{+}$e$^{-}$ colliders}},
  \href{https://doi.org/10.1007/JHEP07(2018)087}{\emph{JHEP} {\bfseries 07}
  (2018) 087} [\href{https://arxiv.org/abs/1802.07616}{{\ttfamily
  1802.07616}}].

\bibitem{Borowka:2018pxx}
S.~Borowka, C.~Duhr, F.~Maltoni, D.~Pagani, A.~Shivaji and X.~Zhao,
  \emph{{Probing the scalar potential via double Higgs boson production at
  hadron colliders}},
  \href{https://doi.org/10.1007/JHEP04(2019)016}{\emph{JHEP} {\bfseries 04}
  (2019) 016} [\href{https://arxiv.org/abs/1811.12366}{{\ttfamily
  1811.12366}}].

\bibitem{Goncalves:2018qas}
D.~Gon\c{c}alves, T.~Han, F.~Kling, T.~Plehn and M.~Takeuchi, \emph{{Higgs
  boson pair production at future hadron colliders: From kinematics to
  dynamics}}, \href{https://doi.org/10.1103/PhysRevD.97.113004}{\emph{Phys.
  Rev. D} {\bfseries 97} (2018) 113004}
  [\href{https://arxiv.org/abs/1802.04319}{{\ttfamily 1802.04319}}].

\bibitem{Chang:2018uwu}
J.~Chang, K.~Cheung, J.S.~Lee, C.-T.~Lu and J.~Park, \emph{{Higgs-boson-pair
  production
  H(\textrightarrow{}bb\textasciimacron{})H(\textrightarrow{}\ensuremath{\gamma}\ensuremath{\gamma})
  from gluon fusion at the HL-LHC and HL-100 TeV hadron collider}},
  \href{https://doi.org/10.1103/PhysRevD.100.096001}{\emph{Phys. Rev. D}
  {\bfseries 100} (2019) 096001}
  [\href{https://arxiv.org/abs/1804.07130}{{\ttfamily 1804.07130}}].

\bibitem{Basler:2018dac}
P.~Basler, S.~Dawson, C.~Englert and M.~M\"uhlleitner, \emph{{Showcasing HH
  production: Benchmarks for the LHC and HL-LHC}},
  \href{https://doi.org/10.1103/PhysRevD.99.055048}{\emph{Phys. Rev. D}
  {\bfseries 99} (2019) 055048}
  [\href{https://arxiv.org/abs/1812.03542}{{\ttfamily 1812.03542}}].

\bibitem{Adhikary:2018ise}
A.~Adhikary, S.~Banerjee, R.~Kumar~Barman and B.~Bhattacherjee, \emph{{Resonant
  heavy Higgs searches at the HL-LHC}},
  \href{https://doi.org/10.1007/JHEP09(2019)068}{\emph{JHEP} {\bfseries 09}
  (2019) 068} [\href{https://arxiv.org/abs/1812.05640}{{\ttfamily
  1812.05640}}].

\bibitem{DiMicco:2019ngk}
J.~Alison et~al., \emph{{Higgs boson potential at colliders: Status and
  perspectives}}, \href{https://doi.org/10.1016/j.revip.2020.100045}{\emph{Rev.
  Phys.} {\bfseries 5} (2020) 100045}
  [\href{https://arxiv.org/abs/1910.00012}{{\ttfamily 1910.00012}}].

\bibitem{Li:2019uyy}
G.~Li, L.-X.~Xu, B.~Yan and C.P.~Yuan, \emph{{Resolving the degeneracy in top
  quark Yukawa coupling with Higgs pair production}},
  \href{https://doi.org/10.1016/j.physletb.2019.135070}{\emph{Phys. Lett. B}
  {\bfseries 800} (2020) 135070}
  [\href{https://arxiv.org/abs/1904.12006}{{\ttfamily 1904.12006}}].

\bibitem{Cheung:2020xij}
K.~Cheung, A.~Jueid, C.-T.~Lu, J.~Song and Y.W.~Yoon, \emph{{Disentangling new
  physics effects on nonresonant Higgs boson pair production from gluon
  fusion}}, \href{https://doi.org/10.1103/PhysRevD.103.015019}{\emph{Phys. Rev.
  D} {\bfseries 103} (2021) 015019}
  [\href{https://arxiv.org/abs/2003.11043}{{\ttfamily 2003.11043}}].

\bibitem{Plehn:2005nk}
T.~Plehn and M.~Rauch, \emph{{The quartic higgs coupling at hadron colliders}},
  \href{https://doi.org/10.1103/PhysRevD.72.053008}{\emph{Phys. Rev. D}
  {\bfseries 72} (2005) 053008}
  [\href{https://arxiv.org/abs/hep-ph/0507321}{{\ttfamily hep-ph/0507321}}].

\bibitem{Papaefstathiou:2015paa}
A.~Papaefstathiou and K.~Sakurai, \emph{{Triple Higgs boson production at a 100
  TeV proton-proton collider}},
  \href{https://doi.org/10.1007/JHEP02(2016)006}{\emph{JHEP} {\bfseries 02}
  (2016) 006} [\href{https://arxiv.org/abs/1508.06524}{{\ttfamily
  1508.06524}}].

\bibitem{Chen:2015gva}
C.-Y.~Chen, Q.-S.~Yan, X.~Zhao, Y.-M.~Zhong and Z.~Zhao, \emph{{Probing
  triple-Higgs productions via 4b2\ensuremath{\gamma} decay channel at a 100
  TeV hadron collider}},
  \href{https://doi.org/10.1103/PhysRevD.93.013007}{\emph{Phys. Rev. D}
  {\bfseries 93} (2016) 013007}
  [\href{https://arxiv.org/abs/1510.04013}{{\ttfamily 1510.04013}}].

\bibitem{Fuks:2015hna}
B.~Fuks, J.H.~Kim and S.J.~Lee, \emph{{Probing Higgs self-interactions in
  proton-proton collisions at a center-of-mass energy of 100 TeV}},
  \href{https://doi.org/10.1103/PhysRevD.93.035026}{\emph{Phys. Rev. D}
  {\bfseries 93} (2016) 035026}
  [\href{https://arxiv.org/abs/1510.07697}{{\ttfamily 1510.07697}}].

\bibitem{Papaefstathiou:2017hsb}
A.~Papaefstathiou, \emph{{Multi-Higgs Boson Production and Self-coupling
  Measurements at Hadron Colliders}},
  \href{https://doi.org/10.5506/APhysPolB.48.1133}{\emph{Acta Phys. Polon. B}
  {\bfseries 48} (2017) 1133}.

\bibitem{Fuks:2017zkg}
B.~Fuks, J.H.~Kim and S.J.~Lee, \emph{{Scrutinizing the Higgs quartic coupling
  at a future 100 TeV proton\textendash{}proton collider with taus and
  b-jets}}, \href{https://doi.org/10.1016/j.physletb.2017.05.075}{\emph{Phys.
  Lett. B} {\bfseries 771} (2017) 354}
  [\href{https://arxiv.org/abs/1704.04298}{{\ttfamily 1704.04298}}].

\bibitem{Liu:2018peg}
T.~Liu, K.-F.~Lyu, J.~Ren and H.X.~Zhu, \emph{{Probing the quartic Higgs boson
  self-interaction}},
  \href{https://doi.org/10.1103/PhysRevD.98.093004}{\emph{Phys. Rev. D}
  {\bfseries 98} (2018) 093004}
  [\href{https://arxiv.org/abs/1803.04359}{{\ttfamily 1803.04359}}].

\bibitem{Papaefstathiou:2019ofh}
A.~Papaefstathiou, G.~Tetlalmatzi-Xolocotzi and M.~Zaro, \emph{{Triple Higgs
  boson production to six $b$-jets at a 100 TeV proton collider}},
  \href{https://doi.org/10.1140/epjc/s10052-019-7457-1}{\emph{Eur. Phys. J. C}
  {\bfseries 79} (2019) 947}
  [\href{https://arxiv.org/abs/1909.09166}{{\ttfamily 1909.09166}}].

\bibitem{deFlorian:2019app}
D.~de~Florian, I.~Fabre and J.~Mazzitelli, \emph{{Triple Higgs production at
  hadron colliders at NNLO in QCD}},
  \href{https://doi.org/10.1007/JHEP03(2020)155}{\emph{JHEP} {\bfseries 03}
  (2020) 155} [\href{https://arxiv.org/abs/1912.02760}{{\ttfamily
  1912.02760}}].

\bibitem{Chiesa:2020awd}
M.~Chiesa, F.~Maltoni, L.~Mantani, B.~Mele, F.~Piccinini and X.~Zhao,
  \emph{{Measuring the quartic Higgs self-coupling at a multi-TeV muon
  collider}}, \href{https://doi.org/10.1007/JHEP09(2020)098}{\emph{JHEP}
  {\bfseries 09} (2020) 098}
  [\href{https://arxiv.org/abs/2003.13628}{{\ttfamily 2003.13628}}].

\bibitem{Abdughani:2020xfo}
M.~Abdughani, D.~Wang, L.~Wu, J.M.~Yang and J.~Zhao, \emph{{Probing the triple
  Higgs boson coupling with machine learning at the LHC}},
  \href{https://doi.org/10.1103/PhysRevD.104.056003}{\emph{Phys. Rev. D}
  {\bfseries 104} (2021) 056003}
  [\href{https://arxiv.org/abs/2005.11086}{{\ttfamily 2005.11086}}].

\bibitem{Papaefstathiou:2020lyp}
A.~Papaefstathiou, T.~Robens and G.~Tetlalmatzi-Xolocotzi, \emph{{Triple Higgs
  Boson Production at the Large Hadron Collider with Two Real Singlet
  Scalars}}, \href{https://doi.org/10.1007/JHEP05(2021)193}{\emph{JHEP}
  {\bfseries 05} (2021) 193}
  [\href{https://arxiv.org/abs/2101.00037}{{\ttfamily 2101.00037}}].

\bibitem{Stylianou:2023xit}
P.~Stylianou and G.~Weiglein, \emph{{Constraints on the trilinear and quartic
  Higgs couplings from triple Higgs production at the LHC and beyond}},
  \href{https://arxiv.org/abs/2312.04646}{{\ttfamily 2312.04646}}.

\bibitem{Degrande:2020evl}
C.~Degrande, G.~Durieux, F.~Maltoni, K.~Mimasu, E.~Vryonidou and C.~Zhang,
  \emph{{Automated one-loop computations in the standard model effective field
  theory}}, \href{https://doi.org/10.1103/PhysRevD.103.096024}{\emph{Phys. Rev.
  D} {\bfseries 103} (2021) 096024}
  [\href{https://arxiv.org/abs/2008.11743}{{\ttfamily 2008.11743}}].

\bibitem{Goertz:2014qta}
F.~Goertz, A.~Papaefstathiou, L.L.~Yang and J.~Zurita, \emph{{Higgs boson pair
  production in the D=6 extension of the SM}},
  \href{https://doi.org/10.1007/JHEP04(2015)167}{\emph{JHEP} {\bfseries 04}
  (2015) 167} [\href{https://arxiv.org/abs/1410.3471}{{\ttfamily 1410.3471}}].

\bibitem{Azatov:2015oxa}
A.~Azatov, R.~Contino, G.~Panico and M.~Son, \emph{{Effective field theory
  analysis of double Higgs boson production via gluon fusion}},
  \href{https://doi.org/10.1103/PhysRevD.92.035001}{\emph{Phys. Rev. D}
  {\bfseries 92} (2015) 035001}
  [\href{https://arxiv.org/abs/1502.00539}{{\ttfamily 1502.00539}}].

\bibitem{Alwall:2011uj}
J.~Alwall, M.~Herquet, F.~Maltoni, O.~Mattelaer and T.~Stelzer, \emph{{MadGraph
  5 : Going Beyond}},
  \href{https://doi.org/10.1007/JHEP06(2011)128}{\emph{JHEP} {\bfseries 06}
  (2011) 128} [\href{https://arxiv.org/abs/1106.0522}{{\ttfamily 1106.0522}}].

\bibitem{Hirschi:2015iia}
V.~Hirschi and O.~Mattelaer, \emph{{Automated event generation for loop-induced
  processes}}, \href{https://doi.org/10.1007/JHEP10(2015)146}{\emph{JHEP}
  {\bfseries 10} (2015) 146}
  [\href{https://arxiv.org/abs/1507.00020}{{\ttfamily 1507.00020}}].

\bibitem{loopxtree}
V.~Hirschi, ``{Computing the interference of loop-induced diagrams with a
  tree-level background with MadEvent in MG5\_aMC}.''
  https://cp3.irmp.ucl.ac.be/projects/madgraph/wiki/LoopInducedTimesTree, 2016.

\bibitem{Bailey:2020ooq}
S.~Bailey, T.~Cridge, L.A.~Harland-Lang, A.D.~Martin and R.S.~Thorne,
  \emph{{Parton distributions from LHC, HERA, Tevatron and fixed target data:
  MSHT20 PDFs}},
  \href{https://doi.org/10.1140/epjc/s10052-021-09057-0}{\emph{Eur. Phys. J. C}
  {\bfseries 81} (2021) 341}
  [\href{https://arxiv.org/abs/2012.04684}{{\ttfamily 2012.04684}}].

\bibitem{deBlas:2019rxi}
J.~de~Blas et~al., \emph{{Higgs Boson Studies at Future Particle Colliders}},
  \href{https://doi.org/10.1007/JHEP01(2020)139}{\emph{JHEP} {\bfseries 01}
  (2020) 139} [\href{https://arxiv.org/abs/1905.03764}{{\ttfamily
  1905.03764}}].

\bibitem{Bahr:2008pv}
M.~Bahr et~al., \emph{{Herwig++ Physics and Manual}},
  \href{https://doi.org/10.1140/epjc/s10052-008-0798-9}{\emph{Eur. Phys. J.}
  {\bfseries C58} (2008) 639}
  [\href{https://arxiv.org/abs/0803.0883}{{\ttfamily 0803.0883}}].

\bibitem{Gieseke:2011na}
S.~Gieseke et~al., \emph{{Herwig++ 2.5 Release Note}},
  \href{https://arxiv.org/abs/1102.1672}{{\ttfamily 1102.1672}}.

\bibitem{Arnold:2012fq}
K.~Arnold et~al., \emph{{Herwig++ 2.6 Release Note}},
  \href{https://arxiv.org/abs/1205.4902}{{\ttfamily 1205.4902}}.

\bibitem{Bellm:2013hwb}
J.~Bellm et~al., \emph{{Herwig++ 2.7 Release Note}},
  \href{https://arxiv.org/abs/1310.6877}{{\ttfamily 1310.6877}}.

\bibitem{Bellm:2015jjp}
J.~Bellm et~al., \emph{{Herwig 7.0/Herwig++ 3.0 release note}},
  \href{https://doi.org/10.1140/epjc/s10052-016-4018-8}{\emph{Eur. Phys. J.}
  {\bfseries C76} (2016) 196}
  [\href{https://arxiv.org/abs/1512.01178}{{\ttfamily 1512.01178}}].

\bibitem{Bellm:2017bvx}
J.~Bellm et~al., \emph{{Herwig 7.1 Release Note}},
  \href{https://arxiv.org/abs/1705.06919}{{\ttfamily 1705.06919}}.

\bibitem{Bellm:2019zci}
J.~Bellm et~al., \emph{{Herwig 7.2 release note}},
  \href{https://doi.org/10.1140/epjc/s10052-020-8011-x}{\emph{Eur. Phys. J. C}
  {\bfseries 80} (2020) 452}
  [\href{https://arxiv.org/abs/1912.06509}{{\ttfamily 1912.06509}}].

\bibitem{Bewick:2023tfi}
G.~Bewick et~al., \emph{{Herwig 7.3 Release Note}},
  \href{https://arxiv.org/abs/2312.05175}{{\ttfamily 2312.05175}}.

\bibitem{hwsim}
{Papaefstathiou, Andreas}, ``{The \texttt{HwSim} analysis package for HERWIG
  7}.'' {https://gitlab.com/apapaefs/hwsim}.

\bibitem{ATLAS:2020cli}
{\scshape ATLAS} collaboration, \emph{{Jet energy scale and resolution measured
  in proton\textendash{}proton collisions at $\sqrt{s}=13$~TeV with the ATLAS
  detector}}, \href{https://doi.org/10.1140/epjc/s10052-021-09402-3}{\emph{Eur.
  Phys. J. C} {\bfseries 81} (2021) 689}
  [\href{https://arxiv.org/abs/2007.02645}{{\ttfamily 2007.02645}}].

\bibitem{CMS-DP-2021-033}
{\scshape CMS} collaboration, \emph{{Jet energy scale and resolution
  measurement with Run~2 Legacy Data Collected by CMS at 13~TeV}}, .

\bibitem{Contino:2014aaa}
{\scshape HDECAY} collaboration, \emph{{eHDECAY: an Implementation of the Higgs
  Effective Lagrangian into HDECAY}},
  \href{https://doi.org/10.1016/j.cpc.2014.06.028}{\emph{Comput. Phys. Commun.}
  {\bfseries 185} (2014) 3412}
  [\href{https://arxiv.org/abs/1403.3381}{{\ttfamily 1403.3381}}].

\bibitem{yr4page}
CERN, ``{Yellow Report webpage}.''
  https://twiki.cern.ch/twiki/bin/view/LHCPhysics/CERNYellowReportPageBR, 2016.

\bibitem{LHCHiggsCrossSectionWorkingGroup:2016ypw}
{\scshape LHC Higgs Cross Section Working Group} collaboration, \emph{{Handbook
  of LHC Higgs Cross Sections: 4. Deciphering the Nature of the Higgs Sector}},
   \href{https://arxiv.org/abs/1610.07922}{{\ttfamily 1610.07922}}.

\bibitem{Pomarol:2013zra}
A.~Pomarol and F.~Riva, \emph{{Towards the Ultimate SM Fit to Close in on Higgs
  Physics}}, \href{https://doi.org/10.1007/JHEP01(2014)151}{\emph{JHEP}
  {\bfseries 01} (2014) 151} [\href{https://arxiv.org/abs/1308.2803}{{\ttfamily
  1308.2803}}].

\bibitem{Dumont:2013wma}
B.~Dumont, S.~Fichet and G.~von Gersdorff, \emph{{A Bayesian view of the Higgs
  sector with higher dimensional operators}},
  \href{https://doi.org/10.1007/JHEP07(2013)065}{\emph{JHEP} {\bfseries 07}
  (2013) 065} [\href{https://arxiv.org/abs/1304.3369}{{\ttfamily 1304.3369}}].

\bibitem{Corbett:2012dm}
T.~Corbett, O.J.P.~Eboli, J.~Gonzalez-Fraile and M.C.~Gonzalez-Garcia,
  \emph{{Constraining anomalous Higgs interactions}},
  \href{https://doi.org/10.1103/PhysRevD.86.075013}{\emph{Phys. Rev. D}
  {\bfseries 86} (2012) 075013}
  [\href{https://arxiv.org/abs/1207.1344}{{\ttfamily 1207.1344}}].

\bibitem{Corbett:2012ja}
T.~Corbett, O.J.P.~\'Eboli, J.~Gonzalez-Fraile and M.C.~Gonzalez-Garcia,
  \emph{{Robust Determination of the Higgs Couplings: Power to the Data}},
  \href{https://doi.org/10.1103/PhysRevD.87.015022}{\emph{Phys. Rev. D}
  {\bfseries 87} (2013) 015022}
  [\href{https://arxiv.org/abs/1211.4580}{{\ttfamily 1211.4580}}].

\bibitem{Corbett:2013pja}
T.~Corbett, O.J.P.~\'Eboli, J.~Gonzalez-Fraile and M.C.~Gonzalez-Garcia,
  \emph{{Determining Triple Gauge Boson Couplings from Higgs Data}},
  \href{https://doi.org/10.1103/PhysRevLett.111.011801}{\emph{Phys. Rev. Lett.}
  {\bfseries 111} 011801} [\href{https://arxiv.org/abs/1304.1151}{{\ttfamily
  1304.1151}}].

\bibitem{Contino:2013kra}
R.~Contino, M.~Ghezzi, C.~Grojean, M.~Muhlleitner and M.~Spira,
  \emph{{Effective Lagrangian for a light Higgs-like scalar}},
  \href{https://doi.org/10.1007/JHEP07(2013)035}{\emph{JHEP} {\bfseries 07}
  (2013) 035} [\href{https://arxiv.org/abs/1303.3876}{{\ttfamily 1303.3876}}].

\bibitem{Giudice:2007fh}
G.F.~Giudice, C.~Grojean, A.~Pomarol and R.~Rattazzi, \emph{{The
  Strongly-Interacting Light Higgs}},
  \href{https://doi.org/10.1088/1126-6708/2007/06/045}{\emph{JHEP} {\bfseries
  06} (2007) 045} [\href{https://arxiv.org/abs/hep-ph/0703164}{{\ttfamily
  hep-ph/0703164}}].

\bibitem{Maltoni:2016yxb}
F.~Maltoni, E.~Vryonidou and C.~Zhang, \emph{{Higgs production in association
  with a top-antitop pair in the Standard Model Effective Field Theory at NLO
  in QCD}}, \href{https://doi.org/10.1007/JHEP10(2016)123}{\emph{JHEP}
  {\bfseries 10} (2016) 123}
  [\href{https://arxiv.org/abs/1607.05330}{{\ttfamily 1607.05330}}].

\bibitem{Deutschmann:2017qum}
N.~Deutschmann, C.~Duhr, F.~Maltoni and E.~Vryonidou, \emph{{Gluon-fusion Higgs
  production in the Standard Model Effective Field Theory}},
  \href{https://doi.org/10.1007/JHEP12(2017)063}{\emph{JHEP} {\bfseries 12}
  (2017) 063} [\href{https://arxiv.org/abs/1708.00460}{{\ttfamily
  1708.00460}}].

\bibitem{DiNoi:2023ygk}
S.~Di~Noi, R.~Gr\"ober, G.~Heinrich, J.~Lang and M.~Vitti, \emph{{On $\gamma_5$
  schemes and the interplay of SMEFT operators in the Higgs-gluon coupling}},
  \href{https://arxiv.org/abs/2310.18221}{{\ttfamily 2310.18221}}.

\bibitem{Heinrich:2023rsd}
G.~Heinrich and J.~Lang, \emph{{Combining chromomagnetic and four-fermion
  operators with leading SMEFT operators for $gg\to hh$ at NLO QCD}},
  \href{https://arxiv.org/abs/2311.15004}{{\ttfamily 2311.15004}}.

\bibitem{Grober:2843280}
R.~Groeber, L.~Alasfar, L.~Cadamuro, C.~Dimitriadi, A.~Ferrari, G.M.~Heinrich
  et~al., \emph{{Effective Field Theory descriptions of Higgs boson pair
  production}},  Tech. Rep.
  \href{https://cds.cern.ch/record/2843280}{LHCHWG-2022-004}, CERN, Geneva
  (2022).

\bibitem{DAmbrosio:2002vsn}
G.~D'Ambrosio, G.F.~Giudice, G.~Isidori and A.~Strumia, \emph{{Minimal flavor
  violation: An Effective field theory approach}},
  \href{https://doi.org/10.1016/S0550-3213(02)00836-2}{\emph{Nucl. Phys. B}
  {\bfseries 645} (2002) 155}
  [\href{https://arxiv.org/abs/hep-ph/0207036}{{\ttfamily hep-ph/0207036}}].

\bibitem{Agashe:2004rs}
K.~Agashe, R.~Contino and A.~Pomarol, \emph{{The Minimal composite Higgs
  model}}, \href{https://doi.org/10.1016/j.nuclphysb.2005.04.035}{\emph{Nucl.
  Phys. B} {\bfseries 719} (2005) 165}
  [\href{https://arxiv.org/abs/hep-ph/0412089}{{\ttfamily hep-ph/0412089}}].

\bibitem{Contino:2006qr}
R.~Contino, L.~Da~Rold and A.~Pomarol, \emph{{Light custodians in natural
  composite Higgs models}},
  \href{https://doi.org/10.1103/PhysRevD.75.055014}{\emph{Phys. Rev. D}
  {\bfseries 75} (2007) 055014}
  [\href{https://arxiv.org/abs/hep-ph/0612048}{{\ttfamily hep-ph/0612048}}].

\bibitem{Contino:2010rs}
R.~Contino, \emph{{The Higgs as a Composite Nambu-Goldstone Boson}},  in
  \emph{{Theoretical Advanced Study Institute in Elementary Particle Physics}:
  {Physics of the Large and the Small}}, pp.~235--306, 2011,
  \href{https://doi.org/10.1142/9789814327183_0005}{DOI}
  [\href{https://arxiv.org/abs/1005.4269}{{\ttfamily 1005.4269}}].

\bibitem{Falkowski:2007hz}
A.~Falkowski, \emph{{Pseudo-goldstone Higgs production via gluon fusion}},
  \href{https://doi.org/10.1103/PhysRevD.77.055018}{\emph{Phys. Rev. D}
  {\bfseries 77} 055018} [\href{https://arxiv.org/abs/0711.0828}{{\ttfamily
  0711.0828}}].

\bibitem{Carena:2014ria}
M.~Carena, L.~Da~Rold and E.~Pont\'on, \emph{{Minimal Composite Higgs Models at
  the LHC}}, \href{https://doi.org/10.1007/JHEP06(2014)159}{\emph{JHEP}
  {\bfseries 06} (2014) 159} [\href{https://arxiv.org/abs/1402.2987}{{\ttfamily
  1402.2987}}].

\bibitem{Buchalla:2014eca}
G.~Buchalla, O.~Cata and C.~Krause, \emph{{A Systematic Approach to the SILH
  Lagrangian}},
  \href{https://doi.org/10.1016/j.nuclphysb.2015.03.024}{\emph{Nucl. Phys. B}
  {\bfseries 894} (2015) 602}
  [\href{https://arxiv.org/abs/1412.6356}{{\ttfamily 1412.6356}}].

\bibitem{Buchalla:2018yce}
G.~Buchalla, M.~Capozi, A.~Celis, G.~Heinrich and L.~Scyboz, \emph{{Higgs boson
  pair production in non-linear Effective Field Theory with full
  $m_t$-dependence at NLO QCD}},
  \href{https://doi.org/10.1007/JHEP09(2018)057}{\emph{JHEP} {\bfseries 09}
  (2018) 057} [\href{https://arxiv.org/abs/1806.05162}{{\ttfamily
  1806.05162}}].

\end{thebibliography}\endgroup

\end{document}